\documentclass[journal]{IEEEtran}
\usepackage{booktabs}
\usepackage{makecell}
\usepackage{multirow}
\usepackage{tabularx}
\usepackage{adjustbox}
\usepackage{amsmath}
\usepackage{array}
\newcolumntype{L}[1]{>{\raggedright\arraybackslash}p{#1}} % ragged-right p-col
\usepackage{graphicx}
\usepackage{caption}
\usepackage{subcaption}
\usepackage{amssymb}
\allowdisplaybreaks
\usepackage[table]{xcolor}
\usepackage{balance}

\usepackage[user]{zref}
\newcommand{\sot}[1]{}

\definecolor{pSec}{RGB}{226,232,240}   % slate  (section number)
\definecolor{pTit}{RGB}{255,236,204}   % apricot (title)
\definecolor{pCon}{RGB}{220,240,222}   % mint    (contents)
\definecolor{pRol}{RGB}{221,229,250}   % periwinkle (role)
\definecolor{pSecH}{RGB}{190,200,215}
\definecolor{pTitH}{RGB}{255,214,153}
\definecolor{pConH}{RGB}{178,222,184}
\definecolor{pRolH}{RGB}{180,196,240}
\newcommand{\cbox}[3]{\fcolorbox{#1!60!black!30}{#1}{\parbox{\dimexpr#2-2\fboxsep-2\fboxrule\relax}{\raggedright\scriptsize #3}}}
\newcounter{revc}
\makeatletter
\zref@newprop{revcontent}{} \zref@addprop{main}{revcontent}
\zref@newprop{revsec}{} \zref@addprop{main}{revsec}
\zref@newprop{revpage}{} \zref@addprop{main}{revpage}
\newcommand{\revi}[2]{%
\zref@setcurrent{revsec}{\thesection}%
\zref@setcurrent{revpage}{\thepage}%
\zref@setcurrent{revcontent}{#2}%
\refstepcounter{revc}%
\label{#1}%
\zlabel{#1}%
\textcolor{black}{#2}}
\newcommand{\revinu}[2]{%
\zref@setcurrent{revsec}{\thesection}%
\zref@setcurrent{revcontent}{#2}%
\refstepcounter{revc}%
\zlabel{#1}%
\label{#1}%
#2 }
\newcommand{\revr}[2]{%
\zref@setcurrent{revsec}{\thesection}%
\zref@setcurrent{revcontent}{#2}%
\refstepcounter{revc}%
\zlabel{#1}%
\label{#1} \sot{#2}}
\newenvironment{revblock}[1]{%
\zref@setcurrent{revsec}{\thesection}%
\zref@setcurrent{revpage}{\thepage}%
\zref@setcurrent{revcontent}{}%
\refstepcounter{revc}%
\label{#1}%
\zlabel{#1}%
\color{black}}{}
\makeatother

\usepackage{cases}
\usepackage{setspace}
\expandafter\def\expandafter\quote\expandafter{\quote\onehalfspacing\fontsize{11}{13}\selectfont}
\usepackage[framemethod=tikz]{mdframed}
\definecolor{mycolor}{rgb}{0.122, 0.435, 0.698}
\newmdenv[innerlinewidth=0.5pt, roundcorner=4pt,linecolor=mycolor,innerleftmargin=6pt,
innerrightmargin=6pt,innertopmargin=6pt,innerbottommargin=6pt]{mybox}
\usepackage{cite}
\usepackage[edges]{forest}
\usepackage{tikz}
\usetikzlibrary{arrows.meta}
\usepackage[hidelinks]{hyperref}
\hypersetup{colorlinks=true,urlcolor=black,linkcolor=black,citecolor=black,filecolor=black,breaklinks=true,bookmarks=true,bookmarksopen=false}

\begin{document}
\title{Joint Communications, Sensing, and Positioning\\in 6G Multi-Functional Satellite Systems:\\Survey and Open Challenges}

\author{Chandan~Kumar~Sheemar,~\IEEEmembership{Member,~IEEE}, Jorge~Querol,~\IEEEmembership{Member,~IEEE},
         Wali Ullah Khan,~\IEEEmembership{Member,~IEEE}, \\ Prabhu Thiruvasagam,~\IEEEmembership{Member,~IEEE}, Sourabh~Solanki,~\IEEEmembership{Member,~IEEE}, Idir Edjekouane,~\IEEEmembership{Senior Member,~IEEE},
        Alejandro~Gonzalez-Garrido,~\IEEEmembership{Student Member,~IEEE},  Mohammed Al-Ansi,~\IEEEmembership{Member,~IEEE}, \\Carla E. Garcia,~\IEEEmembership{Member,~IEEE},
and~Symeon~Chatzinotas,~\IEEEmembership{Fellow,~IEEE}% <-this % stops a space
\thanks{This work was supported by the Framework of Fonds National de la Recherche: SMS2 Project, funded by FNR under Contract
C-C24/IS/18957132/SMS2. The authors are with the SnT, University of Luxembourg, Luxembourg. emails: \{chandankumar.sheemar, jorge.querol, waliullah.khan, prabhu.thiruvasagam, alejandro.gonzalez, sourabh.solanki, idir.edjekouane, mohammed.al-ansi, carla.garcia, symeon.chatzinotas\}@uni.lu. }
}

\markboth{IEEE Communications Surveys \& Tutorials,~Vol.~XX, No.~XX,~quarter~202X}%
{Sheemar \MakeLowercase{\textit{et al.}}: Joint Communications, Sensing and Positioning in 6G Multi-Functional Satellite Systems: Survey and Open Challenges}

\maketitle

\begin{abstract}
Satellite systems are expected to be a cornerstone of sixth-generation (6G) networks, supporting communications, sensing, and positioning, navigation, and timing (PNT) with global coverage. This survey develops a framework for multi-functional satellite systems (MFSS), which combine at least two of these services through shared or jointly designed payload resources. A two-dimensional taxonomy distinguishes the functional configuration---joint communications and sensing (JCAS), joint communications and PNT (JCAP), joint sensing and PNT (JSAP), or joint communications, sensing, and PNT (JCSAP)---from the integration depth---cooperative, integrated, or joint. After consolidating the communication, sensing, and PNT fundamentals and payload architectures into a single background section, the survey examines the three integration levels and then devotes a dedicated section to each configuration, covering its architecture, classification, state of the art, advantages, key challenges, and research directions. The achievable performance of each configuration is illustrated with representative trade-off results from the literature and compared across configurations, the differences between terrestrial multi-functional networks and satellite MFSS are made explicit, and a life-cycle techno-economic framework balances hardware, spectrum, launch, and operational reuse against engineering, qualification, verification, and common-mode risks. The analysis shows that deeper integration can improve resource utilization and functional synergy, but is not automatically the technical or economic optimum for every mission.
\end{abstract}


\IEEEpeerreviewmaketitle
\section{Introduction}
Satellite systems are experiencing a resurgence of interest with the advent of sixth-generation (6G) wireless networks. Unlike previous generations, 6G envisions extensive coverage, ultra-low latency, and extremely high data rates capabilities that terrestrial infrastructure alone cannot deliver efficiently, particularly in remote, rural, or mobile environments \cite{Giordani2021Non-TerrestrialOpportunities}. To meet these ambitious goals, the integration of satellite systems into 6G architectures is being actively pursued, as satellites are uniquely positioned to facilitate seamless global connectivity across land, sea, and air \cite{Azari2022EvolutionSurvey}. Emerging technologies such as high-throughput satellites, non-geostationary orbit (NGSO) mega-constellations \cite{Ravishankar2021NextgenerationMegaconstellations}, and software-defined payloads are increasingly aligned with the performance and flexibility requirements of 6G. This growing interest reflects a broader shift toward integrated space-air-ground networks, offering a scalable and resilient solution to the evolving demands of future communication systems \cite{Khan2024ReconfigurableSky}.

Satellite systems play a fundamental role in modern global infrastructure by delivering critical capabilities across three principal verticals: communications, sensing, and positioning, navigation, and timing (PNT) \cite{Palacios2021ACommunications, 9705087, Prol2022PositionOpportunities}. Each of these domains addresses distinct functional requirements while collectively highlighting the versatility and strategic importance of satellite technologies. Communication satellites support a wide array of data transmission services, enabling long-range connectivity and extending network access to otherwise unreachable areas \cite{Khammassi2024PrecodingSurvey}. Sensing satellites, on the other hand, are pivotal in Earth observation, providing valuable data for weather forecasting, climate monitoring, environmental assessment, and disaster management \cite{Tsokas2022SAROverview}. Meanwhile, PNT satellites form the backbone of global navigation systems, offering highly accurate positioning and timing references essential for transportation, military operations, and synchronization of telecommunication networks \cite{Closas2024EmergingEditorial}.

The scale and distribution of satellite deployments across these verticals underscore the expanding reliance on space-based infrastructure. As of March 2025, public satellite catalogues \cite{UCSDatabase2023,McDowellGCAT} list approximately $8000$ active satellites dedicated to communication services, about $1000$ to Earth observation/sensing applications, and more than $100$ to PNT functions, operating across various orbital regimes. This proliferation reflects not only technological advancements but also the rising global demand for ubiquitous connectivity, high-resolution environmental data, and dependable navigation solutions. As the number of active satellites continues to grow, ensuring efficient coordination, spectrum usage, and orbital management becomes increasingly critical to sustaining the long-term viability of satellite-based services \cite{Xie2021LEOOpportunities}. Furthermore, an increased number of satellites in space also results in significant space debris, making an independent evolution of such technologies unsustainable \cite{Sanchez2017ReliabilityEnvironment}.

In parallel, the paradigms of joint communication and sensing (JCAS) \cite{Sheemar2023Full-Duplex-EnabledSurfaces,sheemar2024duplexjointcommunicationssensing} and joint communication and positioning (JCAP) \cite{Luo2024ResourceNetworks} are evolving rapidly within terrestrial networks, driven by the growing demand for intelligent, spectrum-efficient, and context-aware wireless systems. At the heart of this evolution lies the concept of co-design, where communication waveforms, infrastructure, and processing are reused or jointly optimized to extract physical parameters such as range, velocity, and angle of arrival while transporting data \cite{Liyanaarachchi2021OptimizedExperiments}. Techniques like radar-inspired channel estimation, passive localization using communication signals, and sensing-assisted beamforming are already being trialed in 5G-Advanced and are expected to become key enablers in 6G networks \cite{Feng2020JointSurvey, Mazahir2021ASystems}. Extending this convergence to satellite systems can unlock unprecedented capabilities in global connectivity, Earth observation, space-based situational awareness, and PNT, with shared infrastructure, spectrum, and hardware\footnote{A list of the main acronyms used in this paper is presented in Table \ref{tab:acronyms}.}.

\revi{rev1.5.1}{Since JCAS and JCAP are used side by side throughout this survey, their relationship deserves clarification at the outset. A large part of the integrated sensing and communication (ISAC) literature uses ``sensing'' in a broad sense that includes the localization of connected users, so that JCAS is often regarded as an umbrella term that also covers JCAP \cite{Liu2022ACommunication,Lohan2025EmergingCommunications}. In this survey, the two terms denote distinct functional configurations. JCAS refers to the extraction of properties of objects or of the environment (range, velocity, angle, reflectivity, or imagery) from the transmitted waveform, typically through echoes or reflections, whereas JCAP refers to the estimation of the position, navigation state, and time of a user or platform from downlink (or uplink) signals. The distinction is deliberate in the satellite context: PNT additionally requires precise clocks, ephemerides, navigation messages, integrity, continuity, and availability guarantees that target sensing does not, and its performance is governed by the satellite--user geometry (dilution of precision) rather than by target reflectivity. The two configurations therefore share signal-processing principles but impose different constraints on the waveform, the payload, and the ground segment, and JCSAP denotes the case in which both are combined with communications on the same platform.}

\begin{table*}[!t]
\centering
\caption{List of the main acronyms}
\scriptsize
\setlength{\tabcolsep}{4pt}
\begin{tabular}{l l | l l | l l}
\toprule
\textbf{Acronym} & \textbf{Definition} & \textbf{Acronym} & \textbf{Definition} & \textbf{Acronym} & \textbf{Definition} \\
\midrule
AFS & Atomic Frequency Standard & GDOP & Geometric Dilution of Precision & PNT & Positioning, Navigation, and Timing \\
AI & Artificial Intelligence & GEO & Geostationary Earth Orbit & PRS & Positioning Reference Signal \\
AOA & Angle of Arrival & GNSS & Global Navigation Satellite System & RSMA & Rate-Splitting Multiple Access \\
AOD & Angle of Departure & GSM & Global System for Mobile Communications & RTT & Round Trip Time \\
APEB & Average Position Error Bound & IoRT & Internet of Remote Things & SAINR & Signal-to-Ambiguity-Interference-plus-Noise Ratio \\
BH & Beam Hopping & ISAC & Integrated Sensing and Communication & SAR & Synthetic Aperture Radar \\
CRB & Cram\'{e}r--Rao Bound & ISL & Inter-Satellite Link & SBAS & Satellite-Based Augmentation System \\
CSI & Channel State Information & JCAP & Joint Communication and Positioning & SoO & Signals of Opportunity \\
DAC & Digital to Analog Converter & JCAS & Joint Communication and Sensing & SWaP-C & Size, Weight, Power, and Cost \\
DPU & Digital Processing Unit & JCSAP & Joint Communications, Sensing, and PNT & TDMA & Time Division Multiple Access \\
DSM & Dynamic Spectrum Management & JSAP & Joint Sensing and PNT & TDOA & Time Difference of Arrival \\
DSSS & Direct-Sequence Spread Spectrum & MFSS & Multi-Functional Satellite Systems & TN & Terrestrial Network \\
EO & Earth Observation & MI & Mutual Interference & TOA & Time of Arrival \\
EPFD & Equivalent Power Flux Density & NDU & Navigation Data Unit & TTC & Tracking, Telemetry, and Command \\
FDMA & Frequency Division Multiple Access & NTN & Non-Terrestrial Network & US & User Segment \\
FDOA & Frequency Difference of Arrival & OBDP & On-Board Digital Processor &  &  \\
FOA & Frequency of Arrival & OMA & Orthogonal Multiple Access &  &  \\
\bottomrule
\end{tabular}
\label{tab:acronyms}
\end{table*}

\subsection{Current Satellite Missions and Enabling Technologies}
\label{sec:missions}
\revi{rev1.6.1}{Current communication, sensing, and PNT missions are distinguished not only by the services they advertise but also by their mission objectives and by the technical mechanisms adopted to meet them. This subsection summarizes both aspects because they define the requirements that an MFSS must preserve when resources are shared.}

\revi{rev1.6.2}{\emph{Communication missions:} LEO mega-constellations such as SpaceX's Starlink \cite{Cakaj2021TheShells}, OneWeb \cite{Kozhaya2024APositioning}, Amazon's Project Kuiper \cite{Oughton2023SustainabilityMega-constellations}, the planned European IRIS$^2$ system, and regional efforts such as China's Hongyun \cite{Zhang2022AnConstellations} and Russia's Sphere \cite{Abashidze2022SatelliteAspects} pursue a common objective: high-throughput, low-latency, and geographically ubiquitous connectivity for residential and enterprise broadband, remote communities, maritime and aeronautical users, emergency response, IoT, and direct-to-device services \cite{Kumar20245GOpenAirInterface5G}. Three enabling techniques recur across these missions. Inter-satellite laser links are critical because they route traffic through the constellation when no gateway is in view, thereby extending service over oceans and remote regions, avoiding detours through terrestrial networks, reducing end-to-end latency, and improving resilience; their narrow optical beams also offer very high capacity with negligible interference, at the cost of demanding pointing, acquisition, and tracking. Adaptive RF beamforming allows the satellite to steer capacity toward high-demand regions and to suppress interference as traffic and geometry change. Software-defined networking and regenerative processing enable traffic-aware routing and in-orbit reconfiguration, which is what makes payload resources reusable for other functions.}

\revi{rev1.6.3}{\emph{Sensing missions:} Earth-observation missions optimize spatial resolution, swath, revisit time, coverage, and product latency rather than communication throughput. NASA--ISRO's NISAR provides L- and S-band radar imaging for surface-deformation, ecosystem, cryosphere, and disaster monitoring \cite{Kellogg2020NASA-ISROMission}; ESA's Copernicus programme, through its Sentinel fleet, delivers multispectral and radar data for land, ocean, and atmospheric monitoring \cite{Jutz2020Copernicus:Programme}; commercial constellations such as Planet \cite{Foster2018ConstellationSatellites} and the SAR operator ICEYE \cite{Muff2022TheAchievements} trade individual-platform capability for frequent revisits, supporting precision agriculture, deforestation tracking, emergency response, and urban planning. Finally, constellations such as Spire perform GNSS radio occultation: the bending and delay of navigation signals traversing the atmosphere and ionosphere are measured to retrieve profiles for weather prediction, climate monitoring, and space-weather analysis. Radio occultation is itself an early example of a navigation signal reused for sensing.}

\revi{rev1.6.4}{\emph{PNT missions:} GPS III \cite{Thoelert2019SignalSatellite}, Galileo \cite{Zhou2024InitialReceiver}, BeiDou-3 \cite{Yang2019IntroductionSystem}, NavIC \cite{Shekhar2020SensitivityLand}, and QZSS deliver more than coordinates: they provide navigation continuity, precise timing, integrity and augmentation information, and synchronization for telecommunications, finance, power grids, and autonomous systems. Multi-frequency signals mitigate ionospheric error and improve ambiguity resolution; stable onboard clocks and accurate ephemerides control the ranging bias; regional augmentation (e.g., the highly inclined QZSS orbits that improve availability in urban canyons) and anti-jamming and authentication features improve availability, integrity, and resilience. These mission-level objectives and enabling techniques establish the baseline that any MFSS must at least match.}

\subsection{Motivation}
\label{sec:motivation}
\revi{rev1.7.1}{The case for MFSS follows from specific limitations of today's function-specific missions rather than from a generic demand for scalability. First, separate communication, sensing, and PNT payloads duplicate antennas, RF chains, processors, timing units, power conversion, thermal control, telemetry, ground infrastructure, and launch capacity, even when they fly in similar orbits. Second, their spectrum assignments and observation schedules are planned independently, so that unused resources of one mission (idle beams, guard bands, off-duty radar intervals) cannot serve another. Third, the data they generate are fused only on the ground and with delay, which limits time-critical applications such as disaster response, maritime awareness, or sensing-assisted link adaptation. Fourth, small platforms cannot host several complete payload chains, while large constellations amplify manufacturing, gateway, coordination, collision-avoidance, and debris-mitigation burdens. These limitations do not imply that every mission should maximize integration; they motivate a scalable design space in which functions share only those resources for which the system-level benefit exceeds the added qualification, coupling, and operational risk.}

Formally, we define an MFSS as a satellite system that integrates two or more of the three core satellite services---communications, sensing, and PNT---within a single, coordinated payload or constellation architecture. Four functional configurations result:
\begin{itemize}
\item \emph{Joint Communications and Sensing (JCAS):} integrates wireless data exchange with environmental monitoring/Earth observation or target sensing.
\item \emph{Joint Communications and PNT (JCAP):} integrates data transmission with precise geolocation, navigation support, and timing.
\item \emph{Joint Sensing and PNT (JSAP):} integrates environmental monitoring/Earth observation with precise positioning and timing.
\item \emph{Joint Communications, Sensing, and PNT (JCSAP):} integrates all three services on a single platform.
\end{itemize}
\revi{rev1.7.2}{The integration depth is an independent design dimension: cooperative payloads share selected hardware, integrated payloads additionally coordinate time, frequency, space, code, and power resources, and joint payloads use common waveforms and unified processing. This two-dimensional view prevents ``more functions'' from being conflated with ``deeper integration''.} Such architectures provide a compelling case for the next-generation space era, mainly motivated by the following key performance indicators (KPIs).

\subsubsection{Extremely High Spectral Efficiency} Traditionally, the three services compete for spectrum with orthogonal allocations and guard bands. MFSS can operate several functions in the same band through shared or dynamically multiplexed waveforms: communication signals can serve as radar waveforms, positioning observables such as TDOA and AOA can be derived from communication signals \cite{Taylor2024GNSSMeasurements}, and signals can be jointly designed for a dual purpose, which eliminates spectrum duplication and eases congestion.

\subsubsection{Reduced Mass and Costs}
By consolidating functions into a unified payload, MFSS reduce the duplication of antennas, transceivers, processors, and power systems, and hence the mass, volume, and number of launches required for a given service portfolio, going beyond launcher sharing or payload piggybacking; the savings extend to bus requirements, system integration, and ground-station support. \revi{rev2.1.0}{As discussed in Section~\ref{sec:techno_economic}, these savings are conditional on the added engineering, qualification, and common-mode risk remaining smaller than the avoided duplication.}

\subsubsection{Enhanced Energy and Resource Efficiency}
Shared subsystems and coordinated control reduce power consumption, thermal dissipation, and structural mass compared with independent satellites, and energy-aware beamforming, dynamic power allocation, joint scheduling, and joint transmit-signal design further optimize the limited onboard resources, which is particularly valuable for multi-functional small satellites and CubeSats \cite{Liddle2020SpaceNanosatellites}.

\subsubsection{Functional Synergies and Mutual Enhancement}
Integration enables the services to inform and support one another in real time: sensing can steer beams and avoid interference, PNT can support beam alignment, synchronization, and handovers in NGSO constellations, and communications can disseminate sensing products and PNT corrections. \revi{rev1.2.0}{The configuration-specific synergies are detailed in the advantages subsections of Sections~\ref{sec:jcas}--\ref{sec:jcsap} and compared in Table~\ref{tab:tradeoff_summary}.}

\subsubsection{Space Sustainability}
As constellations grow, orbital congestion and debris have become pressing concerns \cite{NASA2024NASAsSustainability, IndianSpaceResearchOrganisationISRO2024Long-TermActivities}. By integrating several functions on one platform, MFSS can reduce the number of satellites needed for a fixed service portfolio and thereby the risk of debris generation and collisions, provided that reliability and coverage requirements do not demand independent assets.

\subsection{Terrestrial Multi-Functional Networks versus Satellite MFSS}
\label{sec:terrestrial_vs_satellite}
\revi{rev1.3.1}{Multi-functionality has been studied first in terrestrial networks, and much of the signal-processing toolbox of satellite MFSS (joint waveform design, beamforming, delay--Doppler estimation, resource allocation) is inherited from that work. The architectures, implementations, and applications are nevertheless different, as Table~\ref{tab:terrestrial_satellite_comparison} makes explicit. Architecturally, terrestrial multi-functionality extends an existing infrastructure of fixed access points with sensing and positioning, whereas satellite MFSS must converge three historically separate mission families, each with its own payload, waveform, spectrum, ground segment, and metrics, across space, ground, and user segments. In terms of implementation, terrestrial links are short and quasi-static, anchors are easy to deploy, and power, maintenance, and edge computing are accessible, whereas satellite links are long and time-varying, Doppler is large, geometry changes within minutes, hardware must be radiation qualified under strict size, weight, power, and cost (SWaP-C) and thermal limits, and repair after launch is impossible. The applications differ as well: terrestrial sensing and positioning target the local environment and the network users, whereas satellite sensing addresses Earth observation, maritime surveillance, and space situational awareness, and satellite PNT provides wide-area positioning, timing, integrity, and continuity. These differences explain why terrestrial ISAC results cannot be transferred directly, and they shape the challenges identified in Sections~\ref{sec:jcas}--\ref{sec:jcsap}.}

\begin{table*}[!t]
\begin{revblock}{rev1.3.2}
\centering
\caption{Terrestrial multi-functional networks versus satellite MFSS.}
\label{tab:terrestrial_satellite_comparison}
\scriptsize
\begin{tabularx}{\textwidth}{L{2.05cm}XX}
\toprule
\textbf{Dimension} & \textbf{Terrestrial multi-functional networks} & \textbf{Satellite MFSS} \\
\midrule
Architecture & Fixed or slowly changing access points connected to terrestrial transport/core networks; sensing and positioning extend the communication infrastructure. & Space, ground, and user segments; transparent or regenerative payloads; feeder, service, and inter-satellite links; convergence of historically separate communication, Earth-observation, and PNT missions. \\
Implementation & Short links, modest Doppler, accessible maintenance, grid power, dense edge/cloud computing; distributed anchors are easy to deploy. & Long and time-varying links, high Doppler, orbital handovers, radiation-qualified hardware, strict SWaP-C and thermal limits, autonomous operation, precise clocks/ephemerides, no post-launch repair. \\
Sensing objective & Perception of the local environment, user/vehicle tracking, mapping, and situational awareness around the infrastructure. & Earth observation, radar altimetry, scatterometry, atmospheric/ionospheric sounding, maritime surveillance, and space situational awareness; performance governed by orbit, swath, revisit, and (bi)static geometry. \\
PNT objective & Network localization, tracking, and synchronization within regional coverage. & Wide-area positioning, navigation, timing, integrity, and continuity using constellation geometry, clock/ephemeris products, and navigation messages. \\
Dominant impairments & Multipath, blockage, and interference in dense environments. & Path loss, Doppler and Doppler rate, clock and ephemeris errors, geometric dilution of precision, rain attenuation, self-interference in monostatic sensing. \\
Representative applications & Connected mobility, industrial automation, smart environments, local digital twins. & Remote broadband, direct-to-device, maritime/aeronautical links, climate and disaster monitoring, autonomous navigation, precise timing, resilient global services. \\
\bottomrule
\end{tabularx}
\end{revblock}
\end{table*}

\subsection{Related Works}
\label{sec:related}
A wide range of surveys and magazine papers have investigated JCAS or JCAP from multiple dimensions, including system architecture, signal processing, channel modeling, interference mitigation, and the adoption of cutting-edge technologies, focusing mainly on terrestrial networks.

Foundational works on JCAS address waveform design, signal processing, and system architecture: \cite{Zhang2022EnablingSurveyb} introduces perceptive mobile networks for radar-communication coexistence, \cite{Zhou2022IntegratedSurvey} classifies waveforms into communication-centric, sensing-centric, and jointly optimized designs, \cite{Wei2023IntegratedSurvey} reviews signal processing for 5G/6G including PAPR and interference management, \cite{Liu2022ACommunication} proposes a unified performance framework, and \cite{Luo2025ISACTestbeds} presents a layered architecture validated by testbeds and standardization. Channel modeling is addressed in \cite{Zhu2025EnablingNetworks} (JCAS-IoT architecture with intelligent agents), \cite{Wei2025IntegratedSurvey,Liu20246GOpportunities} (radar cross sections, clutter, and AI-assisted modeling), and \cite{Gutierrez2025ChannelChallenges} (modular vehicular channel model); RIS-assisted NOMA is surveyed in \cite{Sur2024ACommunication}, V2X and UAV applications in \cite{Zhong2022EmpoweringOpportunities,Ahmed2025AdvancementsSurvey}, joint communication--sensing--computation in \cite{Wen2024AComputation}, and machine learning for JCAS in \cite{AdeKrisnaRespati2024AApplications}.

In the context of JCAP, \cite{Trevlakis2023LocalizationOutlook} presents the potential of localization in 6G and its applications, \cite{Lohan2025EmergingCommunications} reviews emerging trends such as ultra-massive and XL-MIMO, cell-free networks, machine learning, and RIS, \cite{Luong2021RadioSurvey} partially covers WiFi-based indoor JCAP and resource management (spectrum sharing, power allocation, interference management), and \cite{DeLima2021ConvergentChallenges} overviews opportunities and challenges for both JCAP and JCAS.

For satellite systems, \cite{Stock2025SurveySatellites} surveys opportunistic LEO-PNT using LEO communication signals (error sources, performance limitations, implementation challenges, and some payload modifications that enable PNT), \cite{Prol2022PositionOpportunities} provides a multidisciplinary survey of emerging LEO-PNT systems (design steps, physical-layer parameters, modeling tools, channel models, and commercial prospects), and \cite{Janssen2023ALEO-PNT} surveys large-scale, energy-efficient positioning for IoT, including terrestrial LPWAN methods, GNSS adaptations, and LEO-based solutions; in all three, the multi-functional aspect is discussed only from the signals-of-opportunity perspective rather than as a joint system design. Finally, \cite{Hui2025ADirections} reviews LEO communication payloads and their role in integrated communication, navigation, and remote sensing, covering the evolution of LEO systems, architectures with inter-satellite links, the payload components of constellations such as Iridium, Globalstar, and Starlink, direct-to-cell services, and signals of opportunity for positioning and passive sensing; its perspective remains communication-centric, since sensing and PNT are added to a communication payload rather than co-designed with it.

\begin{table}[!t]
\centering
\scriptsize
\caption{This survey versus related surveys ($\times$: not covered, *: preliminary, **: partial, ***: full coverage). The configuration-specific rows cover architecture, classification, state of the art, advantages, challenges, and research directions of each configuration.}
\label{tab:key_contribs}
\setlength{\tabcolsep}{4pt}
\renewcommand{\arraystretch}{1.25}
\begin{tabular}{|L{1.55cm}|L{2.8cm}|c|c|c|c|c|}
\hline
\multicolumn{2}{|c|}{\textbf{Key Contributions}} & \textbf{\cite{Stock2025SurveySatellites}} & \textbf{\cite{Prol2022PositionOpportunities}} & \textbf{\cite{Janssen2023ALEO-PNT}} & \textbf{\cite{Hui2025ADirections}} & \textbf{Our} \\
\hline
\multirow{4}{*}{\parbox[t]{1.5cm}{Single-function fundamentals}}
  & Communications      & *   & *   & *   & *** & *** \\ \cline{2-7}
  & Sensing             & $\times$ & * & $\times$ & * & *** \\ \cline{2-7}
  & PNT                 & ** & ** & ** & **  & *** \\ \cline{2-7}
  & Payload architectures & * & ** & * & ** & *** \\
\hline
\multirow{3}{*}{\parbox[t]{1.5cm}{MFSS integration levels}}
  & Cooperative payload  & $\times$ & $\times$ & $\times$ & $\times$  & *** \\ \cline{2-7}
  & Integrated payload   & $\times$ & $\times$ & $\times$ & $\times$  & *** \\ \cline{2-7}
  & Joint payload        & $\times$ & $\times$ & $\times$ & * & *** \\
\hline
\multirow{4}{*}{\parbox[t]{1.5cm}{Configuration-specific treatment}}
  & JCAS  & $\times$ & $\times$ & $\times$ & * & *** \\ \cline{2-7}
  & JCAP  & * & * & * & ** & *** \\ \cline{2-7}
  & JSAP  & $\times$ & $\times$ & $\times$ & $\times$ & *** \\ \cline{2-7}
  & JCSAP & $\times$ & $\times$ & $\times$ & * & *** \\
\hline
\multirow{3}{*}{\parbox[t]{1.5cm}{Cross-cutting analyses}}
  & Achievable performance and trade-offs & $\times$ & * & $\times$ & $\times$ & *** \\ \cline{2-7}
  & Terrestrial vs. satellite multi-functionality & $\times$ & $\times$ & * & $\times$ & *** \\ \cline{2-7}
  & Techno-economic assessment & $\times$ & * & $\times$ & * & *** \\
\hline
\end{tabular}
\end{table}

\subsection{Main Contributions and Organization}
\label{sec:contributions}
The works \cite{Stock2025SurveySatellites,Prol2022PositionOpportunities,Janssen2023ALEO-PNT} consider communications and PNT only from the signals-of-opportunity perspective. Similarly, while \cite{Hui2025ADirections} considers the three services, it concentrates on communication payloads and explores how sensing and PNT functionalities can be integrated into such systems, without any treatment of joint functional co-design. A comparison with the related works is shown in Table~\ref{tab:key_contribs}. This paper begins by tracing the independent evolution of payload architectures across the three domains and systematically examines their convergence toward truly integrated, multi-functional satellite systems. Unlike opportunistic approaches centered on communications \cite{Hui2025ADirections}, this work advocates a unified joint design philosophy aimed at enabling native joint functionality. The main contributions are as follows.

\begin{itemize}
\item \emph{A unified two-dimensional taxonomy for MFSS:} \revi{rev1.2.1}{The paper introduces a structured classification of MFSS that separates the functional configuration (JCAS, JCAP, JSAP, or JCSAP) from the integration depth (cooperative, integrated, or joint payloads). The three integration levels are examined in terms of architecture, classification, state of the art, advantages, and level-specific challenges, and a mapping table shows how each configuration is realized at each level.}

\item \emph{Consolidated fundamentals:} \revi{rev1.1.2}{The communication, sensing, and PNT fundamentals---purpose, architectures, classification, applications, and payload architectures---are consolidated into a single background section that provides the baseline from which MFSS payloads evolve, so that the remainder of the survey can focus on multi-functional convergence. The section is complemented by an explicit comparison between terrestrial multi-functional networks and satellite MFSS.}

\item \emph{Dedicated treatment of each MFSS configuration:} \revi{rev1.2.2}{Each of JCAS, JCAP, JSAP, and JCSAP is discussed in its own section, covering its architecture and realizations, classification, state of the art (including the most recent satellite-specific contributions), advantages and functional synergies, configuration-specific key challenges, and tailored research directions spanning hardware--software co-design, signal processing, and AI-based optimization.}

\item \emph{Achievable performance, trade-offs, and techno-economics:} \revi{rev1.4.1}{Within each configuration section, the achievable performance is illustrated through representative trade-off results reported in the literature (communication rate versus sensing accuracy, spectral efficiency versus ranging accuracy, navigation accuracy versus sensing quality, and three-function integration gains); a single cross-configuration table then compares the synergies, dominant trade-offs, and quantitative evidence of the four configurations, and a life-cycle techno-economic framework identifies when deeper integration is, and is not, economically justified.}
\end{itemize}

 \begin{table*}[!t]
\centering
\caption{Structure of this survey.}
\label{tab:structure}
\scriptsize
\setlength{\tabcolsep}{2pt}
\renewcommand{\arraystretch}{1.0}
\newlength{\wSec}\setlength{\wSec}{0.9cm}
\newlength{\wTit}\setlength{\wTit}{3.4cm}
\newlength{\wRol}\setlength{\wRol}{4.6cm}
\newlength{\wCon}\setlength{\wCon}{\dimexpr\textwidth-\wSec-\wTit-\wRol-8\tabcolsep\relax}
\begin{tabular}{@{}p{\wSec}p{\wTit}p{\wCon}p{\wRol}@{}}
\cbox{pSecH}{\wSec}{\centering\textbf{Sec.}} &
\cbox{pTitH}{\wTit}{\textbf{Title}} &
\cbox{pConH}{\wCon}{\textbf{Contents}} &
\cbox{pRolH}{\wRol}{\textbf{Role in the survey}} \\[2pt]
\cbox{pSec}{\wSec}{\centering I} &
\cbox{pTit}{\wTit}{Introduction} &
\cbox{pCon}{\wCon}{Current satellite missions and enabling technologies; motivation; terrestrial multi-functional networks versus satellite MFSS (Table~\ref{tab:terrestrial_satellite_comparison}); related works (Table~\ref{tab:key_contribs}); contributions and organization} &
\cbox{pRol}{\wRol}{Defines MFSS, the four functional configurations (JCAS, JCAP, JSAP, JCSAP), the two-dimensional taxonomy, and the position of the survey with respect to prior work} \\[2pt]
\cbox{pSec}{\wSec}{\centering II} &
\cbox{pTit}{\wTit}{Fundamentals of single-functional satellite systems} &
\cbox{pCon}{\wCon}{Satellite communications, satellite-borne sensing, and satellite PNT, each summarized through purpose, architectures, classification, applications, and payload; convergence boundary} &
\cbox{pRol}{\wRol}{Baseline resources (aperture, RF chains, spectrum, timing, processing) and performance measures from which MFSS payloads evolve} \\[2pt]
\cbox{pSec}{\wSec}{\centering III\\ IV\\ V} &
\cbox{pTit}{\wTit}{Cooperative MFSS payloads\\ Integrated MFSS payloads\\ Joint MFSS payloads} &
\cbox{pCon}{\wCon}{For each level: architecture; classification (realization of the four configurations, Table~\ref{tab:level_config_mapping}); state of the art; advantages; level-specific key challenges} &
\cbox{pRol}{\wRol}{Integration-depth dimension of the taxonomy: shared hardware (cooperative), shared resource grid (integrated), common waveform and unified processing (joint)} \\[2pt]
\cbox{pSec}{\wSec}{\centering VI\\ VII\\ VIII\\ IX} &
\cbox{pTit}{\wTit}{JCAS in satellite systems\\ JCAP in satellite systems\\ JSAP in satellite systems\\ JCSAP in satellite systems} &
\cbox{pCon}{\wCon}{For each configuration: architecture and realizations; classification; state of the art (including the most recent satellite-specific studies); advantages and functional synergies; achievable performance and trade-offs (representative results from the literature, Figs.~\ref{fig:tradeoff_jcas}--\ref{fig:tradeoff_jcsap}); key challenges; research directions} &
\cbox{pRol}{\wRol}{Functional-configuration dimension of the taxonomy: one dedicated section per configuration, from architecture to quantitative evidence and open problems} \\[2pt]
\cbox{pSec}{\wSec}{\centering X} &
\cbox{pTit}{\wTit}{Cross-configuration comparison and techno-economic assessment} &
\cbox{pCon}{\wCon}{Metrics and shared resources; single comparison table of functional coupling and synergies, dominant trade-off and design knob, and quantitative evidence for JCAS, JCAP, JSAP, and JCSAP (Table~\ref{tab:tradeoff_summary}); cross-configuration observations; life-cycle cost model and consolidated table of savings, incremental complexity, and risk by integration depth and by infrastructure entity (Table~\ref{tab:techno_economic})} &
\cbox{pRol}{\wRol}{Side-by-side view of what each configuration achieves and at what cost for the other functions, and of when deeper integration is, and is not, technically and economically justified} \\[2pt]
\cbox{pSec}{\wSec}{\centering XI} &
\cbox{pTit}{\wTit}{Conclusions} &
\cbox{pCon}{\wCon}{Summary of the framework, the evidence, and the open problems} &
\cbox{pRol}{\wRol}{Outlook} \\
\end{tabular}
\end{table*}

\emph{Paper Organization:} \revi{rev1.1.3}{Section~\ref{sezione_fundamentals} consolidates the fundamentals of single-functional satellite communications, sensing, and PNT systems, including their payload architectures. Sections~\ref{sec:cooperative_mfss}--\ref{sec:joint_mfss} present the cooperative, integrated, and joint MFSS integration levels. Sections~\ref{sec:jcas}--\ref{sec:jcsap} are devoted to JCAS, JCAP, JSAP, and JCSAP, respectively, each covering architecture, classification, state of the art, advantages, achievable performance and trade-offs, key challenges, and research directions. Section~\ref{sec:assessment} compares the performance trade-offs of the four configurations in a single table and assesses their techno-economic viability. Section~\ref{sezione_9} concludes the paper. The overall structure is summarized in Table~\ref{tab:structure}.}
\section{Fundamentals of Single-Functional Satellite Systems}
\label{sezione_fundamentals}
\label{sezione_2}\label{sezione_3}\label{sezione_4}\label{sezione_5}

\revi{rev1.1.1}{This section consolidates the background on the three single-functional satellite system classes from which MFSS payloads evolve: satellite communications, satellite-borne sensing, and satellite PNT. Each class is summarized through its purpose, architecture, classification, applications, and payload architecture. These conventional designs define the resources (aperture, RF chains, spectrum, timing, and processing) and the performance measures that MFSS must share or preserve, so that the remainder of the paper can concentrate on multi-functional convergence.}

\begin{figure*}[!t]
   \centering
    \includegraphics[width=0.68\textwidth]{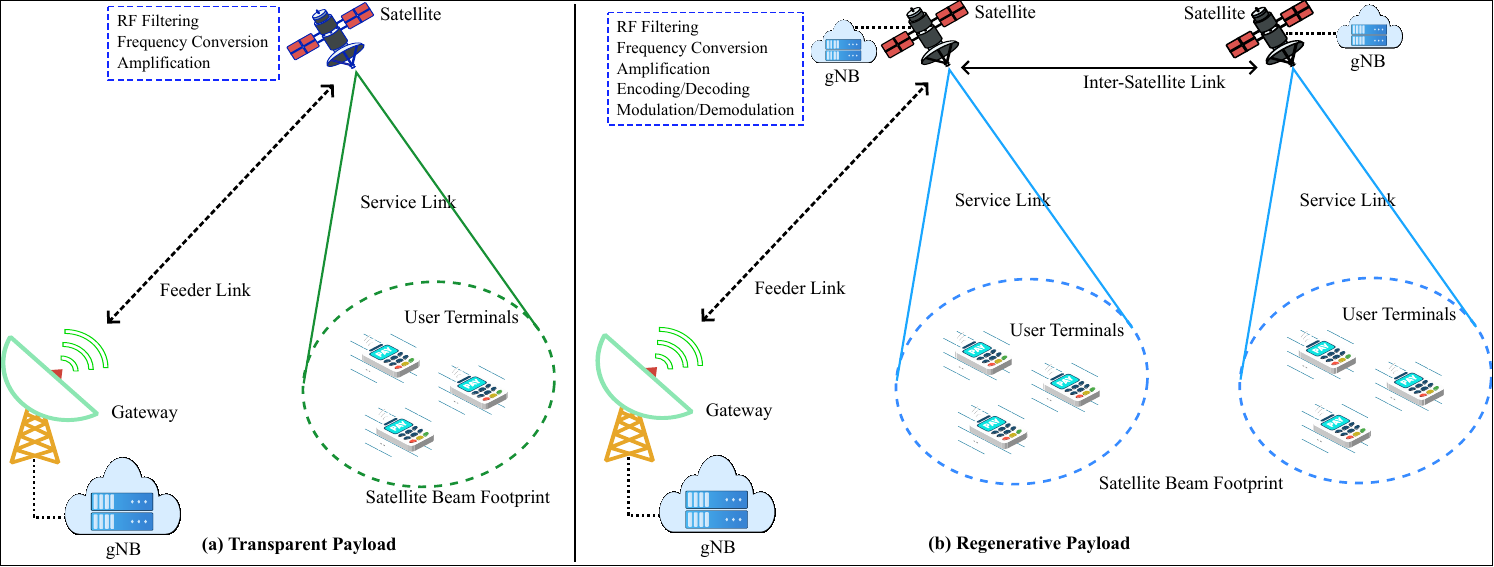}
    \caption{Transparent payload versus regenerative payload.}
   \label{fig:comm_transparent}
\end{figure*}

\subsection{Satellite Communications Systems}
\subsubsection{Purpose}
Global connectivity underpins economic development, education, healthcare, emergency response, direct-to-X services, and international collaboration. Where terrestrial infrastructure is underdeveloped or infeasible---remote, maritime \cite{Wei2021HybridChallenges}, and aeronautical environments \cite{Kodheli2021SatelliteChallenges}---satellite communications are indispensable, enabling real-time data exchange, telemedicine \cite{Alenoghena2023Telemedicine:Challenges}, distance learning, and disaster recovery \cite{Teodoro2022TheManagement,Zhou2021IntegratedTechnology}, and contributing to the United Nations Sustainable Development Goals. Satellites bridge the digital divide \cite{sheemar2025jointcommunicationssensing6g}, provide safety-of-life links for aviation, maritime transport, and disaster management, support scientific research in polar and other hard-to-reach regions, and, with the convergence of satellite communications and the IoT \cite{Centenaro2021AIoT}, enable large-scale smart applications from precision agriculture to environmental monitoring and renewable-energy management \cite{Wu2018InformationPerspectives,Hoyhtya2022SustainableSafety}.

\subsubsection{Architectures}
Satellite communication systems employ two fundamental payload architectures \cite{3gppNTN2024}, illustrated in Fig.~\ref{fig:comm_transparent}: the transparent (bent-pipe) payload performs only RF-level processing (filtering, frequency conversion, amplification), acting as a repeater that preserves the signal structure, whereas the regenerative payload adds onboard demodulation/decoding, routing/switching, and re-modulation, effectively embedding ground-station functionality in the spacecraft. A complete system comprises three segments. The \emph{space segment} contains the satellite with its power and propulsion subsystems, the tracking, telemetry, and command (TT\&C) subsystem that monitors satellite health and executes ground commands \cite{He2025NonterrestrialProspects,Zhan2020ChallengesSystem}, and the communication payload that receives uplink signals through high-gain antennas and transmits the processed downlink toward the coverage areas. The \emph{ground segment} comprises gateway or hub stations with large antennas and high-power amplifiers, together with a network operations center that manages traffic, allocates bandwidth, and ensures quality of service \cite{Al-Hraishawi2023APerspective}. The \emph{user segment} encompasses the end-user equipment, from handheld and broadband terminals to broadcast receivers, emergency devices, direct-to-device (D2D) handsets, and IoT nodes \cite{Al-Hraishawi2023APerspective,Bakhsh2025Multi-SatelliteSurvey}.

\subsubsection{Classification}
Satellite communication systems can be classified according to several criteria. \emph{By orbital altitude} \cite{Azari2022EvolutionSurvey}: geostationary (GEO) satellites at about $35{,}786$~km remain fixed relative to Earth and cover large regions continuously; medium Earth orbit (MEO) satellites ($2000$--$35{,}786$~km) balance latency and coverage; and LEO satellites ($500$--$2000$~km) provide low latency and high throughput, which is crucial for broadband Internet. \emph{By functionality} \cite{Abo-Zeed2019SurveyTrends}: fixed satellite services (FSS) for high-capacity links such as TV distribution or backhaul, mobile satellite services (MSS) for ships, aircraft, and vehicles, and broadcast satellite services (BSS) for one-way transmission to many receivers. \emph{By frequency band} \cite{ESA2025}: the seven main bands and the services and applications assigned to them by the International Telecommunication Union (ITU) are summarized in Table~\ref{Rel_Works}; each country develops its frequency plan within these regulations. \emph{By payload architecture}: transparent or regenerative, as discussed above.

\begin{table}[!t]
\centering
\caption{Communication satellite frequency bands \cite{ESA2025}.}
\label{Rel_Works}
\scriptsize
\begin{tabular}{|p{1cm} | p{1.2cm} | p{1.5cm} | p{3.4cm} |}
  	\hline
{\bf Frequency band } & {\bf Frequency range (GHz)}  & {\bf Satellite service type} & {\bf Applications}  \\
  \hline \hline
L-band  & 1.518--1.675 & MSS & Civil mobile communication services, global positioning system, weather radar system  \\
   \hline
S-band & 1.97--2.69 & MSS & Satellite TV, mobile broadband services, radio broadcasting, inflight connectivity   \\
    \hline
C-band & 3.4--7.025 & FSS & Data services, satellite TV networks, unprocessed satellite feeds   \\
	\hline
X-band & 7.25--8.44 & FSS & Military operations, pulse radar systems, weather monitoring, air traffic control   \\
	\hline
Ku-band & 10.7--14.5 & FSS, BSS & Fixed satellite TV, data services  \\
	\hline
Ka-band & 17.3--30 & FSS, BSS & Two-way broadband services, fixed satellite data services   \\
	\hline
Q/V band & 37.5--51.4 & MSS, BSS & High-speed broadband services, inflight connectivity  \\
	\hline
\end{tabular}
\end{table}

\subsubsection{Applications}
\emph{Global broadband access:} LEO mega-constellations deliver high-speed Internet to remote and underserved regions with data rates of several Gbps and latencies of $30$--$50$~ms, far below the $500$+~ms of GEO systems; approaching the 6G target of Tbps-class peak rates requires Ka-band and Q/V-band operation, high-order modulation, massive MIMO \cite{You2020MassiveCommunications, Fontanesi2025ArtificialSurvey}, and free-space optical ISLs that relay data across the constellation. \emph{IoT communications:} satellites connect massive numbers of low-power devices in oceans, deserts, and polar regions for smart agriculture, environmental monitoring, and industrial automation through NB-IoT or LoRaWAN protocols in L- or S-band with energy-aware scheduling \cite{Centenaro2021AIoT}. \emph{Maritime and aeronautical communications:} satellites provide navigation, safety, operational, and passenger connectivity for ships and aircraft \cite{Wei2021HybridChallenges}, including in-flight broadband \cite{Bilen2022AeronauticalChallenges}. \emph{Emergency and disaster recovery:} when terrestrial infrastructure is damaged or overloaded, satellites support first responders and affected populations through robust bands, redundant ISLs, portable VSATs, and direct-to-device links \cite{Kagai2024RapidlyInfrastructure}. \emph{Military and government communications:} secure tactical networks, surveillance, and data links rely on strong encryption and anti-jamming techniques such as frequency hopping and spread spectrum \cite{TEDESCHI2022109246}. \emph{Direct-to-X (D2X):} smartphones, IoT sensors, vehicles, and drones connect directly to satellites for global messaging, environmental monitoring, disaster recovery, and autonomous transportation \cite{Bakhsh2025Multi-SatelliteSurvey}.

\subsubsection{Communications Payload}
\begin{figure}[!t]
    \centering
    \includegraphics[width=0.86\columnwidth]{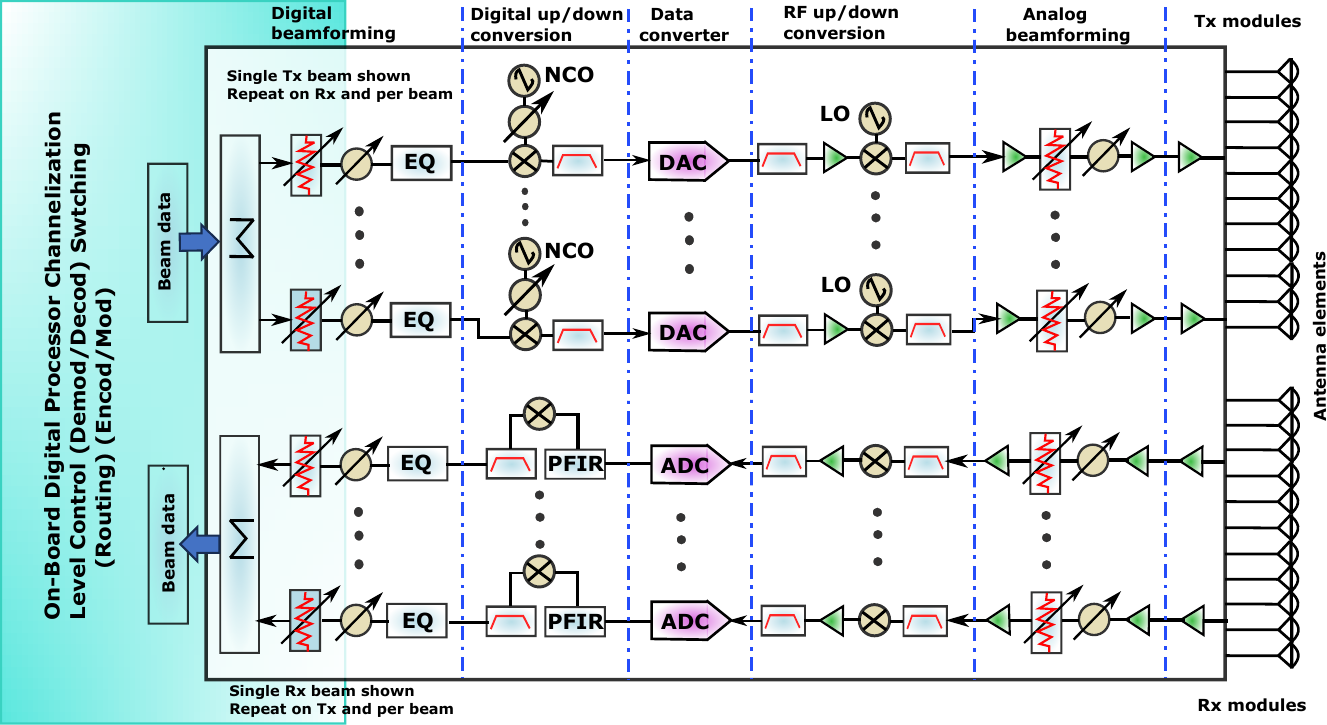}
    \caption{Block diagram of the hybrid beamforming-based communication satellite payload.}
    \label{fig:sc_transponder}
\end{figure}
The communication payload comprises the antenna subsystem and the onboard digital processor (OBDP). State-of-the-art payloads use direct-radiating phased arrays with analog, digital, or hybrid beamforming; hybrid beamforming offers a practical compromise between packaging, power consumption, and digital-processing limits \cite{KumarSheemar2021HybridConsiderations}. Fig.~\ref{fig:sc_transponder} shows a representative hybrid-beamforming payload: the antenna elements radiate the signals; the Tx/Rx modules contain the low-noise and high-power amplifiers and the Tx/Rx switching; analog beamforming merges groups of elements into subarrays; microwave up/down-conversion translates the operating frequency to an intermediate frequency compatible with the ADCs/DACs; digital up/down-conversion reduces the sample rate to the processing bandwidth; and digital beamforming combines the in-phase/quadrature streams into the final beams. The OBDP performs encoding, decoding, modulation, and error correction, and coordinates frequency allocation, power control, and resource optimization.

\subsection{Satellite-Borne Sensing Systems}
\subsubsection{Purpose}
Natural and anthropogenic changes in the Earth system affect the atmosphere, land, oceans, and ecosystems worldwide \cite{Bello2014SatelliteApproach}. Satellite-borne microwave remote sensing observes these changes at large scale, repeatedly, and independently of weather and daylight \cite{Xie2004SatelliteInteraction,Hargreaves2021SatelliteDevelopment}, with consistent coverage of remote regions, land--water boundaries, and oceans \cite{Hargreaves2021SatelliteDevelopment,Champagne2011MonitoringSensing}. Long-term Earth observation (EO) datasets support environmental tracking, policy-making, and planning \cite{2020EARTHIndicators,2017EarthReport}, and are used for poverty reduction, governance, climate-risk assessment, food security \cite{Kogan2019RemoteSecurity}, agriculture \cite{Yang2020RemoteExample}, urban planning \cite{Pham2011AMetrics}, and disaster response \cite{DeLeeuw2010ThePolicy,Thies2011SatellitePerspectives}.

\subsubsection{Architectures}
\begin{figure}[!t]
    \centering
    \includegraphics[width=0.78\columnwidth]{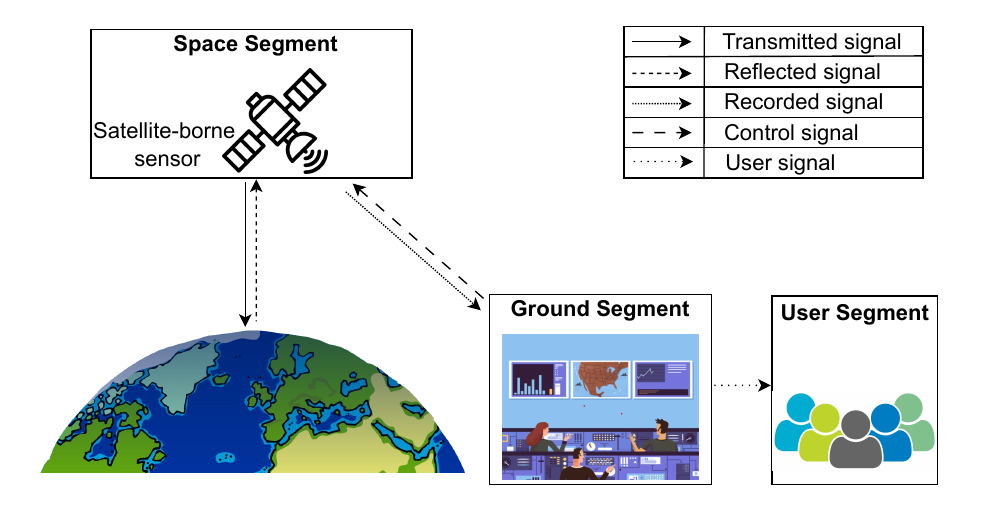}
    \caption{An architecture of monostatic satellite-borne microwave remote sensing.}
    \label{fig:rs_monostatic}
\end{figure}

\begin{figure}[!t]
    \centering
    \includegraphics[width=0.78\columnwidth]{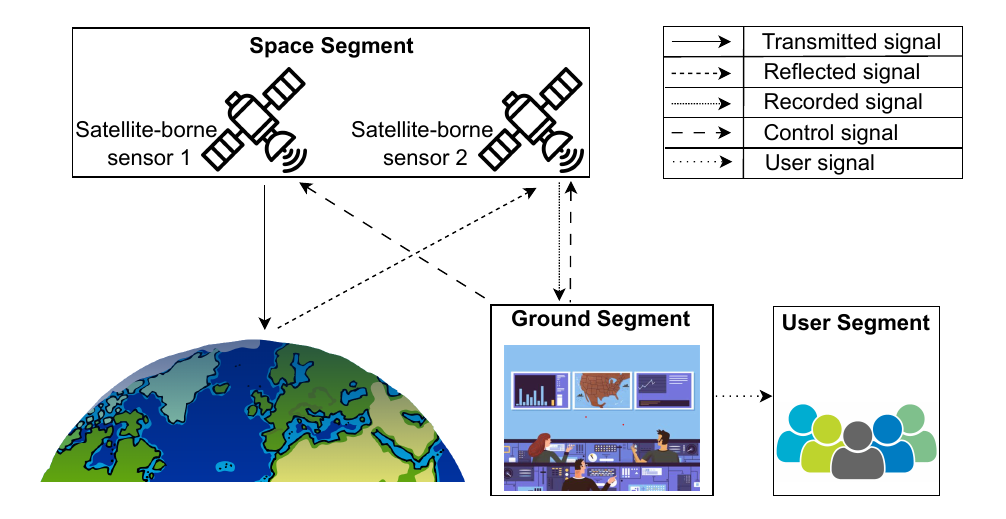}
    \caption{An architecture of bistatic satellite-borne microwave remote sensing.}
    \label{fig:rs_bistatic}
\end{figure}

Satellite-borne microwave remote sensing systems are realized in monostatic or bistatic architectures. In a monostatic system the same satellite transmits the electromagnetic signal and receives the backscattered echoes (Fig.~\ref{fig:rs_monostatic}) \cite{Stojanovic2013CompressedSAR}, whereas in a bistatic system the transmitter and receiver are on separate satellites or platforms (Fig.~\ref{fig:rs_bistatic}) \cite{Shah2012DemonstrationSignals}. The \emph{space segment} (a satellite or constellation) transmits the signals toward the target, receives or records the backscatter, and forwards the measured data to the ground. The \emph{ground segment} comprises ground stations for TTC and payload-data reception, control centers, and ground networks connecting testing, launch, and user facilities; it collects, processes, and distributes value-added sensing products. The \emph{user segment} includes the terminals and services through which the processed products support decision-making and day-to-day applications.

\subsubsection{Classification}
\begin{figure}[!t]
\centering
\resizebox{0.88\columnwidth}{!}{
\begin{forest}
for tree={
    grow=east,
    draw,
    rounded corners,
    node options={align=center, text width=3cm},
    edge={->, >=Stealth},
    parent anchor=east,
    child anchor=west,
    anchor=west,
    l sep+=10pt,
    s sep+=10pt,
    font=\sffamily\small,
}
[Satellite-based Microwave\\Remote Sensing
  [Active Sensing
    [Imaging
      [Real Aperture Radar]
      [Synthetic Aperture Radar (SAR)
      ]
    ]
    [Non-imaging
      [Radar Altimetry]
      [Spectrometry \& Reflectometry]
      [Point-to-Point Link\\Monitoring]
    ]
  ]
  [Passive Sensing
    [Imaging
      [Real Aperture Radiometry]
      [Synthetic Aperture Radiometry]
    ]
    [Non-imaging
      [Total Power Radiometry]
      [Opportunistic Reflectometry]
      [Opportunistic Radiometry]
    ]
    [Ground-to-Space]
    [Space-to-Space]
  ]
]
\end{forest}}
\caption{Classification of satellite-based microwave remote sensing.}
\label{fig:rs_classification}
\end{figure}
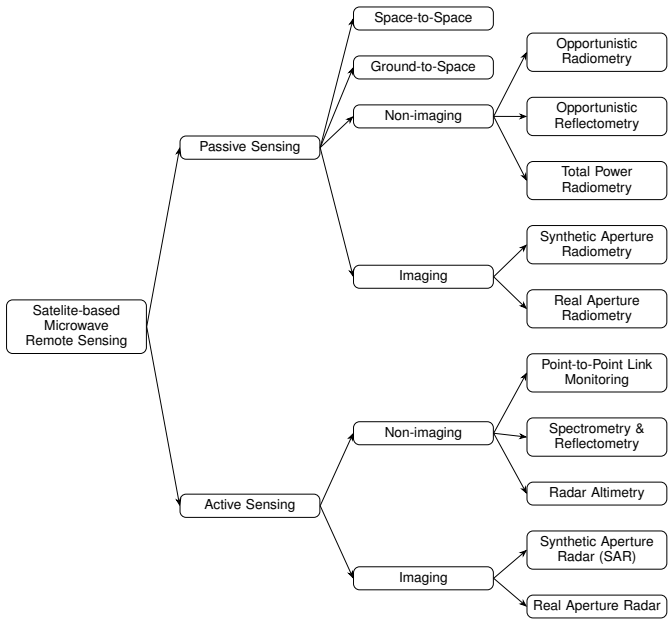

As shown in Fig.~\ref{fig:rs_classification}, satellite-borne microwave remote sensing is divided into passive sensing \cite{Njoku1982PassiveReview}, in which radiometers and scanners receive naturally emitted, reflected, or scattered radiation, and active sensing \cite{Tuia2009ActiveClassification}, in which radars transmit their own signals and infer surface or atmospheric properties from the echoes. This work focuses on active microwave sensing, whose controllable transmission parameters can be jointly exploited with other functions; passive methods appear in Fig.~\ref{fig:rs_classification} for completeness, and the less common active techniques (point-to-point link monitoring, spectrometry, and reflectometry) are covered in \cite{Bu2024LandOverview,Hassan2022RadioactiveEgypt}. Altimeters transmit short nadir-pointing pulses and estimate the satellite-to-target distance from the round-trip delay \cite{Srinivasan2023SatelliteReview}, supporting the retrieval of ocean topography, currents, wave height, ice-sheet elevation, sea-ice extent, inland-water levels, and oil-slick indicators. Scatterometers measure the normalized radar cross section of the surface \cite{Isleifson2023ASensing}, from which ocean winds, sea ice, soil moisture, vegetation indices, and crop conditions are retrieved \cite{Hauser2023SatelliteNow, Singh2022ThePerspective}. Real aperture radars image with the physical antenna \cite{Werner2008AInterferometry}, but high resolution would require impractically large antennas at orbital ranges; SAR exploits platform motion to synthesize a large effective aperture \cite{Brown1967SyntheticRadar,Moreira2013ARadar} and enables high-resolution 2D imaging and 3D reconstruction. Polarimetric SAR extracts physical characteristics of land, ocean, snow, ice, and urban surfaces from multiple polarizations \cite{Shokr2023PolarimetricReview}, interferometric SAR derives topography and digital elevation models from multiple images \cite{Wang2024InterferometricReview}, and differential SAR detects surface motion from temporally separated images for earthquake, volcano, landslide, deforestation, glacier, and disaster monitoring \cite{Zhang2022AInterferometry}.

\subsubsection{Applications}
The EO data collected from space support the applications summarized in Table~\ref{tab:rs_applications}. \emph{Environment monitoring} covers land (land-cover and land-use mapping, geological hazards, urban planning, forestry, biomass and greenhouse-gas accounting, and precision agriculture) \cite{2020EARTHIndicators,NationalAcademiesofSciences2015ASpectrum}, the atmosphere (cloud properties, precipitation, and winds for storm, air-quality, and climate monitoring) \cite{NationalAcademiesofSciences2015ASpectrum,Sentinel-3:Services}, the oceans (winds, wave height, currents, temperature, salinity, and sea-surface height for climate, sea-level, and circulation studies) \cite{NationalAcademiesofSciences2015ASpectrum,Sentinel-3:Services}, and the cryosphere (ice thickness, extent, and motion, snow cover, glaciers, and frozen ground for climate-change, sea-level, and hazard studies) \cite{YeRemoteReview}. \emph{Object detection and tracking} exploits radar imagery for near-real-time maritime surveillance (unreported vessels, illegal fishing, oil spills, shipping routes, sea ice), coastal monitoring, search and rescue, and security \cite{2020EARTHIndicators,NationalAcademiesofSciences2015ASpectrum,MaritimeMonitoring}, and supports disaster management through the detection of environmental threats, millimetre-level ground deformation, and flood extent \cite{Bonafilia2020Sen1Floods11:Sentinel-1}. \emph{Space situational awareness (SSA)} addresses the tens of thousands of objects already in orbit: the detection, tracking, characterization, and prediction of satellites and debris support collision avoidance, space-traffic coordination, autonomous maneuver planning, planetary defense, and space-weather monitoring.

\begin{table}[!t]
\centering \scriptsize
\setlength{\tabcolsep}{2.5pt}
\caption{Applications and sensors used for microwave remote sensing}
\label{tab:rs_applications}
\begin{tabular}{|l|l|c|c|c|}
\hline
\multicolumn{2}{|c|}{\textbf{Application}} & \textbf{SAR} & \textbf{Altimeter} & \textbf{Scatterometer} \\ \hline
\multirow{7}{*}{\rotatebox{90}{\textit{Land}}} & Land use and cover dynamics & \checkmark & & \\ \cline{2-5}
 & Forest and vegetation & \checkmark & & \checkmark \\ \cline{2-5}
 & Agriculture & \checkmark & & \checkmark \\ \cline{2-5}
 & Urban planning & \checkmark & & \\ \cline{2-5}
 & Topography and geology & \checkmark & \checkmark & \\ \cline{2-5}
 & Wetlands and soil moisture & \checkmark & & \checkmark \\ \cline{2-5}
 & Inland waters & \checkmark & \checkmark & \checkmark \\ \hline
\multirow{3}{*}{\rotatebox{90}{\textit{Atmos.}}} & Precipitation and clouds & \checkmark & \checkmark & \checkmark \\ \cline{2-5}
 & Hurricane, cyclone, and typhoon monitoring & \checkmark & \checkmark & \checkmark \\ \cline{2-5}
 & Weather prediction & & \checkmark & \checkmark \\ \hline
\multirow{4}{*}{\rotatebox{90}{\textit{Ocean}}} & Surface winds, waves, and currents & \checkmark & \checkmark & \checkmark \\ \cline{2-5}
 & Temperature and salinity & & & \checkmark \\ \cline{2-5}
 & Ocean altimetry & & \checkmark & \\ \cline{2-5}
 & Coastal hazards & \checkmark & \checkmark & \checkmark \\ \hline
\multirow{2}{*}{\rotatebox{90}{\textit{Cryo.}}} & Snow cover, sea ice, glaciers, permafrost & \checkmark & \checkmark & \checkmark \\ \cline{2-5}
 & Snowpack and ice-sheet characterization & & \checkmark & \checkmark \\ \hline
\multirow{4}{*}{\rotatebox{90}{\textit{Objects}}} & Ground moving target indication & \checkmark & & \\ \cline{2-5}
 & Ship detection and maritime surveillance & \checkmark & & \checkmark \\ \cline{2-5}
 & Oil spill detection & \checkmark & & \checkmark \\ \cline{2-5}
 & Disaster monitoring and management & \checkmark & & \checkmark \\ \hline
\textit{SSA} & Navigation aids & \checkmark & & \checkmark \\ \hline
\end{tabular}
\end{table}

\subsubsection{Active Remote Sensing Payload}
\begin{figure}[!t]
    \centering
    \includegraphics[width=0.88\columnwidth]{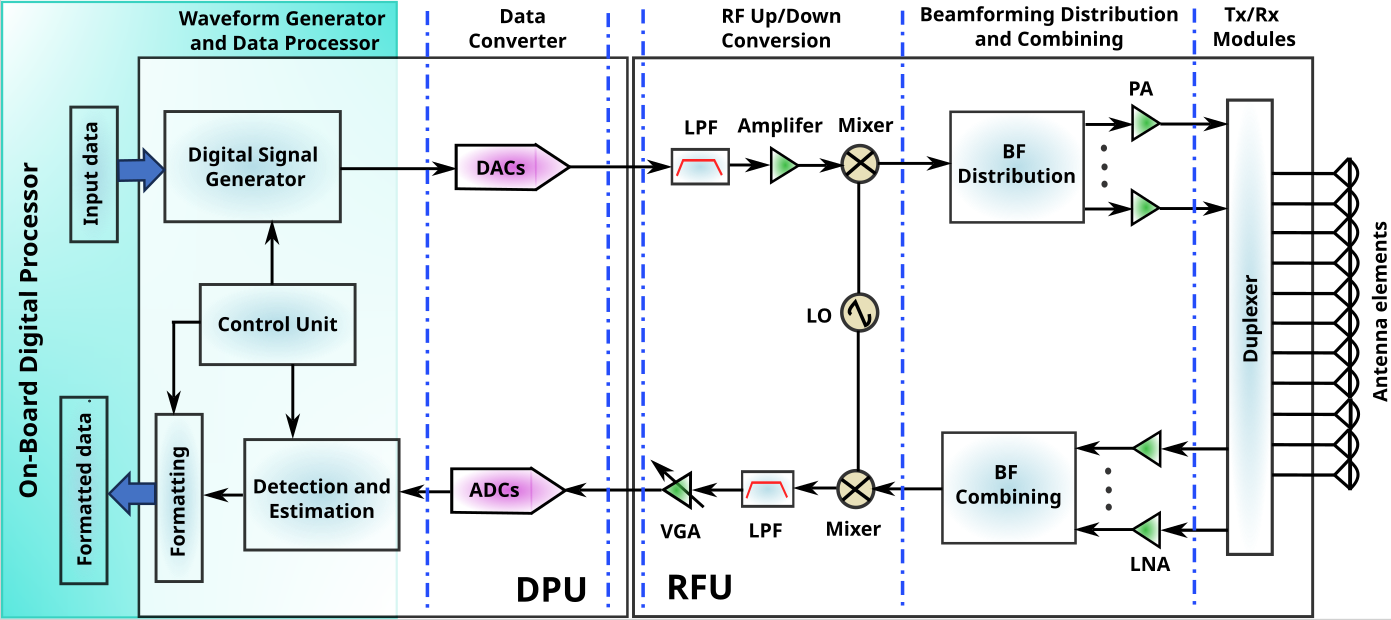}
    \caption{Satellite radar sensing payload.}
    \label{fig:radar_arch}
\end{figure}
Fig.~\ref{fig:radar_arch} shows the architecture of an active sensing payload, which consists of a radio frequency unit (RFU) and a digital processing unit (DPU) \cite{Sentinel-3:Services, DonlonRSE}. The DPU generates the radar waveform (digital generator, DACs), runs the detection and estimation algorithms on the digitized echoes (ADCs, digital processing), constructs the images (formatting), and controls the unit; the RFU contains the filters, the power and low-noise amplifiers, the mixer and local oscillator for frequency conversion, the RF chains and beamforming network, and the duplexer that supports bi-directional operation on a single antenna. The formatted EO data are transmitted to the ground for further processing, and today this chain is hosted on a payload entirely separate from the communication payload. Table~\ref{Table:Frequency Band Radar} lists the bands allocated to active sensing and the bandwidth each sensor type requires to meet its resolution requirements \cite{Rec.ITU-RRS.577-72009FrequencyServices, NationalAcademiesofSciences2015ASpectrum}: scatterometers, altimeters, SARs, precipitation radars, and cloud-profile radars measure different physical properties (soil moisture, vegetation, snow, ice boundaries, ocean wind, rain rate) and therefore use different frequencies; about $100$~MHz is appropriate for most applications, while high-resolution imaging requires more \cite{Rec.ITU-RRS.577-72009FrequencyServices}.

\begin{table*}[!t]
\centering
\caption{Frequency bands allocated for spaceborne radars \cite{NationalAcademiesofSciences2015ASpectrum} \cite{Rec.ITU-RRS.577-72009FrequencyServices}} \footnotesize
\label{Table:Frequency Band Radar}
\begin{tabular}{|c|l|lllll|}
\hline
\multirow{2}{*}{\begin{tabular}[c]{@{}c@{}}Satellite Frequency \\ Bands\end{tabular}} & \multicolumn{1}{c|}{\multirow{2}{*}{\begin{tabular}[c]{@{}c@{}}Allocated Frequency\\ for Active Sensing\end{tabular}}} & \multicolumn{5}{c|}{Assigned Bandwidth for Sensors}                                                                                                                                                                                                                               \\ \cline{3-7}
                                                                                      & \multicolumn{1}{c|}{}                                                                                                  & \multicolumn{1}{c|}{Scatterometer} & \multicolumn{1}{c|}{Altimeter} & \multicolumn{1}{c|}{SAR}        & \multicolumn{1}{c|}{\begin{tabular}[c]{@{}c@{}}Precipitation \\ Radar\end{tabular}} & \multicolumn{1}{c|}{\begin{tabular}[c]{@{}c@{}}Cloud Profile\\  Radar\end{tabular}} \\ \hline
P (0.3-1 GHz)                                                                         & 432-438 MHz                                                                                                            & \multicolumn{1}{l|}{}              & \multicolumn{1}{l|}{}          & \multicolumn{1}{l|}{6 MHz}      & \multicolumn{1}{l|}{}                                                               &                                                                                     \\ \hline
L (1-2 GHz)                                                                           & 1215-1300 MHz                                                                                                          & \multicolumn{1}{l|}{5-500 kHz}     & \multicolumn{1}{l|}{}          & \multicolumn{1}{l|}{20-85 MHz}  & \multicolumn{1}{l|}{}                                                               &                                                                                     \\ \hline
S (2-4 GHz)                                                                           & 3100-3300 MHz                                                                                                          & \multicolumn{1}{l|}{}              & \multicolumn{1}{l|}{200 MHz}   & \multicolumn{1}{l|}{20-200 MHz} & \multicolumn{1}{l|}{}                                                               &                                                                                     \\ \hline
C (4-8 GHz)                                                                           & 5250-5570 MHz                                                                                                          & \multicolumn{1}{l|}{5-500 kHz}     & \multicolumn{1}{l|}{320 MHz}   & \multicolumn{1}{l|}{20-320 MHz} & \multicolumn{1}{l|}{}                                                               &                                                                                     \\ \hline
\multirow{2}{*}{X (8-12 GHz)}                                                         & 8550-8650 MHz                                                                                                          & \multicolumn{1}{l|}{5-500 kHz}     & \multicolumn{1}{l|}{100 MHz}   & \multicolumn{1}{l|}{20-100 MHz} & \multicolumn{1}{l|}{}                                                               &                                                                                     \\ \cline{2-7}
                                                                                      & 9300-9900 MHz                                                                                                          & \multicolumn{1}{l|}{5-500 kHz}     & \multicolumn{1}{l|}{300 MHz}   & \multicolumn{1}{l|}{20-600 MHz} & \multicolumn{1}{l|}{}                                                               &                                                                                     \\ \hline
\multirow{2}{*}{Ku (12-18 GHz)}                                                       & 13.25-13.75 GHz                                                                                                        & \multicolumn{1}{l|}{5-500 kHz}     & \multicolumn{1}{l|}{500 MHz}   & \multicolumn{1}{l|}{}           & \multicolumn{1}{l|}{0.6-14 MHz}                                                     &                                                                                     \\ \cline{2-7}
                                                                                      & 17.2-17.3 GHz                                                                                                          & \multicolumn{1}{l|}{5-500 kHz}     & \multicolumn{1}{l|}{}          & \multicolumn{1}{l|}{}           & \multicolumn{1}{l|}{0.6-14 MHz}                                                     &                                                                                     \\ \hline
K (18-27 GHz)                                                                         & 24.05-24.25 GHz                                                                                                        & \multicolumn{1}{l|}{}              & \multicolumn{1}{l|}{}          & \multicolumn{1}{l|}{}           & \multicolumn{1}{l|}{0.6-14 MHz}                                                     &                                                                                     \\ \hline
Ka (27-40 GHz)                                                                        & 35.5-36 GHz                                                                                                            & \multicolumn{1}{l|}{5-500 kHz}     & \multicolumn{1}{l|}{500 MHz}   & \multicolumn{1}{l|}{}           & \multicolumn{1}{l|}{0.6-14 MHz}                                                     &                                                                                     \\ \hline
\multirow{2}{*}{W (75-110 GHz)}                                                       & 78-79 GHz                                                                                                              & \multicolumn{1}{l|}{}              & \multicolumn{1}{l|}{}          & \multicolumn{1}{l|}{}           & \multicolumn{1}{l|}{}                                                               & 0.3-10 MHz                                                                          \\ \cline{2-7}
                                                                                      & 94-94.1 GHz                                                                                                            & \multicolumn{1}{l|}{}              & \multicolumn{1}{l|}{}          & \multicolumn{1}{l|}{}           & \multicolumn{1}{l|}{}                                                               & 0.3-10 MHz                                                                          \\ \hline
\multicolumn{1}{|l|}{\multirow{2}{*}{G (110-300 GHz)}}                                & 133.5-134 GHz                                                                                                          & \multicolumn{1}{l|}{}              & \multicolumn{1}{l|}{}          & \multicolumn{1}{l|}{}           & \multicolumn{1}{l|}{}                                                               & 0.3-10 MHz                                                                          \\ \cline{2-7}
\multicolumn{1}{|l|}{}                                                                & 237.9-238 GHz                                                                                                          & \multicolumn{1}{l|}{}              & \multicolumn{1}{l|}{}          & \multicolumn{1}{l|}{}           & \multicolumn{1}{l|}{}                                                               & 0.3-10 MHz                                                                          \\ \hline
\end{tabular}
\end{table*}

\subsection{Satellite-Borne PNT Systems}
\subsubsection{Purpose}
PNT systems allow users to determine their location, navigate, and synchronize operations in civilian and military domains \cite{Prol2022PositionOpportunities,Li2019LEOServicesb}. They underpin aviation, maritime, road, and rail transportation, emergency response, autonomous-vehicle guidance, asset tracking, precision farming, and the synchronization of telecommunications, broadcasting, power grids, and financial transactions \cite{spilker_jr_global_1996, misra_global_2011}, and in defense they support force coordination, targeting, and mobility. Modern satellite PNT therefore provides global, continuous, all-weather coverage with high accuracy, availability, and integrity, forming the backbone of critical infrastructure \cite{Kaplan2017UnderstandingEdition}, with growing demand for resilience in contested, obstructed, or degraded environments.

\subsubsection{Architectures}
\begin{figure}[!t]
    \centering
\includegraphics[width=0.85\linewidth]{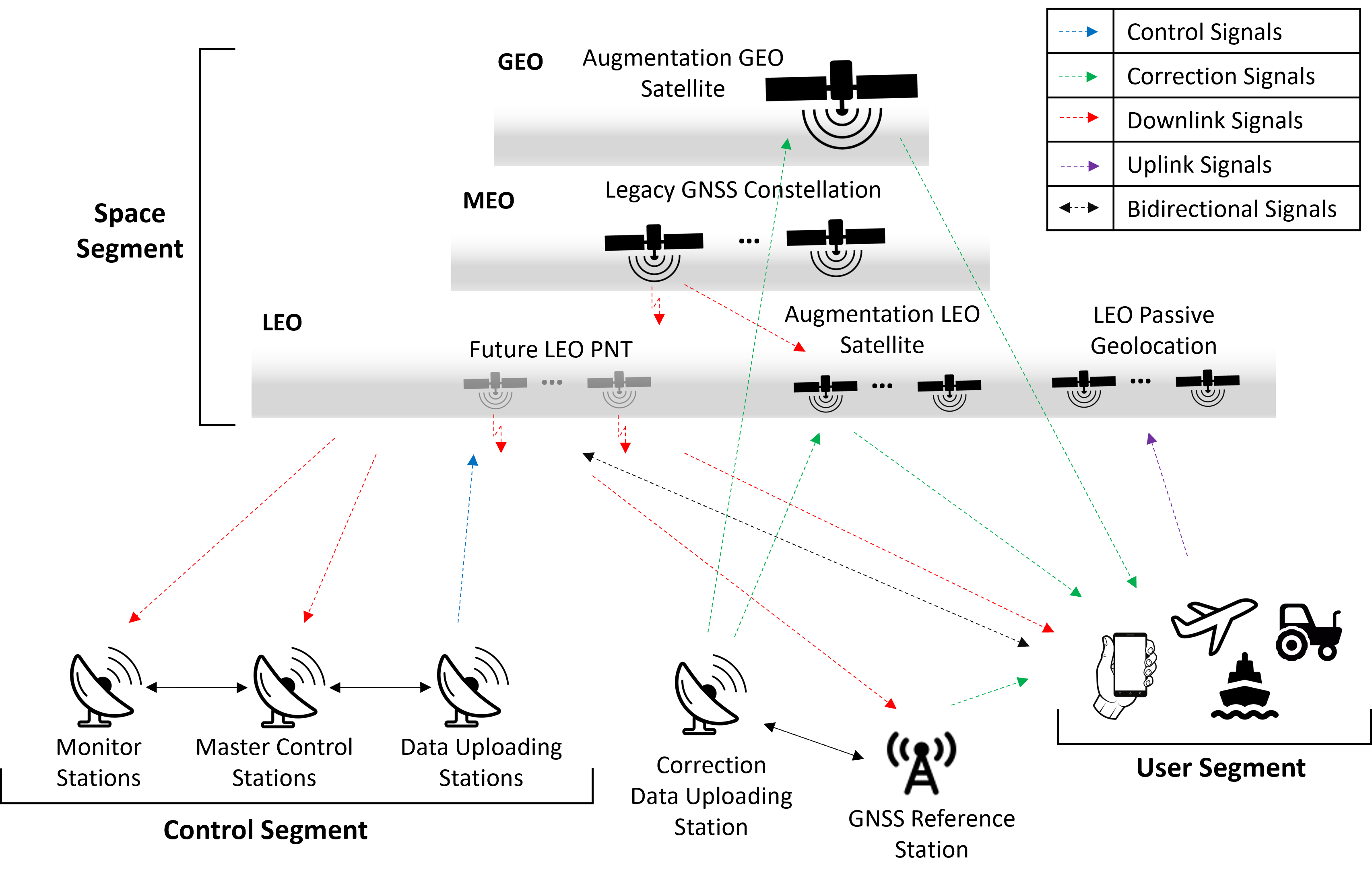}
    \caption{Legacy GNSS architecture.}
\label{fig:Legacy_GNSS_Architecture}
\end{figure}

A satellite geolocation system comprises the space, control, and user segments, whose details differ between legacy GNSS and the LEO-PNT architectures under development. \emph{Legacy GNSS:} a constellation of MEO satellites from which a receiver with at least four satellites in view computes its position by trilateration \cite{Vincent2017AsymptoticallyTrilateration}; four global constellations exist (e.g., GPS and Galileo), complemented by regional systems such as IRNSS/NavIC. As shown in Fig.~\ref{fig:Legacy_GNSS_Architecture} \cite{Kaplan2017UnderstandingEdition}, the \emph{space segment} transmits radio-navigation signals with specific structures, governed by precise onboard atomic clocks, retransmits the navigation messages received from the control segment, and may include a satellite-based augmentation system (SBAS) providing differential corrections, integrity parameters, and ionospheric data; the \emph{control segment} is a global network of master control, data-uploading, and monitoring stations that estimates satellite orbits (ephemerides) and clock errors, generates the navigation message, and uploads it to the satellites; and the \emph{user segment} is the receiver, which forms pseudoranges and other observables and solves the navigation equations for position and time. \emph{LEO-PNT:} the potential of LEO satellites for PNT has led to the LEO-PNT concept \cite{Prol2022PositionOpportunities}, with designs ranging from the full reuse of NTN broadband infrastructure to dedicated systems whose architecture depends on the positioning method. Dedicated systems customize the payload (possibly multi-frequency, as in GNSS) and the segments to replicate MEO-GNSS capabilities while exploiting the favorable LEO geometry and lower transmit power, at the cost of larger Doppler and higher sensitivity to ephemeris errors; examples are Centispace \cite{Xu2024Multi-GNSSMission} ($120$ satellites in $12$ planes at $975$~km), the $300$-satellite Xona constellation for secure augmentation and independent high-precision PNT \cite{Miller2023SNAP:Receiver}, and ARGOS (operated by CLS and Kin\'eis), an uplink-based system in which a single satellite performs time and frequency measurements of user messages that are relayed to a ground solver for Doppler-based positioning \cite{Lopez2014ImprovingFiltering}.

\subsubsection{Classification}
Satellite geolocation systems can be classified by the measurement technique. \emph{Time measurements:} the travel time between transmitter and receiver yields the distance used for trilateration; time of arrival (TOA) \cite{Kaplan2017UnderstandingEdition} is used by GNSS, its variant time difference of arrival (TDOA) \cite{Ho1993SolutionTDOA} measures the arrival-time difference between synchronized receivers for multilateration and is considered for LEO-PNT with 5G NTN signals \cite{Gonzalez-Garrido20235GScenario}, and round-trip time (RTT) \cite{Nawaz2024RoundSingle-Satellite} measures the two-way travel time. \emph{Doppler measurements:} the relative motion between transmitter and receiver shifts the carrier frequency, and the frequency of arrival (FOA) is mapped to a position \cite{Shi2023RevisitingSatellites}; FOA is used by ARGOS in the uplink and by most signals-of-opportunity (SoO) attempts in the downlink \cite{Benzerrouk2019LEOEstimation,Neinavaie2022AcquisitionResults}, possibly jointly with the Doppler rate \cite{Ellis2020UseSatellite}, while frequency difference of arrival (FDOA) between synchronized receivers is used for multilateration \cite{LI2012AOnly}, e.g., with Iridium signals \cite{Leng2015JointSignal}. \emph{Angle measurements:} angles of arrival (AOA) at several receivers of known position enable triangulation \cite{Peng2006AngleNetworks}, and angle of departure (AOD) measures the emission angle toward several receivers \cite{Wu2016SingleEnvironment}; no commercial satellite system uses AOA yet, but research is exploring it \cite{Florio2024LEO-BasedOpportunity}. \emph{Power measurements:} received signal strength (RSS) decreases with distance and can be combined at several known points to estimate position, e.g., jointly with FOA for localizing ground devices from a LEO constellation \cite{Hashim2022SatelliteMeasurements}.

\subsubsection{Applications}
Satellite PNT enables maritime and aerial navigation (route planning, real-time tracking, collision avoidance, and synchronization of communication, sensing, and control systems); autonomous-vehicle and drone navigation with lane-level accuracy and timing for cooperative maneuvers, supported by ultra-dense LEO satellite--terrestrial architectures that guarantee continuity in urban, rural, and remote areas \cite{Qin2023Service-AwareApproach}; precision farming with centimeter-level guidance of machinery and synchronized IoT devices \cite{Qin2023Service-AwareApproach}; environmental monitoring and climate research through georeferenced measurements and synchronized sensing networks \cite{Leyva-Mayorga2020LeoCommunications}; infrastructure and asset monitoring \cite{Su2019BroadbandTechnologies}; and defense and security, where resilient PNT supports troop movement, asset tracking, and mission coordination, with collaborative satellite architectures improving robustness against jamming and spoofing \cite{Yue2023CollaborativeThings}.

\subsubsection{PNT Payload}
\begin{figure}[!t]
    \centering
    \includegraphics[width=0.88\linewidth]{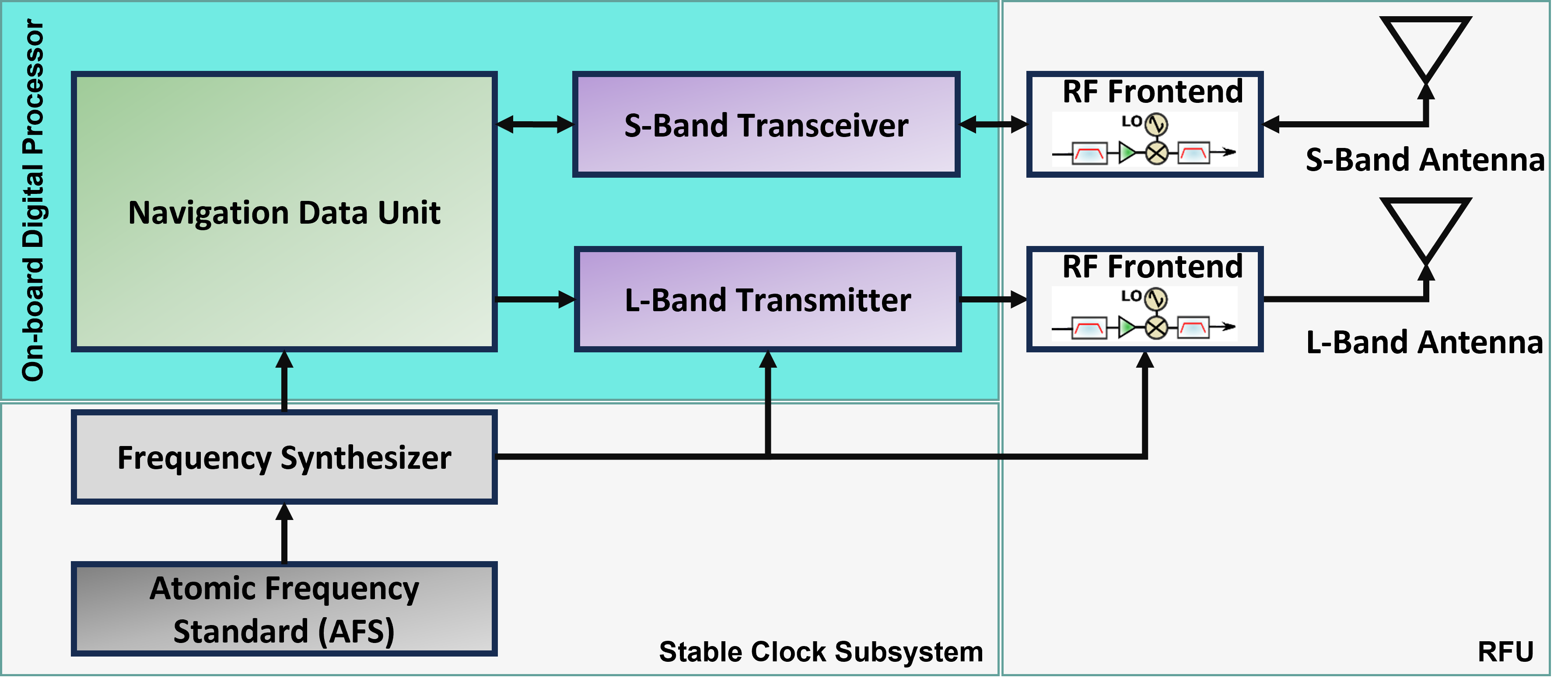}
    \caption{GNSS satellite payload.}
    \label{fig:gnss_payload}
\end{figure}
A GNSS payload continuously transmits navigation signals on two or more L-band frequencies; Fig.~\ref{fig:gnss_payload} illustrates a typical payload \cite{Kaplan2017UnderstandingEdition}. The navigation data unit (NDU) receives the orbit and clock-correction information through an S-band link and generates the navigation message; the L-band subsystem adds the spreading sequences (ranging codes), modulates the signal, and feeds the L-band RF front-end. The key enabler of GNSS performance is the atomic frequency standard (AFS) that keeps each satellite synchronized to the system time (e.g., Galileo System Time), and modern constellations add inter-satellite ranging and communication links to improve onboard orbit determination and reduce the reliance on the ground segment. GNSS operates in the L-band allocated to the radionavigation satellite service (RNSS), except NavIC, which also uses an S-band signal; Table~\ref{tab:gnss_bandwidths} summarizes the carrier frequencies, signal types, and bandwidths of the main constellations.

\begin{table*}[!t]
\scriptsize
\caption{GNSS Carrier Frequencies, Signal Types, and Bandwidths}
\centering
\begin{tabular}{|l|l|p{11.5cm}|}
\hline
\textbf{GNSS System (Country)} & \textbf{Band [GHz]} & \textbf{Signal Types and Approximate Bandwidths [MHz]} \\ \hline

\multirow{3}{*}{GPS (US)}
&   L1: 1.57542 & C/A: 2.046 MHz, P(Y): 20.46 MHz, L1C: 24 MHz (modernized) \\
& L2: 1.22760 & P(Y): 20.46 MHz, L2C (civil): 24 MHz \\
& L5: 1.17645 & L5 (safety-of-life): 24 MHz, BPSK(10) \\ \hline

\multirow{4}{*}{Galileo (EU)}
& E1: 1.57542 & E1-B/C (Open Service): 15 MHz, CBOC modulation \\
& E5a: 1.17645 & E5a-I/Q (Open Service): 20 MHz, AltBOC(15,10) \\
& E5b: 1.20714 & E5b-I/Q (Open/Commercial): 20 MHz, AltBOC(15,10) \\
& E6: 1.27875 & E6-B/C: 40 MHz (Commercial Service), BPSK(10) or QPSK \\ \hline

\multirow{3}{*}{Glonass (Russia)}
& G1: 1.598--1.609 & L1OF (legacy FDMA): 0.511 MHz per channel, L1OC (modern): 10 MHz BPSK/QPSK \\
& G2: 1.243--1.251 & L2OF (legacy FDMA): 0.511 MHz, L2OC (modern): 10 MHz \\
& G3: 1.189--1.214 & L3OC (Glonass-K): up to 25 MHz, QPSK \\ \hline

\multirow{3}{*}{Beidou (China)}
& B1: 1.561098 & B1I: 4.092 MHz (legacy), B1C: 24 MHz (modern, similar to GPS L1C) \\
& B2: 1.20714 & B2a (Open Service): 20 MHz, AltBOC(15,10); B2b (Commercial): 20 MHz \\
& B3: 1.26852 & B3I: 10 MHz (legacy), B3A (BD-3): up to 40 MHz \\ \hline

\multirow{4}{*}{QZSS (Japan)}
& L1: 1.57542 & L1C/A: 2.046 MHz, L1C: 24 MHz \\
& L2: 1.22760 & L2C: 24 MHz \\
& L5: 1.17645 & L5: 24 MHz, BPSK(10) \\
& L6: 1.27875 & L6-D (MADOCA augmentation): 40 MHz \\ \hline

\multirow{2}{*}{NavIC / IRNSS (India)}
& L5: 1.17645 & SPS (Standard Positioning Service): 24 MHz \\
& S-band: 2.492028 & SPS and RS (Restricted Service): $\sim$16 MHz; BPSK or spread-spectrum (classified details) \\ \hline

\end{tabular}
\label{tab:gnss_bandwidths}
\end{table*}

\revi{rev1.1.4}{\emph{Convergence boundary.} The three classes share the same segment structure but inherit different design centers: communications is driven by rate, latency, reliability, and coverage and consumes spectrum, EIRP, beams, gateway capacity, and processing; sensing is driven by resolution, detection and estimation quality, swath, and revisit and consumes bandwidth, aperture, peak power, coherent time, and processing; PNT is driven by accuracy, integrity, continuity, availability, and timing stability and depends on clock stability, ranging bandwidth and codes, ephemeris quality, geometry, and monitoring. In this survey, a payload is not multi-functional merely because communication telemetry serves a sensing mission or because a sensor product is geotagged with an independent GNSS receiver: MFSS requires deliberate sharing or joint coordination of payload hardware, spectrum, waveforms, timing, processing, or mission resources across at least two service functions, which is the subject of the following sections.}
\section{Cooperative MFSS Payloads}
\label{sec:cooperative_mfss}\label{sezione_6}

\begin{figure}[!t]
   \centering
    \includegraphics[width=0.86\columnwidth]{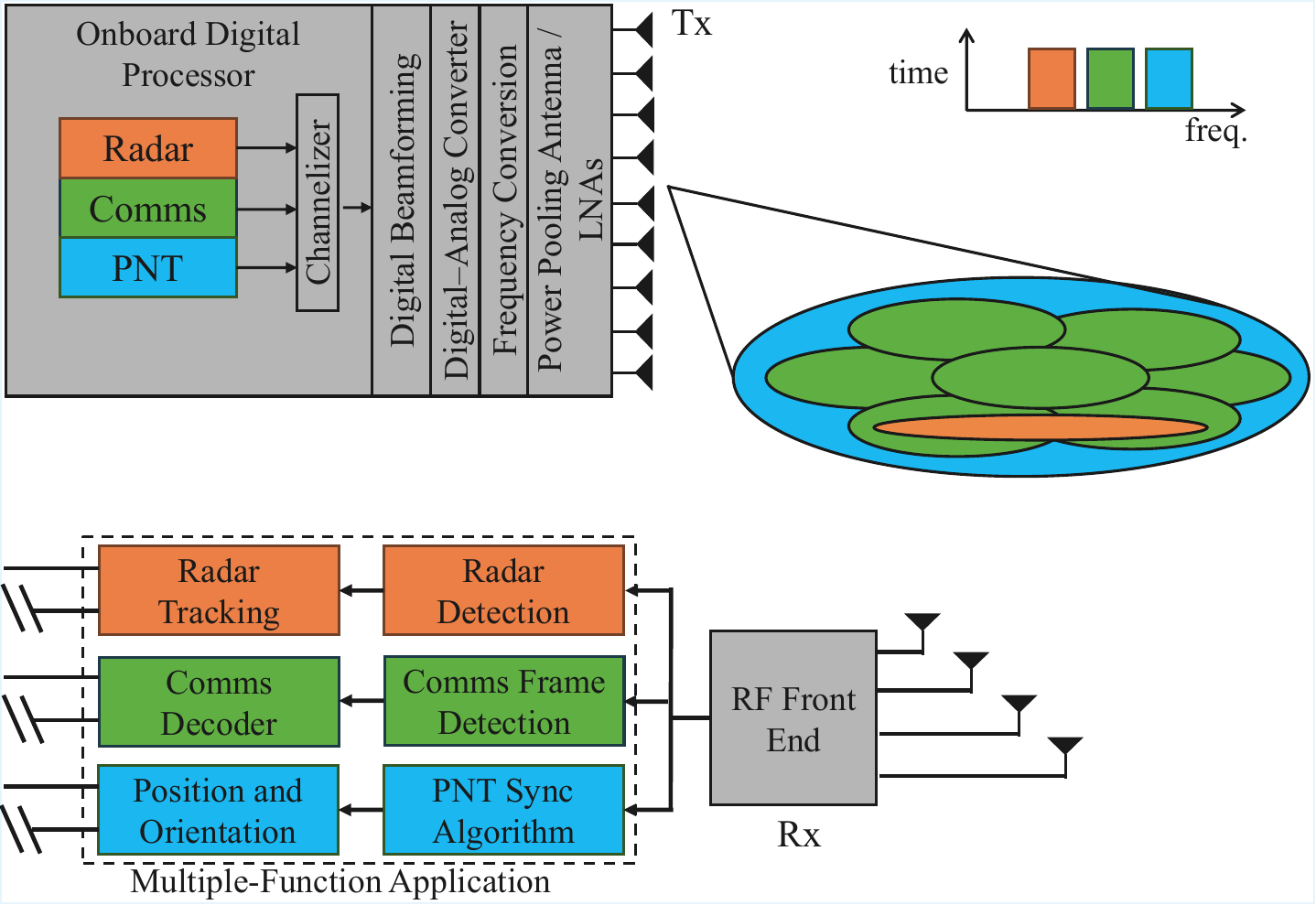}
    \caption{Cooperative MFSS payload.}
    \label{fig:cooperativepayload}
\end{figure}

Cooperative MFSS represent the first and most practical level of integration. Multiple functions are hosted on a common platform and share selected payload resources (antenna aperture, RF front-end, power amplifiers, frequency converters, channelizers, digital beamforming units, and onboard processors), while each function preserves its own waveform, frame structure, processing chain, and performance objectives. As illustrated in Fig.~\ref{fig:cooperativepayload}, communications, radar/sensing, and PNT remain logically separated inside the onboard processor while sharing the RF, antenna, and digital-processing resources.

\subsection{Architecture}
The cooperative architecture relies on hardware sharing and coordinated resource management rather than waveform-level fusion. Communications may employ conventional satellite waveforms (DVB-S2X-like frames, 5G/6G NTN waveforms), sensing may rely on dedicated radar signals (pulsed, FMCW, chirp, scatterometer, or SAR waveforms), and PNT may use GNSS-like ranging codes, synchronization pilots, Doppler observables, or timing beacons; the signals occupy adjacent bands, separate time slots, different beams, or independently scheduled service windows. The onboard processor assigns frequency channels, time slots, beams, transmit power, and processing resources to each service while limiting cross-function interference: a radar burst may be scheduled in a low-traffic communication interval, PNT synchronization signals inserted periodically, and communication traffic routed through beams spatially separated from active sensing beams. Conventional payloads can thus be consolidated without a radical redesign of their signal structures.

\subsection{Classification}
\revi{rev1.2.3}{Cooperative MFSS are classified according to the functions that coexist on the shared platform, i.e., cooperative JCAS, JCAP, JSAP, or JCSAP. In cooperative JCAS, communications and sensing share the satellite hardware but retain separate waveforms and receiver chains (e.g., a communication satellite that schedules radar bursts, or a sensing mission that reuses the communication subsystem for high-rate downlink). In cooperative JCAP, PNT is supported by pilots, synchronization sequences, Doppler observables, or signals of opportunity of the communication payload, or by a hosted PNT payload that shares only the bus and selected resources \cite{Iannucci2020EconomicalGNSS,Ardito2019PerformancePositions,GPSWorldStaff2016IridiumService}. In cooperative JSAP, PNT provides georeferencing, orbit determination, time tagging, and attitude information to the sensing chain, while sensing products assist navigation and situational awareness. In cooperative JCSAP, the three functions coexist and a payload controller coordinates them through scheduling, calibration exchange, timing alignment, and interference-aware resource allocation. Table~\ref{tab:level_config_mapping} summarizes how each configuration is realized at this and at the two deeper integration levels; the configuration-specific details are given in Sections~\ref{sec:jcas}--\ref{sec:jcsap}.}

\begin{table*}[!t]
\begin{revblock}{rev1.2.4}
\centering
\caption{Realization of the four MFSS configurations at the three integration levels.}
\label{tab:level_config_mapping}
\scriptsize
\begin{tabularx}{\textwidth}{L{1.2cm}XXX}
\toprule
\textbf{Config.} & \textbf{Cooperative (shared hardware, separate waveforms)} & \textbf{Integrated (shared hardware and resource grid, separate waveforms)} & \textbf{Joint (common waveform, unified processing)} \\
\midrule
JCAS & Radar bursts scheduled in idle communication intervals or adjacent bands; communication subsystem reused for sensing-data downlink. & Communication frames and sensing pulses multiplexed in time, frequency, space, or code on a common resource mapper; blanking and space-time processing suppress leakage. & One waveform carries data and provides radar-friendly ambiguity properties (OFDM/ODDM radar-communication, RSMA common stream as radar sequence); joint decoding and delay--Doppler--angle estimation. \\
\midrule
JCAP & Signals of opportunity or hosted PNT payload (e.g., Iridium STL) sharing bus, aperture, and power; navigation processing at the receiver. & Fused LEO-PNT: pilots/PRS, ephemeris and clock dissemination, and timing references managed on the communication resource grid. & Waveform designed for ranging (e.g., DSSS overlaid on OFDM with a controlled power split); joint synchronization, data, and PNT extraction at the receiver. \\
\midrule
JSAP & PNT receiver provides time, orbit, and attitude to a separate sensing chain; sensing products used for terrain-aided navigation and integrity on the ground. & Sensing measurements time-tagged and georeferenced in the resource grid; range/Doppler/angle observables feed navigation and integrity; navigation and sensing scheduled on shared beams. & Dual-function frames in which navigation ranging and sensing illumination share beams and power (joint PVT/SAINR design); reflected navigation signals used for reflectometry. \\
\midrule
JCSAP & All three functions on one bus with coordinated scheduling and calibration exchange, each keeping its waveform. & Data, sensing pulses, and PNT synchronization scheduled in a unified resource grid with interference-aware power, beam, and time control. & A single waveform supports data, sensing observables, and ranging/timing; constellation-level cooperation through ISLs (distributed sensing, localization, and beamforming). \\
\bottomrule
\end{tabularx}
\end{revblock}
\end{table*}

\subsection{State of the Art}
Cooperative MFSS reuse existing single-functional technologies with limited modifications. In JCAS, communications and sensing remain distinct services coordinated through shared resources, e.g., learning-based waveform and receive-filter optimization \cite{Lukito2025LearningPerspective}, beam-squint-aware massive MIMO LEO JCAS \cite{You2022BeamSystems}, and secure NOMA-enabled JCAS \cite{Huang2024SecureCommunication}. In JCAP, signals-of-opportunity positioning exploits existing LEO communication signals without operator cooperation, shifting the burden of estimating Doppler, angle, power, and delay observables to the receiver \cite{Neinavaie2022AcquisitionResults,Orabi2021OpportunisticSatellites}, cooperative signals of opportunity add operator-provided clock and ephemeris information \cite{Ardito2019PerformancePositions}, and hosted PNT payloads add dedicated hardware to communication satellites, as in the Iridium-based satellite time and location service \cite{GPSWorldStaff2016IridiumService}. For JCSAP, the cooperative architecture is the realistic near-term path for LEO mega-constellations, whose dense coverage, inter-satellite links, and onboard processing can host the three functions without a unified waveform; cooperative JSAP and JCSAP nevertheless remain less mature than JCAS and JCAP (Sections~\ref{sec:jsap} and~\ref{sec:jcsap}).

\subsection{Advantages}
The primary advantage of cooperative MFSS is low implementation risk: each service retains its own waveform and processing chain, so that mature communication, radar, and PNT technologies are reused with limited modification, each function can be validated independently before coexistence is assessed, and the path to deployment and certification is short. Consolidating payloads on a shared bus reduces SWaP-C through common antennas, RF front-ends, processors, timing references, and thermal control. Finally, cross-function assistance is available without deep waveform integration: PNT supplies timing, position, attitude, and orbit information for sensing geolocation and beam pointing, sensing supplies environmental awareness and propagation information for adaptive communications, and communications distribute sensing products and PNT corrections, all while preserving modularity.

\subsection{Cooperative Configuration-Specific Key Challenges}
The main challenge is interference management among heterogeneous waveforms: high-power radar transmissions can saturate communication or PNT receivers, broadband communication signals raise the noise floor of sensing and weak PNT measurements, and independently designed sidelobes, timing structures, and power levels are not naturally compatible. Resource scheduling is a second challenge, since communications require continuous service, sensing coherent dwell time, and PNT stable periodic references, which conflict in LEO where visibility windows are short and geometry changes rapidly. Hardware compromise limits performance, because antennas optimized for communications may lack the beamwidth, polarization, or phase stability required for radar imaging, RF chains optimized for radar peak power may lack the linearity needed for high-throughput communications, and communication-centric payloads may not offer the clock stability and predictable signal structures that PNT requires. Finally, coexistence must be verified at system level, since subsystems that individually meet their requirements may still fail through cross-function coupling, thermal interactions, calibration drift, or scheduling conflicts.

\section{Integrated MFSS Payloads}
\label{sec:integrated_mfss}

\begin{figure}[!t]
    \centering
    \includegraphics[width=0.86\columnwidth]{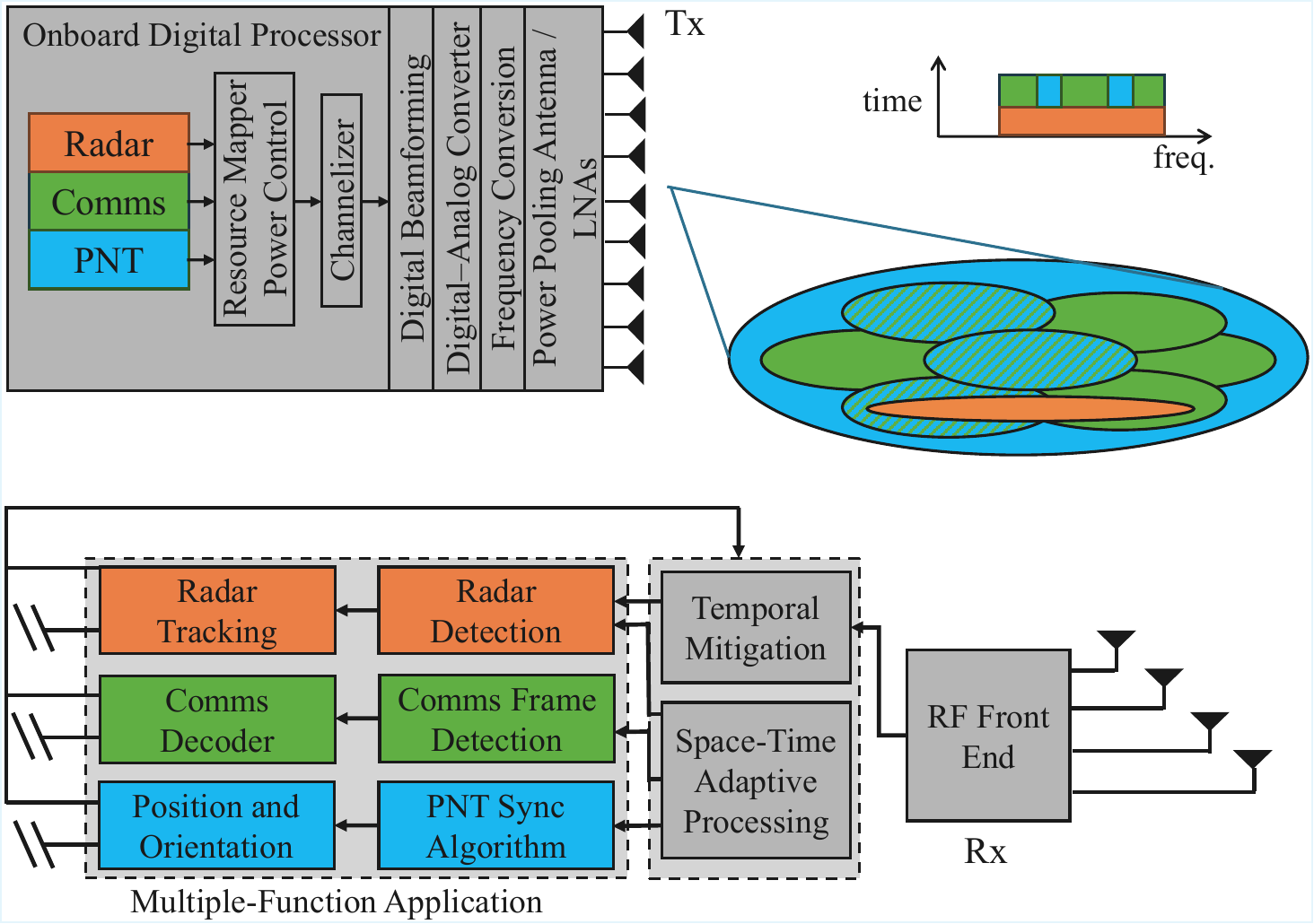}
   \caption{Integrated MFSS payload.}
    \label{fig:integratedpayload}
\end{figure}

Integrated MFSS represent the second level of multi-functionality. Unlike cooperative payloads, where functions share hardware but are scheduled separately, integrated payloads coordinate the use of common spectral, temporal, spatial, code, and power resources through a resource mapper, a power-control module, a temporal-mitigation block, and a space-time adaptive processing block (Fig.~\ref{fig:integratedpayload}), which allow the functions to coexist in overlapping or shared resources.

\subsection{Architecture}
The integrated payload consists of a shared aperture, RF front-end, channelizer, frequency-conversion chain, digital beamforming unit, and onboard processor, with a cross-functional resource mapper that jointly controls time, frequency, beam, code, and power resources. The functions may still use distinct waveforms, but their resource occupation is planned jointly: time-division multiplexing assigns intervals to communication frames, radar pulses, and PNT reference signals; frequency-domain multiplexing allocates subbands or subcarriers with adaptive guard bands; spatial multiplexing separates communication users, sensing targets, and PNT reference directions with phased arrays, MIMO processing, and digital beamforming; and code-domain multiplexing uses (quasi-)orthogonal sequences for services sharing the same time-frequency resources. The receive chain needs stronger interference suppression than in the cooperative case: temporal mitigation blanks, cancels, or suppresses high-power transmissions that overlap sensitive reception intervals, and space-time adaptive processing suppresses sidelobe leakage, clutter, jamming, and cross-function interference.

\subsection{Classification}
\revi{rev1.2.5}{Integrated MFSS are classified by the functions that are jointly coordinated on the common resource grid (Table~\ref{tab:level_config_mapping}). Integrated JCAS jointly schedules communication frames and sensing pulses, or uses beamforming to maintain user links while illuminating targets, since both throughput and sensing accuracy depend on bandwidth, power, beam direction, and dwell time. Integrated JCAP coordinates communications and PNT through pilots, synchronization signals, ephemeris dissemination, Doppler tracking, and timing references, managing the frame structure and pilot resources so that positioning and timing observables are supported without a full waveform redesign. Integrated JSAP time-tags, georeferences, and aligns sensing measurements with satellite attitude and orbit information, while range, Doppler, angle, and environmental maps improve navigation, tracking, and integrity monitoring. Integrated JCSAP schedules communication data, radar sensing, and PNT synchronization in a unified resource grid while mitigating interference through time-domain blanking, beam control, power control, and adaptive processing.}

\subsection{State of the Art}
Integrated MFSS are most developed for JCAS and JCAP. Sensing-assisted random access in satellite-terrestrial IoT relay networks \cite{Zhao2023IntegratedNetworks}, jointly optimized beamforming and sensing waveforms in cooperative multi-satellite constellations \cite{Wang2025Multiple-SatelliteConstellations}, secure NOMA-enabled JCAS \cite{Huang2024SecureCommunication}, and beam-squint-aware massive MIMO JCAS \cite{You2022BeamSystems} show that integrated JCAS must jointly address payload architecture, resource allocation, beamforming, and security. For JCAP, fused LEO-PNT concepts intentionally use the communication signal and infrastructure for navigation: cooperative signals of opportunity assume that clock and orbit corrections are delivered through the communication signal or an auxiliary link \cite{Ardito2019PerformancePositions}, and fused LEO-PNT slightly modifies the broadband downlink so that it supports navigation observables \cite{Iannucci2020EconomicalGNSS}, at lower cost than a hosted payload. Integrated JSAP and JCSAP are natural extensions of these trends and are discussed in Sections~\ref{sec:jsap} and~\ref{sec:jcsap}.

\subsection{Advantages}
Integrated MFSS provide higher spectral efficiency than cooperative payloads, because the functions reuse the same band, aperture, and processing through time, space, code, and power coordination rather than strict spectral separation, which matters where spectrum is scarce, mass is limited, and antenna real estate is expensive. The resource mapper can explicitly optimize the functions together, which strengthens the cross-function gains: sensing provides environmental maps, blockage awareness, mobility information, and channel prediction to communications; communications support data fusion, cooperative control, and rapid product delivery; and PNT provides timing, synchronization, orbit information, and georeferencing to both. Integrated payloads are also operationally flexible, since resources can be shifted toward sensing and emergency communications during disasters, toward broadband in normal operation, or toward resilient PNT and situational awareness in contested environments.

\subsection{Integrated Configuration-Specific Key Challenges}
Integrated MFSS need tight synchronization and real-time coordination: time-division integration requires precise timing alignment, spatial integration accurate beam pointing and sidelobe control, and code-domain integration low cross-correlation among waveforms with different bandwidth, power, and processing gain. Residual interference persists, because non-ideal filters, beam sidelobes, Doppler, phase noise, and hardware nonlinearities create leakage even when resources are nominally separated, which is critical when high-power transmissions coexist with weak PNT signals; cancellation, adaptive filtering, and space-time processing mitigate it at the cost of computation and power. Resource allocation is a multi-objective problem that balances rate, sensing resolution, detection probability, positioning accuracy, latency, reliability, and power under conflicting dependencies (bandwidth improves range resolution and delay estimation but may reduce spectral efficiency; radar dwell time improves sensing but reduces throughput) and rapidly changing geometry. Finally, centralized optimization may be impractical for mega-constellations with thousands of satellites, beams, users, and targets, distributed coordination adds signaling overhead and convergence issues, and traditional communication, radar, and PNT test procedures do not assess cross-functional performance in shared resources.

\section{Joint MFSS Payloads}
\label{sec:joint_mfss}

\begin{figure}[!t]
    \centering
    \includegraphics[width=0.86\columnwidth]{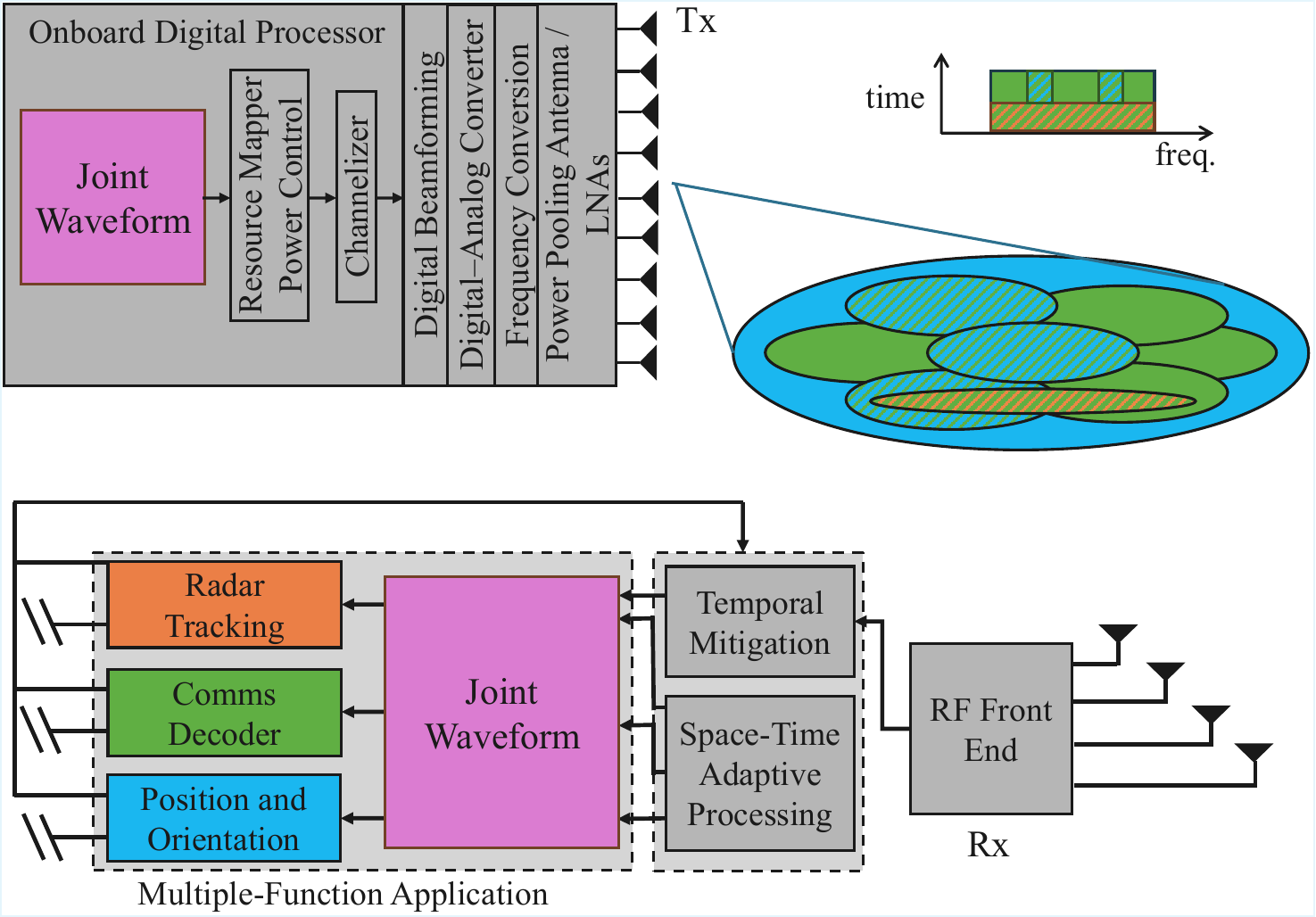}
    \caption{Joint MFSS payload.}
    \label{fig:jointpayload}
\end{figure}

Joint MFSS represent the highest level of integration: communications, sensing, and PNT share the same hardware, spectrum, and waveform. A common waveform transmits information, probes the environment, and provides timing or ranging observables; as shown in Fig.~\ref{fig:jointpayload}, the separate waveform-generation chains are replaced by a joint waveform block, and the same signal is processed at the receiver for data decoding, radar tracking, and position/orientation estimation.

\subsection{Architecture}
The joint payload is built around a unified waveform generator, resource mapper, power-control unit, digital beamforming stage, channelizer, frequency-conversion chain, shared aperture, and RF front-end, and waveform design, pilot placement, data embedding, beamforming, power allocation, synchronization, and receiver processing are optimized together rather than merely coordinated. A joint waveform may embed communication symbols while preserving radar-friendly ambiguity properties, and may include pilots or synchronization structures enabling delay and Doppler estimation, clock synchronization, and positioning; multi-carrier radar-communication signals, spread-spectrum meta-signals, chirp-embedded data signals, and pilot-rich broadband waveforms are examples. In the satellite context the waveform must also cope with Doppler dynamics, long propagation delay, atmospheric effects, beam squint, nonlinear high-power amplifiers, and limited onboard processing. At the receiver, a common RF front-end captures the signal and joint processing (interference cancellation, matched filtering, channel estimation, delay-Doppler-angle estimation, synchronization, and decoding) extracts data, sensing observables, and PNT measurements; in full JCSAP systems the receiver must jointly handle direct-path signals, echoes, multipath, clutter, user data, timing information, and Doppler signatures from the same waveform.

\subsection{Classification}
\revi{rev1.2.6}{Joint MFSS are classified according to the functions embedded in the common waveform (Table~\ref{tab:level_config_mapping}). A joint JCAS waveform simultaneously supports data transmission and sensing, with communication symbols, pilots, preambles, or subcarriers used to estimate target range, Doppler, angle, or environmental parameters; this is the most widely studied category. A joint JCAP waveform is explicitly designed to support PNT observables such as TOA, TDOA, FOA, Doppler, RSS, or carrier phase together with data delivery. A joint JSAP waveform is a sensing waveform that also supports PNT, so that radar echoes, delay-Doppler measurements, and angular observables assist positioning and orientation estimation while PNT information improves coherent sensing and geolocation. A joint JCSAP waveform supports all three functions simultaneously and requires the joint optimization of waveform structure, beamforming, synchronization, power allocation, coding, and receiver processing.}

\subsection{State of the Art}
Joint MFSS are dominated by JCAS: learning-based joint waveform and receive-filter design \cite{Lukito2025LearningPerspective}, beam-squint-aware hybrid beamforming with statistical CSI \cite{You2022BeamSystems}, secure NOMA-based precoding \cite{Huang2024SecureCommunication}, sensing-assisted random access \cite{Zhao2023IntegratedNetworks}, and jointly optimized beamforming and sensing waveforms in multi-satellite constellations \cite{Wang2025Multiple-SatelliteConstellations} illustrate that joint satellite waveforms must be evaluated together with the antenna architecture, the resource allocation, and the security requirements, while joint sensing, communication, and computing with UAVs and LEO satellites \cite{Zhang2025IntegratedAllocation,Zuo2024IntegratingOpportunities} points beyond the four configurations toward onboard computing and AI. For JCAP, the joint waveform concept corresponds to fused LEO-PNT with an intentionally adapted communication signal \cite{Iannucci2020EconomicalGNSS}, which provides better control of bandwidth, pilots, synchronization, and timing observability than signals of opportunity but must satisfy strict timing, clock-stability, and integrity requirements. Joint JSAP and JCSAP are the least developed, although the first explicit designs have recently appeared (Sections~\ref{sec:jsap} and~\ref{sec:jcsap}).

\subsection{Advantages}
Joint MFSS provide the highest spectral efficiency, because one waveform serves all functions in the same bandwidth, time interval, beam, and transmit power, and the strongest hardware efficiency, because separate waveform generators and functional chains are replaced by a unified waveform and shared processing, which lowers SWaP-C, lets small satellites support functions that would otherwise require multiple payloads, and improves energy efficiency since one transmission yields data, sensing measurements, and PNT observables. Mutual enhancement is embedded in the signal and processing design: sensing provides channel-state awareness, blockage prediction, target mobility, and clutter maps; communications distribute sensing data, corrections, and cooperative control messages; and PNT provides the time and position references for coherent sensing, beam alignment, inter-satellite coordination, and georeferencing. Joint MFSS also support low-latency autonomous operation with reduced signaling overhead, and joint waveforms in constellations enable distributed sensing, cooperative localization, and coordinated beamforming.

\subsection{Joint Configuration-Specific Key Challenges}
The most fundamental challenge is the conflict among waveform requirements: communications favor spectral efficiency, adaptive modulation, coding gain, and flexible scheduling; sensing favors wide bandwidth, high transmit power, coherent processing, low sidelobes, and favorable ambiguity functions; and PNT favors stable timing, predictable signal structures, ranging accuracy, integrity monitoring, and robust synchronization, so that a waveform optimized for one function degrades the others. Joint waveform design is therefore a multi-objective problem in which data randomness improves throughput but degrades ambiguity properties, pilot density improves sensing and PNT accuracy but reduces spectral efficiency, transmit power improves sensing range but worsens interference and consumption, and bandwidth improves resolution and timing accuracy but increases hardware complexity. Receiver design is equally demanding, since data, echoes, and PNT observables must be extracted from the same signal under clutter, multipath, Doppler, oscillator instability, nonlinear distortion, and, for near-simultaneous transmit-receive operation, self-interference. Real-time joint decoding, delay-Doppler-angle estimation, beamforming, tracking, positioning, and interference mitigation can exceed the capacity of space-qualified processors, and AI-based alternatives raise reliability, interpretability, and certification questions. Finally, unified evaluation frameworks are missing: communications are evaluated by rate, BER, latency, and availability, radar by resolution, detection and false-alarm probabilities, and estimation accuracy, and PNT by accuracy, integrity, continuity, and availability, so that designs cannot yet be compared, certified, or standardized across functions. Joint MFSS thus offer the highest potential gain but require the most substantial advances in waveform theory, payload architecture, signal processing, optimization, and validation.
\section{JCAS in Satellite Systems}
\label{sec:jcas}

\revi{rev1.2.7}{JCAS is the MFSS configuration in which the satellite payload transports data and, at the same time, extracts properties of targets or of the environment from the transmitted signal. This section describes its architecture and possible realizations, classifies the existing designs, reviews the state of the art, summarizes its advantages, illustrates its achievable performance with results from the literature, and then identifies its key challenges and research directions.}

\subsection{Architecture and Realizations}
\revi{rev1.2.8}{A satellite JCAS system reuses the communication payload of Fig.~\ref{fig:sc_transponder} as a radar transmitter and, depending on the geometry, as a radar receiver. In the \emph{monostatic} realization the same satellite transmits the dual-function signal and receives the echoes (Fig.~\ref{fig:rs_monostatic}); aperture, RF chains, and OBDP are fully shared, but transmit and receive operate simultaneously on the same platform, so that self-interference and the large dynamic range between direct and reflected signals become the dominant hardware constraints. In the \emph{bistatic} and \emph{multistatic} realizations the echoes are collected by a second satellite, a gateway, or a terrestrial node (Fig.~\ref{fig:rs_bistatic}), which removes self-interference and adds geometric diversity at the cost of nanosecond-level synchronization between separated platforms and of a link to the fusion point. In all cases sensing competes with communications for bandwidth, radiated power, beams, dwell time, and processing, and the two-way path loss of the echo is far larger than the one-way loss of the communication link, so that a sensing beam must dwell longer, radiate more power, or integrate coherently over longer intervals than a communication beam of the same aperture. JCAS can be realized cooperatively (radar bursts or SAR acquisitions in idle communication intervals or adjacent bands), in an integrated manner (communication frames and sensing pulses multiplexed in time, frequency, space, or code), or jointly (one waveform carries the data and provides radar-friendly ambiguity properties, e.g., OFDM or orthogonal delay--Doppler multiplexing waveforms whose pilots serve channel estimation and imaging, or an RSMA common stream reused as a radar sequence).}

\subsection{Classification}
\revi{rev1.2.9}{Satellite JCAS designs can be classified along four axes: \emph{by geometry}, into monostatic, bistatic, and multistatic systems whose receiver may be a cooperating satellite, a gateway, or a terrestrial base station; \emph{by design philosophy}, into communication-centric designs that reuse an existing waveform (OFDM, DVB-S2X, 5G NTN) for sensing with minimal changes, sensing-centric designs in which a radar or SAR waveform carries embedded data, and jointly optimized designs in which waveform, beamforming, and receiver are co-designed under both sets of constraints; \emph{by sensing objective}, into Earth-surface imaging and environmental monitoring, detection and tracking of maritime, aerial, or ground targets, atmospheric sensing (e.g., rain attenuation) exploited for link adaptation, and space situational awareness such as debris detection; and \emph{by integration level}, into the realizations of Table~\ref{tab:level_config_mapping}. Most published satellite JCAS studies belong to the jointly optimized, monostatic, target-detection class, while bistatic and imaging-oriented designs have emerged only recently.}

\subsection{State of the Art}
\label{sec:jcas_sota}
\revi{rev1.2.10}{The recent literature on satellite JCAS has grown rapidly. We classify a study as satellite JCAS when communications and sensing are jointly designed or supported by a shared waveform, RF chain, aperture, processing chain, or resource-allocation mechanism; surveys, generic terrestrial studies, and works in which sensing is used only as side information are not counted.} The work \cite{Lukito2025LearningPerspective} proposed a learning-assisted transceiver that jointly optimizes the transmit waveform and receive filter without explicit channel estimation. Reference \cite{You2022BeamSystems} studied massive-MIMO LEO JCAS under statistical propagation and beam-squint effects, proposing a beam-squint-aware hybrid analog/digital design. Secure NOMA-based precoding for perfect and imperfect eavesdropper CSI was developed in \cite{Huang2024SecureCommunication}. Dynamic JCAS-assisted random access for satellite--terrestrial IoT relay networks was considered in \cite{Zhao2023IntegratedNetworks}, and \cite{Wang2025Multiple-SatelliteConstellations} jointly optimized communication beamforming and sensing waveforms for a dual-function multi-satellite constellation.

\revi{rev1.2.11}{Subsequent work has moved from general feasibility toward impairment-aware and resource-aware designs. In \cite{Yin2024LEOISAC}, the authors distinguished opportunistic from optimized LEO ISAC and showed how rate-splitting multiple access (RSMA) exposes the communication--sensing trade-off. The authors of \cite{Liu2024QuantizedLEOISAC} then included low-resolution digital-to-analog converters and formulated a max--min fair energy-efficiency design under sensing constraints. Other studies used sensing to estimate atmospheric (rain) attenuation and improve NTN association and resource allocation \cite{LeyvaMayorga2025AtmosphericSensing}, separated the LEO transmitter from a ground-based radar receiver in a bistatic RSMA architecture \cite{Park2025BistaticISAC}, and developed partially connected hybrid precoding for RSMA-enabled, multi-target LEO ISAC serving power-grid IoT (PIoT) devices \cite{Liu2025HybridPrecodingLEOISAC}.}

\revi{rev1.2.12}{The architectural scope has also broadened. An integrated communication-and-remote-sensing design combines an orthogonal delay--Doppler multiplexing waveform, a 5G-NR-compatible protocol, and a software-defined-radio prototype \cite{Xu2025IntegratedRemoteSensing}; another work embeds debris sensing into space-shift-keyed communication signals \cite{Ngoufo2025SSKDebris}. Joint satellite--terrestrial beamforming \cite{Wang2026SatelliteTerrestrialISAC} and gateway-assisted bistatic LEO sensing \cite{Zhang2026BistaticLEOISAC} distribute the sensing receiver across the ground segment. Recent works additionally address robust design under channel-phase uncertainty \cite{Yang2026RobustLEOISAC}, maritime satellite--terrestrial sensing \cite{Xiong2026MaritimeISAC}, energy-efficient satellite--UAV MIMO ISAC \cite{Thu2026SatelliteUAVISAC}, multistatic coexistence between satellite and terrestrial networks \cite{Jee2025IntegratedCoexistence}, and centralized versus distributed sensing in cooperative multi-satellite networks \cite{Kim2026CooperativeMultisatellite}. Except for the software-defined-radio prototype of \cite{Xu2025IntegratedRemoteSensing}, these contributions are analytical or simulation-based, so that the flight validation of a unified JCAS payload remains an open step.}

\revi{rev1.2.13}{\emph{Adjacent extension---joint sensing, communication, and computing.} Satellite research has also extended JCAS to include computation: \cite{Zhang2025IntegratedAllocation} developed a NOMA--MIMO NTN in which UAV sensing and offloading are supported by LEO computing resources, with beamforming, offloading, and resource allocation decomposed to minimize energy consumption, and \cite{Zuo2024IntegratingOpportunities} discussed the joint design of communication, sensing, and computation for satellite IoT. These studies are an adjacent extension because computation lies outside the four-function taxonomy adopted here.}

\subsection{Advantages and Functional Synergies}
\revi{rev1.2.14}{The advantages of satellite JCAS are of three kinds. First, resource reuse: communication signals, pilots, beams, spectrum, and aperture are reused for sensing, so that a constellation deployed for connectivity also delivers environmental or target observations without a dedicated radar payload, and a radar mission can reuse its transmitter and downlink chain for connectivity. Second, sensing-to-communication assistance: sensing supports beam alignment, blockage and rain-attenuation awareness \cite{LeyvaMayorga2025AtmosphericSensing}, propagation maps, mobility prediction, user tracking, and interference management, which improves link robustness and throughput in the rapidly changing NGSO geometry. Third, communication-to-sensing assistance: the links disseminate sensing products with low latency and enable cooperative processing across satellites, gateways, and terrestrial nodes, turning a constellation into a distributed multistatic sensor \cite{Kim2026CooperativeMultisatellite,Park2025BistaticISAC}. These synergies are strongest in the joint realization, where they are embedded in waveform and receiver design, but they are already available in cooperative and integrated payloads through the exchange of sensing products and scheduling information.}

\subsection{Achievable Performance and Trade-Offs}
\label{sec:tradeoffs_jcas}
\begin{figure}[!t]
\begin{revblock}{rev1.4.4}
\centering
\includegraphics[width=\columnwidth]{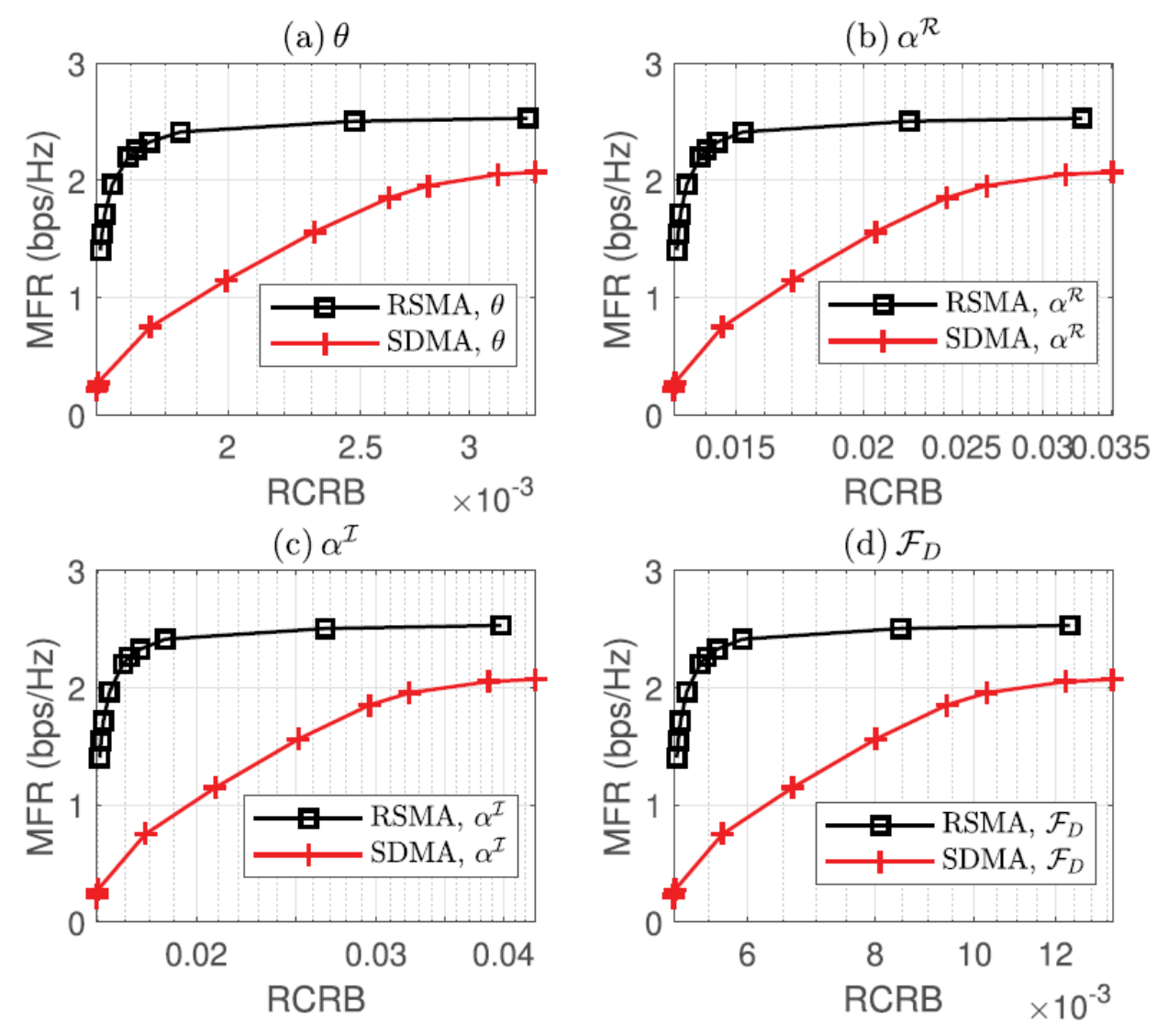}
\caption{Achievable trade-off between the minimum fairness rate (MFR) among $K=16$ single-antenna users and the root CRB (RCRB) of the target parameters (angle, complex reflection coefficient, and Doppler) for an $N_t=8$-feed LEO JCAS satellite with RSMA- and SDMA-assisted joint beamforming \cite[Fig.~3]{Yin2024LEOISAC}.}
\label{fig:tradeoff_jcas}
\end{revblock}
\end{figure}

\revi{rev1.4.2a}{The communication--sensing coupling described above has been quantified in several satellite studies. The representative result is reproduced in Fig.~\ref{fig:tradeoff_jcas}; the setups and operating points of all JCAS results are also collected in Table~\ref{tab:tradeoff_summary} (Section~\ref{sec:tradeoffs}).}

\revi{rev1.4.5}{Fig.~\ref{fig:tradeoff_jcas}, reproduced from \cite{Yin2024LEOISAC}, is the canonical illustration of the JCAS trade-off: the minimum fairness rate (MFR) guaranteed to every user of a LEO satellite with $N_t=8$ antenna feeds and $K=16$ users (two per beam, i.e., an overloaded system) is plotted against the RCRB of the angle, reflection coefficient, and Doppler of a target illuminated by the same beams, for joint beamformers optimized under RSMA and under conventional SDMA. At the sensing-priority end both schemes achieve nearly the same estimation accuracy, because the precoders radiate maximum power toward the target; moving toward the communication-priority end, SDMA loses MFR much faster than RSMA, because with $K>N_t$ it has exhausted its spatial degrees of freedom, whereas the RSMA common stream provides an additional degree of freedom that serves the users and simultaneously acts as a radar sequence. The RSMA frontier therefore dominates the SDMA frontier over the whole plane: the price of adding sensing to a communication satellite depends critically on the multiple-access scheme and on the ratio between users and antennas.}

\revi{rev1.4.6}{Two further results refine this picture. In \cite{Liu2024QuantizedLEOISAC}, the authors considered a $20$-GHz LEO satellite at $600$~km with $25$~MHz bandwidth, $N_t=N_r=4$ antennas, $K=5$ users, and low-resolution ($2$-bit) DACs, and plotted the max--min fairness EE of the users against the CRB threshold imposed on the target-angle estimate (\cite[Fig.~6(b)]{Liu2024QuantizedLEOISAC}): the EE increases monotonically as the sensing constraint is relaxed, RSMA with a dedicated radar sequence and successive interference cancellation attains the best trade-off, and a sum-rate objective silently switches the radar sequence off, so that fairness- or EE-oriented objectives are needed to keep both functions alive. The authors in \cite{You2022BeamSystems} traded the two functions through the weighting coefficient $\zeta$ of a hybrid precoder for a $20$-GHz, $800$-MHz, $1000$-km massive-MIMO satellite with up to $2304$ antennas and $16$ RF chains: at a false-alarm probability of $10^{-7}$, a $90\%$ detection probability for four targets requires a larger power budget with $\zeta=0.9$ (communication priority) than with $\zeta=0.4$, the communication EE grows with $\zeta$, and a beam-squint-aware design achieves a higher detection probability than a squint-unaware one for the same power. Bistatic designs with the radar receiver on the ground \cite{Park2025BistaticISAC} or at a gateway \cite{Zhang2026BistaticLEOISAC} remove the self-interference cost of the monostatic geometry at the price of an additional synchronized receiver.}

\subsection{Key Challenges}
\label{sec:jcas_challenges}
\begin{itemize}
    \item \textit{Antenna architecture:} dual-function antennas and beamforming must fit the size, thermal, and mechanical constraints of spacecraft, which limit array size and complexity and hence the achievable communication and sensing performance.
    \item \textit{Self-interference:} in monostatic configurations, SI between transmit and receive chains is the most significant challenge \cite{Sheemar2022PracticalRange}; the compact hardware limits antenna separation, orbital dynamics complicate SI management, and the power budget restricts digital cancellation \cite{Sheemar2021HybridChannel}, calling for algorithms optimized for space.
    \item \textit{Security:} dual-function signals create spoofing and jamming vulnerabilities that can target both functions at once \cite{TEDESCHI2022109246}; the interdependent processing chains complicate spoofing detection, and long propagation times delay countermeasures.
    \item \textit{Waveform and modulation design:} spectrally efficient waveforms supporting high-rate communications and accurate sensing without mutual interference, while remaining compatible with legacy waveforms such as DVB-S2X, are complex to design.
    \item \textit{Onboard data processing:} sensing generates large raw-data volumes that must be processed in near real time (detection, Doppler estimation) to support adaptive routing or beam management on LEO satellites with limited processing and energy; balancing onboard processing with downlinking for edge/cloud inference remains unsolved.
\end{itemize}

\subsection{Research Directions}
\label{sec:jcas_directions}
\begin{itemize}
\item \textit{Compact and reconfigurable antennas:} lightweight shared-aperture arrays, multifunctional metasurfaces \cite{Sheemar2025JointConstraints,Sheemar2025MinimumMaximization}, and electronically steerable multi-band phased arrays with dynamic beam control, robust to thermal stress, radiation, and vibration.
\item \textit{Self-interference cancellation for space:} lightweight analog--digital hybrid cancellation under hardware nonlinearities \cite{Sheemar2023Full-Duplex-EnabledSurfaces}, learning-based estimators with space-compatible footprints, and coordinated beam nulling and frequency hopping against residual interference.
\item \textit{Security frameworks for dual-function signals:} physical-layer authentication, waveform obfuscation, and spread spectrum against spoofing and jamming, anomaly-based intrusion detection tailored to satellite processors, and secure command and control of dual-function payloads.
\item \textit{Joint beamforming and scheduling:} predictive and cooperative beamforming accounting for orbital dynamics and user mobility, optimized for both link quality and angular resolution jointly with multi-beam scheduling, spatial multiplexing, and power allocation.
\item \textit{Flexible waveform design:} multi-carrier schemes with radar-friendly subcarriers or chirp-OFDM combinations, robust to Doppler, multipath, and synchronization errors, compatible with DVB-S2X or 5G NR, and adaptive to the orbital zone, user density, and sensing task.
\item \textit{AI-enhanced onboard processing:} compressed models (quantization-aware training, distillation, pruning), federated learning across satellite clusters, and adaptive switching between onboard and ground inference.
\end{itemize}
\section{JCAP in Satellite Systems}
\label{sec:jcap}

\revi{rev1.2.15}{JCAP is the MFSS configuration in which a communication constellation also provides positioning, navigation, or timing, or in which PNT information is used to operate the communication system. Compared with JCAS, the second function is not the observation of a target but the estimation of the user's own state, which brings the clock, ephemeris, geometry, and integrity requirements of Section~\ref{sezione_fundamentals} into the design of the communication payload.}

\subsection{Architecture and Realizations}
\revi{rev1.2.16}{A satellite JCAP system uses the downlink (or uplink) signals of a communication constellation as ranging sources. The user extracts one or more observables---TOA/TDOA from the signal timing, FOA/Doppler and Doppler rate from the carrier, AOA from array processing, RSS from the received power, and carrier phase for high-precision solutions---from several satellites and combines them with the satellite orbit and clock states to solve for position, velocity, and time. Three ingredients determine the architecture: (i) the signal structure, i.e., whether the communication waveform is used as is, complemented with pilots or positioning reference signals (PRSs), or overlaid with a dedicated ranging component such as a direct-sequence spread-spectrum (DSSS) code; (ii) the source of orbit and clock information, i.e., whether the user must estimate it, the operator disseminates ephemeris and clock corrections through the communication link, or a control segment performs precise orbit determination as in GNSS; and (iii) the geometry, since LEO constellations offer many satellites in view and fast-changing geometry that lowers the dilution of precision but produces large Doppler and short visibility windows. In the cooperative realization the communication payload is untouched and PNT relies on signals of opportunity or on a hosted payload; in the integrated realization the pilot, PRS, beam, and power resources of the frame are managed jointly with the traffic; and in the joint realization the waveform itself is designed for ranging, e.g., by superimposing a DSSS navigation signal on the 3GPP OFDM downlink with a controlled power split \cite{DeGaudenzi2025IntegratedLEOPNT,GonzalezGarrido2025JointNavigation}. In the reverse direction, PNT information serves the communication system for beam pointing, Doppler pre-compensation, handover prediction, and network synchronization.}

\subsection{Classification}
\revi{rev1.2.17}{JCAP can be organized into three principal architectural approaches that differ in operator cooperation, space- and user-segment complexity, infrastructure cost, and achievable accuracy.}
\begin{itemize}
    \item \emph{Signals-of-opportunity approach:} PNT is derived from existing LEO communication transmissions without operator cooperation, with acquisition, signal characterization, and navigation processing in the user terminal, which derives Doppler, angle, delay, and received-power observables from signals not designed for navigation. Its limitations are specialized receivers, unsynchronized satellite and receiver clocks, incomplete waveform knowledge, and orbit/clock uncertainty; Starlink, OneWeb, Telesat, and Kuiper have motivated many SoO studies \cite{Neinavaie2022AcquisitionResults,Orabi2021OpportunisticSatellites}, and more visible satellites improve geometry and availability but do not remove correlated ephemeris and clock errors.
    \item \emph{Fused LEO-PNT approach:} introduced in \cite{Iannucci2020EconomicalGNSS}, PNT is derived from the primary communication signal without a dedicated on-orbit PNT transmitter, but the operator cooperates with the user: in the cooperative signal-of-opportunity variant \cite{Ardito2019PerformancePositions}, precise orbit and clock corrections are carried by the communication link or a separate data service, whereas a more tightly integrated variant modifies the broadband downlink so that its spectrum and waveform support both communication and PNT \cite{Iannucci2020EconomicalGNSS}.
    \item \emph{Hosted-PNT-payload approach:} dedicated navigation hardware is hosted on each communication platform but operates largely independently of the primary mission, which permits customization of the navigation frequency, bandwidth, signal, clock, and message at the cost of additional hardware, accommodation, power, qualification, and integration; performance still depends on orbit determination, time synchronization, broadcast ephemeris and clock models, and a geometry designed for communications. Satellite Time and Location (STL) over Iridium is a representative hosted service providing strong timing and location signals, including indoors \cite{GPSWorldStaff2016IridiumService}.
\end{itemize}
\revi{rev1.2.18}{The three approaches map onto the cooperative (signals of opportunity and hosted payloads), integrated (fused LEO-PNT with corrections and managed pilots), and joint (modified or overlaid waveform) levels of Table~\ref{tab:level_config_mapping}. A complementary classification follows the observables used (time, Doppler, angle, power, or carrier phase) and the link direction (downlink positioning at the user, as in GNSS, or uplink positioning at the ground segment, as in ARGOS).}

\subsection{State of the Art}
\label{sec:jcap_sota}
\revi{rev1.2.19}{At the waveform and link levels, we modelled the aggregate interference among 5G NTN positioning reference signals transmitted by multiple satellites in the same bandwidth part and provided a stochastic (generalized extreme value) interference model \cite{GonzalezGarrido2024PRSInterference}. Massive-MIMO LEO studies jointly optimized hybrid precoding for communication and localization \cite{You2024IntegratedLocalization} and made that design robust to angular uncertainty using a worst-case localization bound \cite{Zhu2025RobustICAL}. A complementary line of work superimposed a DSSS navigation signal on a 3GPP-compatible OFDM communication waveform \cite{DeGaudenzi2025IntegratedLEOPNT}, where we evaluated joint communication and navigation from a multi-beam LEO satellite with shared and independent beam models \cite{GonzalezGarrido2025JointNavigation}, and designed a receiver that extracts both services from the overlaid waveform \cite{GonzalezGarrido2025Receiver}. Tracking a GNSS-like overlay with a minimally modified GNSS receiver under LEO Doppler was subsequently examined by us in our recent work \cite{Edjekouane2025Tracking}.}

\revi{rev1.2.20}{At the system level, a two-part direct-to-device study compared three integrated LEO communication/PNT architectures and evaluated their service performance \cite{DeGaudenzi2026Beyond5GPartI,DeGaudenzi2026Beyond5GPartII}. More recent preprints use terrestrial reconfigurable intelligent surfaces to improve the communication-SNR/position-error trade-off in blocked urban NTN links \cite{Khalil2026RISJCAP}, quantify the effect of PRS periodicity and differential satellite delays on the communication service \cite{GonzalezGarrido2026PRSImpact}, and coordinate beam scheduling and power allocation to preserve both communication capacity and PNT availability \cite{GonzalezGarrido2026BeamScheduling}. Collectively, these works shift JCAP from opportunistic signal exploitation toward explicit waveform, receiver, and network co-design.}

\subsection{Advantages and Functional Synergies}
\revi{rev1.2.21}{The advantages of JCAP follow from the complementarity of the two functions. Communication waveforms and infrastructure provide ranging, timing, Doppler, angular, and synchronization observables without a fully independent navigation payload, and the dense LEO geometry, the stronger received power, and the fast geometry change of communication constellations improve availability, convergence time, and resilience with respect to MEO GNSS, particularly in urban and contested environments. In the reverse direction, PNT information enables beam pointing, user localization, Doppler pre-compensation, network synchronization, mobility prediction, and handover management, which are especially valuable in NGSO constellations where the satellite--user geometry changes within minutes, and the communication link carries ephemeris, clock, integrity, and authentication information, which removes the main limitation of pure signals-of-opportunity positioning. The cost is the pilot, power, and beam overhead devoted to ranging and the tighter clock and ephemeris requirements imposed on a payload originally designed for communications.}

\subsection{Achievable Performance and Trade-Offs}
\label{sec:tradeoffs_jcap}
\begin{figure}[!t]
\begin{revblock}{rev1.4.7}
\centering
\includegraphics[width=\columnwidth]{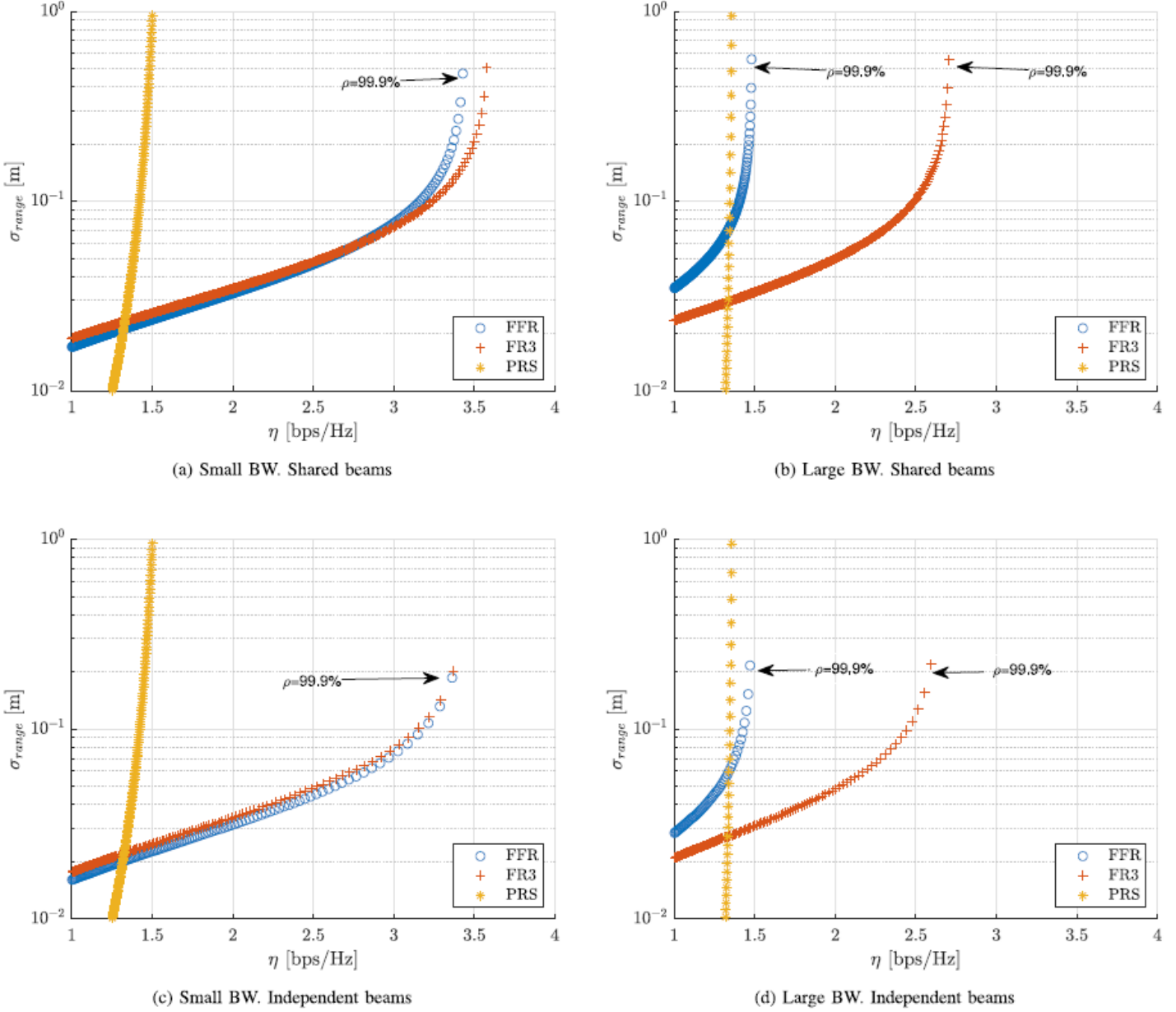}
\caption{Achievable trade-off between the communication spectral efficiency $\eta$ and the ranging CRLB of a DSSS navigation signal overlaid on the 5G NTN OFDM downlink of a $600$-km multi-beam LEO satellite ($16\times16$ array, seven beams, S-band), for different data/navigation power splits $\rho$, beam models, and frequency-reuse scenarios \cite[Fig.~14]{GonzalezGarrido2025JointNavigation}.}
\label{fig:tradeoff_jcap} \vspace{-5mm}
\end{revblock}
\end{figure}

\revi{rev1.4.2b}{The price of adding navigation to a communication constellation has been quantified at waveform, link, and system level. The representative result is reproduced in Fig.~\ref{fig:tradeoff_jcap} and summarized with the other JCAP results in Table~\ref{tab:tradeoff_summary}.}

\revi{rev1.4.8}{Fig.~\ref{fig:tradeoff_jcap}, reproduced from \cite{GonzalezGarrido2025JointNavigation}, shows the JCAP trade-off for a joint waveform in which a DSSS navigation signal is superimposed on the 3GPP OFDM downlink of a LEO-600 satellite (3GPP TR~38.821 Set-2 parameters with a $16\times16$ planar array, seven beams, and the S-band n256 NTN band). Each curve is the Pareto front obtained by sweeping the power split $\rho$ between the data signal (fraction $\rho$ of the beam power) and the navigation signal (fraction $1-\rho$): the spectral efficiency of the data service is plotted against the CRLB of the range estimate, for a shared-beam model, in which the navigation signal rides on the seven data beams, and an independent-beam model, in which one wide navigation beam is radiated separately, under full frequency reuse with $30$~MHz for data and $30$~MHz for navigation (FFR 30/30) or three-color reuse with $15$/$5$~MHz (FR3 15/5) and $10$/$20$~MHz (FR3 10/20). The reported operating points quantify the price of navigation: with FFR 30/30 the shared-beam model must devote $\rho\approx0.78$ of the power to data to keep $\eta\geq1$~bit/s/Hz, which leaves a ranging CRLB of about $3.5$~cm, whereas in the less interference-limited FR3 15/5 scenario the balance is found at $\rho\approx0.54$ with $1.9$~cm at the same spectral efficiency; the independent-beam model attains a lower CRLB than the shared-beam model at every $\eta$, and curves for the 5G DL-PRS used as positioning signal are included as a reference. Prioritizing communications ($\rho=0.8$) yields spectral efficiencies of the order of $2$~bit/s/Hz with a coarser range estimate, whereas prioritizing navigation ($\rho=0.2$) reduces $\eta$ to about $0.3$~bit/s/Hz for a range error that can be two to four times smaller.}

\revi{rev1.4.9}{Massive-MIMO and system-level studies confirm that the JCAP trade-off is comparatively benign when the array offers enough degrees of freedom. In \cite{You2024IntegratedLocalization}, the authors plotted the downlink sum spectral efficiency against the average position error bound (APEB) of nine terminals served by a $200$-km, $2$-GHz, $15.36$-MHz satellite with a $24\times24$ array and $36$ RF chains (\cite[Fig.~6]{You2024IntegratedLocalization}): at $26$~dBW the fully connected hybrid precoder delivers about $85$~bit/s/Hz in total while keeping the localization error bound below $5$~m for every terminal, and more transmit power improves both functions at once. The two-part direct-to-device study of \cite{DeGaudenzi2026Beyond5GPartI,DeGaudenzi2026Beyond5GPartII}, for a $1200$-km constellation of $263$ satellites in the 2-GHz MSS band, allocates $20\%$ of the RF power to the positioning overlay and reports $95$th-percentile horizontal errors of $21$--$38$~m for an NB-IoT carrier with a narrowband PRS overlay and of $3.0$--$3.4$~m for NR with a $5$-MHz DL-PRS; the overlay at $-10$ to $-15$~dB relative power has a limited, Gaussian-like impact on the NB-IoT bit error rate, and reducing the DL-PRS share to $10\%$ increases the ranging error standard deviation by only about $7\%$. The JCAP trade-off is thus primarily a power and pilot-overhead split whose communication cost is a fraction of a decibel, whereas the achievable accuracy is governed by the ranging bandwidth and by the constellation geometry.}

\subsection{Key Challenges}
\label{sec:jcap_challenges}
\begin{itemize}
    \item \emph{Waveform and protocol trade-offs:} communication waveforms are bandwidth-efficient, whereas positioning favors wide bandwidths and sharp autocorrelation; a unified waveform without significant loss on either side remains a major challenge.
    \item \emph{Synchronization and clock bias:} positioning is highly sensitive to the timing errors and clock offsets that communications tolerate, which translate into large localization errors in fast-changing LEO geometry.
    \item \emph{Channel estimation and Doppler:} TOA, AOA, and Doppler measurements need accurate CSI in channels with fading, blockage, atmospheric impairments, and large Doppler that must be compensated for both coherent communication and positioning, hardest for fast-moving platforms.
    \item \emph{Resource sharing and coupling:} bandwidth, power, antennas, and computation are shared between coupled tasks, so that, e.g., beamforming focused on throughput may not provide a favorable geometry for positioning.
    \item \emph{Geometric dilution of precision:} in low-visibility or urban environments, unfavorable satellite geometry yields a high GDOP and degrades localization even when signal quality is high.
    \item \emph{Security and privacy:} positioning signals are susceptible to spoofing and replay attacks, and communication signals embedding location information raise privacy concerns.
\end{itemize}

\subsection{Research Directions}
\label{sec:jcap_directions}
\begin{itemize}
\item \textit{Multi-objective waveform co-design:} unified waveform families balancing spectral efficiency and autocorrelation sharpness, optimized jointly for positioning accuracy, interference resilience, and throughput, with backward compatibility and adaptive waveform switching for phased deployment.
\item \textit{Timing and synchronization:} protocols robust to extreme Doppler, clock drift, and asymmetric delays, including satellite-aided two-way ranging, hybrid GNSS/JCAP clock correction, AI-assisted clock-bias prediction, and cooperative synchronization among satellites and ground relays.
\item \textit{Joint channel estimation and Doppler-resilient tracking:} extraction of TOA, AOA, and Doppler from communication pilots without extra overhead, sparsity- or learning-based multi-resolution estimation, joint delay-Doppler processing, and predictive tracking of user and satellite trajectories.
\item \textit{Joint resource allocation and beam management:} cross-layer assignment of power, bandwidth, and beams according to service priority, user geometry, and traffic, with beam management that minimizes GDOP while preserving communication capacity \cite{GonzalezGarrido2026BeamScheduling}.
\item \textit{Interference mitigation and coexistence:} orthogonal waveform sets, adaptive spatial filtering, cooperative interference prediction, and coordinated multi-point techniques adapted to satellite mobility.
\item \textit{Secure, scalable, and context-aware architectures:} secure time-stamping, encrypted ranging, location obfuscation, and physical-layer authentication, together with service slicing and priority-driven scheduling that reconfigure JCAP from massive IoT to mission-critical robotics.
\end{itemize}
\section{JSAP in Satellite Systems}
\label{sec:jsap}

\revi{rev1.2.22}{JSAP combines Earth observation or target sensing with positioning, navigation, and timing on a shared platform, waveform, or signal. It is the least explored configuration, because sensing and PNT have historically been served by missions with opposite design centers: sensing radars use wide bandwidths and high peak power on dedicated satellites, whereas navigation signals are narrowband, low power, and continuously broadcast from a constellation whose main asset is its clock and orbit knowledge. The synergy lies precisely in this complementarity, as detailed below.}

\subsection{Architecture and Realizations}
\revi{rev1.2.23}{A JSAP system can be built in four ways. In \emph{PNT-assisted sensing}, the onboard PNT receiver or navigation payload provides the precise time, orbit, and attitude that a sensing chain needs for time tagging, georeferencing, motion compensation, and coherent multi-platform processing; this loosest coupling is standard practice in SAR missions and becomes an MFSS in the sense of Section~\ref{sezione_fundamentals} only when the timing reference, aperture, or processing are shared. In \emph{sensing-assisted PNT}, radar, altimeter, or imaging observables (range, Doppler, angle, terrain features, environmental maps) feed the navigation filter for terrain- or feature-aided navigation, multipath and blockage mapping, and integrity monitoring. In \emph{dual-use navigation signals}, the navigation transmission itself is the sensing source: direct and Earth-reflected navigation signals are received by a satellite, an aircraft, or a ground node and processed as a bistatic radar, as in GNSS reflectometry and radio occultation (the passive opportunistic classes of Fig.~\ref{fig:rs_classification}) and, more recently, in receivers that perform navigation and passive-radar sensing from the same signal \cite{Sachdeva2026NavIC}. In \emph{co-designed dual-function frames}, a LEO constellation transmits navigation ranging and remote-sensing illumination in the same frame and beam budget, and multi-satellite beamforming is optimized jointly for the navigation error bound and the sensing quality \cite{Wang2026NavigationRemoteSensing}. The first two options correspond to the cooperative and integrated levels of Table~\ref{tab:level_config_mapping}, the last two to the joint level, with the transmitter (PNT) and the receiver (sensing) possibly on different platforms.}

\subsection{Classification}
\revi{rev1.2.24}{JSAP designs can accordingly be classified by the direction of assistance (PNT-to-sensing, sensing-to-PNT, or bidirectional), by the sensing modality (active radar or SAR, passive reflectometry, or radio occultation), by the geometry (monostatic radar with an onboard navigation reference, or bistatic/multistatic with the navigation transmitter and the sensing receiver on separate platforms), and by the integration level. The distinguishing feature with respect to JCAS is that the shared signal must preserve the ranging accuracy, clock stability, and predictable structure required for navigation, and the distinguishing feature with respect to JCAP is that the sensing function needs bandwidth, power, and coherent integration that navigation signals were not designed to provide.}

\subsection{State of the Art}
\label{sec:jsap_sota}
\revi{rev1.2.25}{Explicit satellite JSAP is now emerging. In \cite{Wang2026NavigationRemoteSensing}, the authors  proposed a dual-function LEO transmission frame for navigation and remote sensing and jointly designed multi-satellite beamforming using a position--velocity--time Cram\'{e}r--Rao bound (CRB) and a signal-to-ambiguity-interference-plus-noise ratio (SAINR) metric for the sensing function. At the hardware level, an RF-system-on-chip prototype used direct and reflected NavIC signals for simultaneous navigation and passive-radar sensing in the delay--Doppler domain \cite{Sachdeva2026NavIC}. NavIC is not a LEO constellation; consequently, the latter result is evidence for a shared satellite navigation/sensing receiver chain rather than a validation of LEO JSAP geometry or coverage. Beyond these two works, the JSAP evidence base consists of the mature GNSS-reflectometry and radio-occultation literature, in which navigation signals are reused opportunistically for sensing, and of the cooperative use of onboard GNSS receivers for SAR georeferencing; both remain outside the joint co-design that JSAP targets.}

\subsection{Advantages and Functional Synergies}
\revi{rev1.2.26}{PNT provides precise timing, orbit, attitude, and position for georeferencing, coherent sensing, and multi-platform fusion, and a constellation whose timing is disciplined for navigation is also an ideal multistatic sensing network, because the nanosecond-level synchronization that bistatic sensing needs is already available. In the reverse direction, sensing measurements (range, Doppler, angle, terrain features, environmental maps) support navigation, integrity monitoring, localization, and situational awareness where satellite visibility is poor, and dual-use navigation signals give global, continuous, and free illumination for reflectometry and occultation without an additional transmitter. The costs are that conventional navigation signals are narrowband and low power, which limits the achievable sensing resolution and detection performance, and that multipath and reflected components, informative for sensing, are error sources for positioning.}

\subsection{Achievable Performance and Trade-Offs}
\label{sec:tradeoffs_jsap}
\begin{figure}[!t]
\begin{revblock}{rev1.4.10}
\centering
\includegraphics[width=\columnwidth,height=4cm]{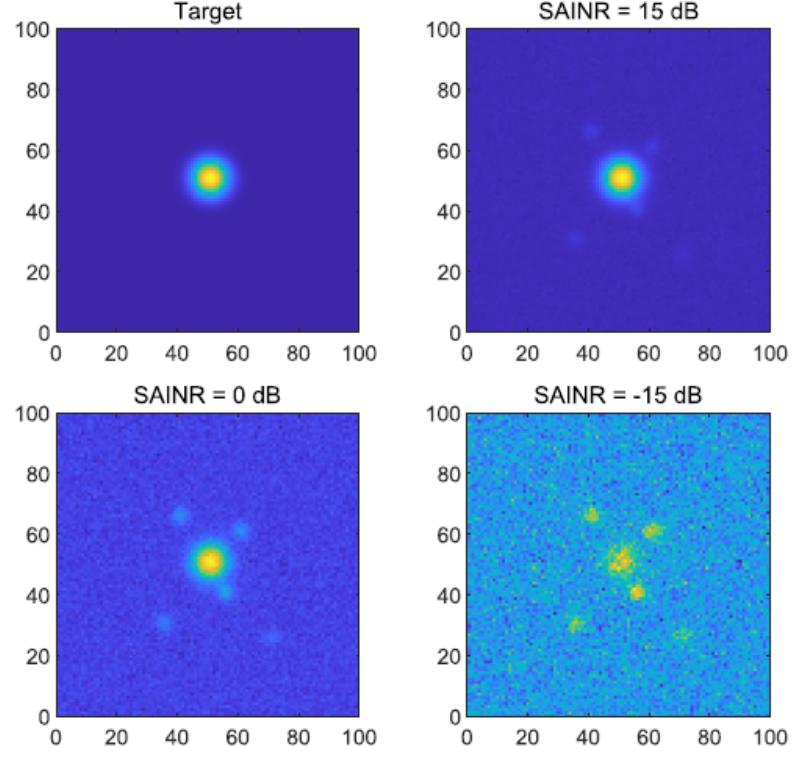}
\caption{Remote-sensing imaging quality obtained at the central satellite of a cooperating LEO group as a function of the SAINR, in the dual-function navigation/remote-sensing frame in which the SAINR requirement is traded against the position--velocity--time error bound of the navigation users \cite[Fig.~5]{Wang2026NavigationRemoteSensing}.} \vspace{-5mm}
\label{fig:tradeoff_jsap}
\end{revblock}
\end{figure}

\revi{rev1.4.2c}{Quantitative JSAP evidence is still scarce; the co-designed frame of \cite{Wang2026NavigationRemoteSensing} and the hardware prototype of \cite{Sachdeva2026NavIC} are the two available data points, reproduced and summarized in Fig.~\ref{fig:tradeoff_jsap} and Table~\ref{tab:tradeoff_summary}.}

\revi{rev1.4.11}{The only quantitative navigation--sensing co-design available for LEO constellations is that of Wang \emph{et al.} \cite{Wang2026NavigationRemoteSensing}. A service group of $K$ satellites (one central and $K-1$ auxiliary satellites, each with an $N$-element planar array) transmits navigation and remote-sensing signals in a common dual-function frame; $M$ single-antenna navigation users solve their position, velocity, and time from pseudoranges (by weighted least squares, shown to outperform unweighted least squares for every number of visible satellites) and from maximum-likelihood Doppler estimates, whereas the echoes of the remote-sensing illumination are collected at the central satellite as a MIMO-SAR. The sensing quality is measured by the SAINR of the received echo, whose effect on the reconstructed image is shown in Fig.~\ref{fig:tradeoff_jsap}, reproduced from \cite{Wang2026NavigationRemoteSensing}: the imaging accuracy improves markedly as the SAINR at the central satellite increases. The trade-off arises because the multi-satellite beamformers minimize the weighted average PVT error bound of the navigation users subject to a minimum-SAINR constraint for remote sensing and to a per-satellite power budget: a stricter SAINR requirement diverts power and spatial degrees of freedom from the navigation beams toward the illuminated scene and raises the achievable PVT error bound, and vice versa. The non-convex problem is solved with a penalty-based semidefinite-relaxation and successive-convex-approximation algorithm with monotonically decreasing objective, and the numerical results of \cite{Wang2026NavigationRemoteSensing} validate the cooperative design against conventional beamforming benchmarks and show how the achievable PVT accuracy depends on the SAINR requirement, the array size, the transmit power, the number of users, and the constellation parameters.}

\revi{rev1.4.12}{The hardware evidence quantifies the other side of JSAP, namely how much sensing a navigation signal can deliver. The RF-system-on-chip receiver of \cite{Sachdeva2026NavIC} processes the direct and reflected NavIC L5 signal ($1.176$~GHz, $1.023$~Mchip/s, about $2$~MHz bandwidth) as a bistatic radar in the delay--Doppler domain: the native range resolution of the $1023$-chip code is about $293$~m (refined to about $40$~m by oversampling), two targets separated by $4$~km in range and $500$~Hz in Doppler are resolved at a received SNR of $-10$~dB thanks to the correlation gain of the ranging code, and the estimates remain within the resolution limits down to $-12$~dB. These numbers illustrate the JSAP challenge of Section~\ref{sec:jsap_challenges}: the narrow bandwidth and low power that make a navigation signal robust also bound its sensing resolution and detection range, so that fine-resolution sensing requires either co-designed wideband frames \cite{Wang2026NavigationRemoteSensing} or fusion with wideband communication or radar payloads.}

\subsection{Key Challenges}
\label{sec:jsap_challenges}
\begin{itemize}
    \item \emph{Accuracy and resolution:} achieving the decimeter-to-centimeter accuracy required by many sensing applications from reflected or scattered navigation signals is challenging.
    \item \emph{Coverage:} sensing requires high signal strength and frequent observations, while secure PNT demands uninterrupted availability; blockage in urban areas, terrain masking, or limited visibility at high latitudes degrade both.
    \item \emph{Frequency bands and transmission power:} PNT spectrum is narrow and PNT signals are low power, optimized for positioning rather than for high-resolution sensing, which requires large contiguous bandwidths, multi-band operation, and high received power; this mismatch is a fundamental challenge.
    \item \emph{Coordination among nodes:} JSAP requires tight synchronization between navigation transmitters and distributed sensing receivers; without it, accuracy degrades, sensing is limited, and latency increases.
    \item \emph{Interference and beam design:} multipath helps PNT but is undesired for sensing, and balancing this trade-off under dynamic service requirements is complex.
\end{itemize}

\subsection{Research Directions}
\label{sec:jsap_directions}
\begin{itemize}
\item \textit{Signal design for dual utility:} GNSS-compatible structures with sharper autocorrelation, adaptive pulse shaping, multiplexing for joint reflectivity and positioning extraction, and sidebands or dynamic subcarriers that allocate partial bandwidth to sensing without disrupting legacy PNT.
\item \textit{Multi-band and wideband architectures:} wideband, multi-frequency designs resilient to distortion and atmospheric impairments, fusion of GNSS/LEO-PNT signals with wideband communication channels, and spectrum switching driven by sensing requirements and regulation.
\item \textit{Calibration and synchronization:} sub-nanosecond timing under orbital motion through clock-drift compensation, inter-satellite timing relay, cooperative ground beacons, and decentralized synchronization of dispersed nodes.
\item \textit{Signal enhancement and resolution:} coherent accumulation, adaptive matched filtering, and AI-based denoising of weak reflected signals; super-resolution, multipath deconvolution, and sparse recovery integrated with terrain models and auxiliary radar or optical data.
\item \textit{Context-adaptive processing:} identification and classification of multipath components traded off adaptively between navigation and sensing, with cross-layer coordination of beam steering and spatial filtering and lightweight onboard inference of terrain, motion state, and obstructions.
\item \textit{Benchmarks and testbeds:} standardized metrics for sensing accuracy, resolution, robustness, and synchronization fidelity, with open simulators and hardware-in-the-loop emulators of orbital dynamics, surface variability, and atmospheric effects.
\end{itemize}
\section{JCSAP in Satellite Systems}
\label{sec:jcsap}

\revi{rev1.2.27}{JCSAP is the full convergence of the three services on a single platform or constellation. It inherits the couplings of JCAS, JCAP, and JSAP simultaneously and adds a third dimension to every design decision: the same bandwidth, power, beams, timing reference, and processing must be divided among data delivery, environment perception, and navigation. JCSAP is therefore the upper end of the MFSS design space rather than the default target of every mission, as the techno-economic assessment of Section~\ref{sec:techno_economic} makes explicit.}

\subsection{Architecture and Realizations}
\revi{rev1.2.28}{Two architectural families can be distinguished. In the \emph{platform-level} family, one satellite hosts a tri-functional payload: at the cooperative level the three chains coexist on the bus with coordinated scheduling and calibration exchange; at the integrated level data, sensing pulses, and PNT synchronization signals are scheduled in a unified resource grid with interference-aware power, beam, and time control (Fig.~\ref{fig:integratedpayload}); and at the joint level a single waveform supports data, sensing observables, and ranging/timing simultaneously (Fig.~\ref{fig:jointpayload}). In the \emph{constellation-level} family, the three functions are distributed across cooperating satellites connected by ISLs: multi-satellite cooperative beamforming serves users, multistatic sensing exploits the spatial diversity of the constellation, and coordinated multi-satellite localization exploits its geometry, with the ISLs providing the time transfer, data exchange, and cooperative processing that tie the functions together \cite{Zhang2026DISLAC}. A complementary view distinguishes the depth at which the services are merged \cite{Sheng2025ICNR}: function-level integration hosts all functions on every satellite on separate hardware, signal-level integration uses one multi-purpose waveform and shared spectrum, data-level integration fuses the products of the three services, and application-level integration exposes them as one service; the first two correspond to the cooperative/integrated and joint payload levels of this survey, the last two concern the ground and application segments.}

\subsection{Classification}
\revi{rev1.2.29}{JCSAP designs can thus be classified by the locus of integration (single platform versus constellation), by the integration level (cooperative, integrated, or joint), by the depth of merging (function, signal, data, or application level), and by the dominant service that drives the design: communication-centric JCSAP, in which sensing and PNT are added to a broadband constellation (the most likely near-term path given the size of current LEO communication fleets), sensing-centric JCSAP, in which an EO or SSA mission adds connectivity and navigation, and navigation-centric JCSAP, in which a LEO-PNT constellation adds sensing and data services.}

\subsection{State of the Art}
\label{sec:jcsap_sota}
\revi{rev1.2.30}{Explicit tri-functional satellite studies have appeared only recently. In \cite{Sheng2025ICNR}, the authors introduced an integrated communication, navigation, and remote-sensing framework for LEO vehicular applications and examined the sharing of orbit constellations, wireless resources, and data at four integration levels, with a numerical case study. The work \cite{Zhang2026DISLAC} subsequently proposed distributed integrated sensing, localization, and communications over LEO constellations, using inter-satellite links and cooperative processing to share sensing, positioning, and communication resources, and discussed synchronization, antenna reconfigurability, and ISL design as the enabling system-level aspects. These publications show a progression from architecture-level integration to constellation-level cooperative design; both remain analytical or simulation-based, and the reviewed literature does not yet report a flight-demonstrated payload that unifies all three functions. Communication-centric JCSAP additionally builds on the JCAS and JCAP results of Sections~\ref{sec:jcas_sota} and~\ref{sec:jcap_sota}, and on the LEO communication payload review of \cite{Hui2025ADirections}.}

\subsection{Advantages and Functional Synergies}
\revi{rev1.2.31}{A common payload jointly exploits communication data, sensing observables, and PNT information, which yields the maximum hardware and spectrum reuse and enables mutual assistance among the three functions: sensing and PNT improve communication resource allocation and beam management, communication links support distributed sensing and the dissemination of corrections and products, and PNT provides the common spatial and temporal reference for coherent multi-satellite operation, distributed sensing, and cooperative localization. For autonomous mobility, maritime surveillance, disaster response, and defense, the value of JCSAP lies in delivering connectivity, situational awareness, and resilient positioning from the same asset with low latency and without cross-mission data fusion on the ground. Against these benefits stand a three-way trade-off among communication rate and reliability, sensing resolution and detection, and positioning and timing accuracy, and the highest waveform, synchronization, resource-allocation, processing, and validation complexity of all configurations.}

\subsection{Achievable Performance and Trade-Offs}
\label{sec:tradeoffs_jcsap}
\begin{figure}[!t]
\begin{revblock}{rev1.4.13}
\centering
 \includegraphics[width=0.9\columnwidth]{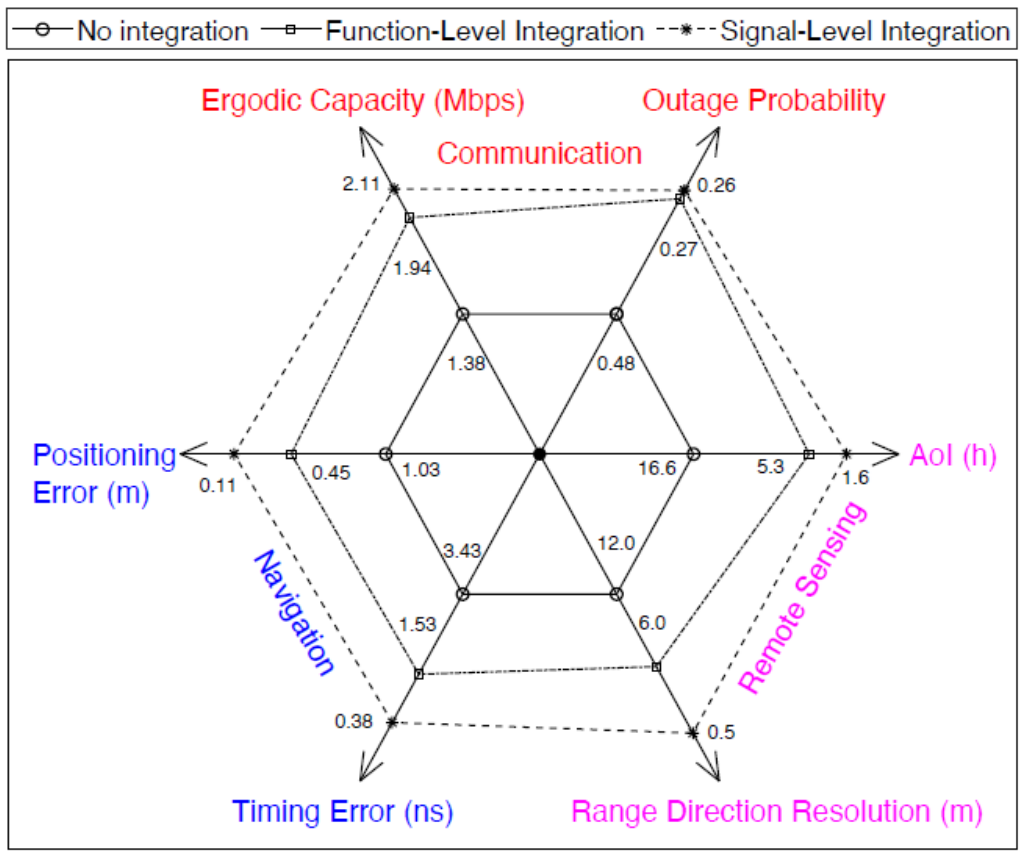}
\caption{Estimated gains of function-level and signal-level integration with respect to separate communication, navigation, and remote-sensing satellites in terms of capacity-outage probability and ergodic capacity, positioning accuracy and timing error, and range resolution and age of information, for a $500$-km LEO case study with $4500$/$250$/$250$ communication/navigation/remote-sensing satellites, $250$/$25$/$25$~MHz, and $20$/$20$/$80$~W \cite[Fig.~4]{Sheng2025ICNR}.}
\label{fig:tradeoff_jcsap}
\end{revblock}
\end{figure}

\revi{rev1.4.2d}{For the tri-functional configuration, the available evidence consists of integration-gain case studies and pairwise waveform trade-offs rather than a three-dimensional performance frontier; the representative result is reproduced in Fig.~\ref{fig:tradeoff_jcsap} and summarized in Table~\ref{tab:tradeoff_summary}.}

\revi{rev1.4.14}{For JCSAP, no study yet reports a Pareto surface over the three functions; the most complete quantitative evidence is the case study of \cite{Sheng2025ICNR}, reproduced in Fig.~\ref{fig:tradeoff_jcsap}. A rural vehicular scenario with $1400$ vehicles reporting sensing and maneuver data is served by a $500$-km LEO network in three ways: by the traditional separate fleets of $4500$ communication, $250$ navigation, and $250$ remote-sensing satellites with $250$, $25$, and $25$~MHz and $20$, $20$, and $80$~W, respectively; by function-level integration, in which every satellite hosts the three functions on separate hardware so that all satellites serve every function; and by signal-level integration, in which one multi-purpose waveform shares the whole spectrum. The radar chart compares the capacity-outage probability and the mean ergodic capacity (communication), the positioning accuracy and the timing error (navigation), and the range resolution and the age of information (remote sensing). Both integration levels improve all six metrics. The gain of function-level integration comes from the larger number of satellites available to each function (geometry, coverage, revisit), whereas the gain of signal-level integration comes from the larger spectrum available to each function, which benefits navigation and remote sensing (whose dedicated bandwidths were small) far more than communications: the authors report near-centimeter-level positioning and decimeter-level Earth-observation resolution under signal-level integration, roughly one order of magnitude better than with separate satellites, while acknowledging that a fundamental trade-off remains in the design of the multi-purpose signal itself.}

\revi{rev1.4.15}{That residual trade-off is made explicit by the constellation-level design of \cite{Zhang2026DISLAC}, which quantifies the effect of one shared waveform parameter, the OFDM subcarrier spacing, on multistatic sensing: a larger spacing improves the maximum unambiguous velocity (needed for the $7.5$~km/s relative velocities of LEO) but reduces the maximum unambiguous range and the range resolution, so that the large spacings favored by communication-centric numerologies (e.g., $60$~kHz at $2$~GHz for a $600$-km orbit) can cause range aliasing for long-range sensing (\cite[Fig.~4(a)]{Zhang2026DISLAC}); in the same framework, decentralized ring- and star-topology cooperative beamforming over ISLs achieves a sum rate close to centralized processing with far lower signaling, and, for multistatic sensing with four cooperating satellites, a low-complexity fusion of local estimates reaches a performance comparable to the fusion of the raw observations when each satellite collects enough antenna observations. JCSAP performance is thus currently characterized by integration gains and pairwise waveform trade-offs rather than by a single three-dimensional frontier, and none of the reported results has been validated in flight.}

\subsection{Key Challenges}
\label{sec:jcsap_challenges}
\begin{itemize}
\item \textit{Multi-objective waveform design and cross-domain trade-offs:} a unified waveform must reconcile spectral efficiency, time-frequency resolution, and sharp autocorrelation, and optimizing one function may degrade the others (e.g., beamforming that maximizes a user's SNR may yield poor angular resolution for sensing or an unfavorable geometry for positioning).
\item \textit{Joint channel estimation and synchronization:} the three functions rely on overlapping delay, Doppler, and angle features whose joint estimation in fast-fading, long-delay-spread channels is intensive, and timing for positioning, coherence for communication, and phase alignment for sensing must be kept despite clock drift, delays, and Doppler.
\item \textit{Limited onboard resources and interference:} concurrent operation on power-, computation-, and RF-constrained platforms requires low-latency hardware/software co-design, while co-channel interference from the integrated functions, neighboring satellites, or terrestrial systems can degrade all three services.
\item \textit{Security and privacy:} spoofing or jamming can simultaneously affect localization, detection, and data delivery, and shared waveforms increase interception and inference risks.
\item \textit{Geometry, coverage, and heterogeneous QoS:} unfavorable satellite--target--user geometry degrades GDOP, angular resolution, and sensing accuracy even with stable links, and low latency, high resolution, and high accuracy must be satisfied jointly by reconfigurable architectures adapted to mission profiles from IoT to disaster response.
\end{itemize}

\subsection{Research Directions}
\label{sec:jcsap_directions}
\begin{itemize}
\item \textit{Unified waveform and cross-domain optimization:} parameterizable waveform frameworks (multi-layered OFDM, sparse time-frequency structures, dual-purpose spread spectrum) and context-aware multi-objective optimization with AI-based decision engines that reconfigure power, beams, and waveform parameters to meet multi-domain QoS.
\item \textit{Integrated channel modeling and synchronization:} unified delay-Doppler-angle models with joint estimation at reduced pilot overhead, and clock dissemination through ISLs, cooperative beaconing, or quantum-based references assisted by GNSS or terrestrial anchors.
\item \textit{Onboard resource co-optimization:} energy-efficient processors for real-time fusion, inference, and scheduling, with task offloading, compression-aware computing, and low-power hardware (FPGAs, ASICs) co-designed with modular real-time software.
\item \textit{Interference management and security:} interference-aware scheduling and beam coordination across spatial, temporal, and spectral dimensions, coordinated multi-point processing, AI-driven interference prediction, secure waveform generation, spoofing-resistant positioning, and authenticated sensing data chains.
\item \textit{Geometry, QoS-aware scheduling, and benchmarks:} geometry-aware beam management and handover for favorable GDOP and resolution, allocation of time, frequency, and compute by task criticality under service-level agreements, and common benchmarks and in-orbit demonstrators that allow the four configurations to be compared quantitatively.
\end{itemize}
\section{Cross-Configuration Comparison and Techno-Economic Assessment}
\label{sec:assessment}
\revi{rev1.4.2z}{This section closes the configuration-by-configuration treatment with two cross-cutting views: Section~\ref{sec:tradeoffs} compares, in a single table, the performance trade-offs reported for the four configurations, and Section~\ref{sec:techno_economic} assesses when the additional complexity of integration is economically justified.}

\subsection{Comparison of Performance Trade-Offs}
\label{sec:tradeoffs}
\revi{rev1.4.2}{Sections~\ref{sec:jcas}--\ref{sec:jcsap} have illustrated, configuration by configuration, what the recent satellite literature reports as achievable performance and at what cost for the other function. Table~\ref{tab:tradeoff_summary} brings this evidence together: for each configuration it lists the functional coupling and synergies, the dominant trade-off and the design knob that controls it, and the representative studies with their setups and reported operating points. The communication metrics involved are rate or spectral efficiency, outage, energy efficiency (EE), latency, and availability; the sensing metrics are the accuracy and resolution of the estimated parameters (usually the Cram\'{e}r--Rao bound, CRB, of angle, reflection coefficient, delay, or Doppler), the detection and false-alarm probabilities, and, for imaging, the SAINR; and the PNT metrics are the position, velocity, and time error bounds (CRB, position error bound and its average, APEB, or ranging CRLB) together with integrity, continuity, and availability. All of them depend on the same finite resources---transmit power, bandwidth, beams and spatial degrees of freedom, dwell or integration time, pilot overhead, and onboard computation---and on the satellite geometry, which fixes path loss, Doppler, and dilution of precision. The results were obtained under different assumptions (orbit, band, bandwidth, array size, number of users, channel models), so that only the trends within each study are directly comparable.}

\begin{table*}[!t]
\begin{revblock}{rev1.4.16}
\centering
\caption{Cross-configuration comparison of the four MFSS configurations: functional coupling and synergies, dominant trade-off and design knob, and representative quantitative evidence from the literature (numbers as reported in the cited studies; setups are not directly comparable).}
\label{tab:tradeoff_summary}
\scriptsize
\setlength{\tabcolsep}{3pt}
\begin{tabularx}{\textwidth}{L{0.9cm}L{3.9cm}L{3.6cm}X}
\toprule
\textbf{Config.} & \textbf{Functional coupling and principal synergies} & \textbf{Dominant trade-off and design knob} & \textbf{Representative evidence: study (figure), setup, reported operating points} \\
\midrule
\textbf{JCAS} &
Communications $\leftrightarrow$ sensing. Communication signals, pilots, beams, spectrum, and aperture reused for target or environmental sensing; sensing supports beam alignment, blockage/rain awareness, mobility prediction, and interference management; links disseminate sensing products and enable multistatic cooperation. &
Throughput and sensing accuracy compete for bandwidth, power, beams, dwell time, and processing; monostatic operation adds self-interference. Knob: beamforming weight between communication and sensing objectives, or a CRB threshold. &
Yin \emph{et al.} \cite{Yin2024LEOISAC} (Fig.~3): $N_t=8$ feeds, $K=16$ users, RSMA vs. SDMA; MFR vs. RCRB of angle, reflection coefficient, and Doppler---the RSMA frontier dominates SDMA over the whole plane (equal sensing accuracy at the sensing end, much higher MFR at the communication end). Liu \emph{et al.} \cite{Liu2024QuantizedLEOISAC} (Fig.~6(b)): $20$~GHz, $600$~km, $25$~MHz, $4\times4$ antennas, $5$ users, 2-bit DACs; max--min EE increases monotonically as the CRB threshold is relaxed; RSMA with radar sequence and SIC best. You \emph{et al.} \cite{You2022BeamSystems} (Figs.~3, 9): $20$~GHz, $800$~MHz, $1000$~km, up to $2304$ antennas, $16$ RF chains; $90\%$ detection at $P_{FA}=10^{-7}$ needs more power for $\zeta=0.9$ than for $\zeta=0.4$; EE grows with $\zeta$. \\
\midrule
\textbf{JCAP} &
Communications $\leftrightarrow$ PNT. Communication waveforms and infrastructure provide ranging, timing, Doppler, angular, and synchronization observables, with ephemeris, clock, integrity, and authentication data on the link; dense LEO geometry improves availability; PNT enables beam pointing, Doppler pre-compensation, synchronization, and handover prediction. &
Spectral efficiency vs. positioning accuracy: pilot, PRS, or overlay power improves ranging but reduces data capacity; accuracy governed by ranging bandwidth and geometry. Knob: data/navigation power split or pilot density. &
Gonzalez-Garrido \emph{et al.} \cite{GonzalezGarrido2025JointNavigation} (Fig.~14): LEO-600, $16\times16$ array, $7$ beams, S-band, DSSS over OFDM; FFR 30/30 needs $\rho\approx0.78$ for $\eta\ge1$~bit/s/Hz with $\approx3.5$~cm ranging CRLB, FR3 15/5 gives $\rho\approx0.54$ with $1.9$~cm; independent beams better than shared beams. You \emph{et al.} \cite{You2024IntegratedLocalization} (Fig.~6): $2$~GHz, $15.36$~MHz, $200$~km, $24\times24$ array, $36$ RF chains, $9$ users; $\approx85$~bit/s/Hz with APEB $<5$~m per user at $26$~dBW. De Gaudenzi \emph{et al.} \cite{DeGaudenzi2026Beyond5GPartII}: $1200$~km, $263$ satellites, $20\%$ power to the PRS overlay; $95$th-percentile horizontal error $21$--$38$~m (NB-IoT + NPRS) and $3.0$--$3.4$~m (NR + 5-MHz DL-PRS); a $10\%$ share costs $\approx7\%$ ranging accuracy. \\
\midrule
\textbf{JSAP} &
Sensing $\leftrightarrow$ PNT. PNT provides timing, orbit, attitude, and position for georeferencing, coherent sensing, and multi-platform fusion; a navigation-disciplined constellation is a ready-made multistatic sensor; sensing observables support navigation, integrity, and situational awareness; navigation signals give free illumination for reflectometry and occultation. &
Navigation signals are narrowband and low power, whereas high-resolution sensing needs bandwidth and received power; multipath helps sensing but degrades positioning. Knob: SAINR threshold imposed on the sensing function in the joint beamforming design. &
Wang \emph{et al.} \cite{Wang2026NavigationRemoteSensing} (Fig.~5): group of $K$ LEO satellites with $N$-element arrays, $M$ navigation users, MIMO-SAR at the central satellite; imaging quality improves with SAINR, and a stricter SAINR requirement raises the PVT error bound. Sachdeva \emph{et al.} \cite{Sachdeva2026NavIC} (Fig.~6, Table~I): NavIC L5, $1.023$~Mchip/s, RFSoC prototype; $\approx293$~m native ($\approx40$~m oversampled) range resolution; two targets $4$~km/$500$~Hz apart resolved at $-10$~dB SNR. \\
\midrule
\textbf{JCSAP} &
Communications $\leftrightarrow$ sensing $\leftrightarrow$ PNT. Maximum hardware and spectrum reuse; sensing and PNT improve resource allocation and beam management, communications support distributed sensing and correction dissemination, PNT provides the common space--time reference for coherent multi-satellite operation. &
Three-way trade-off among rate/reliability, sensing resolution/detection, and positioning/timing accuracy on one waveform and resource grid; highest validation complexity. Knobs: integration level and shared waveform parameters (e.g., subcarrier spacing). &
Sheng \emph{et al.} \cite{Sheng2025ICNR} (Fig.~4): $500$~km; $4500$/$250$/$250$ satellites; $250$/$25$/$25$~MHz; $20$/$20$/$80$~W; $1400$ vehicles; function- and signal-level integration improve all six metrics, with near-cm positioning and dm-level EO under signal-level integration (about one order of magnitude). Zhang \emph{et al.} \cite{Zhang2026DISLAC} (Figs.~2, 4): $600$~km, $2$~GHz, $60$~kHz subcarriers, $4$ cooperating satellites via ISLs; larger subcarrier spacing improves unambiguous velocity but degrades unambiguous range and resolution; decentralized ring/star beamforming approaches the centralized sum rate. \\
\bottomrule
\end{tabularx}
\end{revblock}
\end{table*}

\revi{rev1.2.32}{Table~\ref{tab:tradeoff_summary} shows that, although every configuration can be realized through cooperative, integrated, or joint payloads, the synergies and the dominant trade-offs differ: JCAS exploits the coupling between data transmission and environmental perception, JCAP leverages communication signals and infrastructure for positioning and timing while using PNT information to operate the communication system, JSAP combines sensing observables with accurate spatial and temporal referencing, and JCSAP extends these interactions to all three functions, offering the highest synergy at the expense of the greatest waveform, synchronization, resource-allocation, processing, and validation complexity. Moving from cooperative to integrated and joint implementations increases the reuse of information, spectrum, beams, timing references, processing, and signals across functions, but also makes the performance objectives more interdependent, so that the architecture should be selected according to the mission-specific balance among communication throughput and reliability, sensing accuracy and coverage, PNT accuracy and integrity, onboard resources, and implementation complexity rather than according to the achievable level of integration alone.}

\revi{rev1.4.17}{\emph{Cross-configuration observations.} Four conclusions emerge from Table~\ref{tab:tradeoff_summary} and from the configuration-specific results of Sections~\ref{sec:tradeoffs_jcas}, \ref{sec:tradeoffs_jcap}, \ref{sec:tradeoffs_jsap}, and \ref{sec:tradeoffs_jcsap}. First, in all configurations the trade-off is controlled by a scalar design knob (weighting coefficient, power split, pilot density, or constraint threshold), so that the same payload can be re-tasked between functions in orbit, which is what makes integrated and joint MFSS operationally flexible. Second, the price of the second function is small when spatial degrees of freedom are abundant (massive arrays, RSMA common streams, independent navigation beams) and grows sharply in overloaded or power-limited regimes, which is why the multiple-access scheme, the array size, and the DAC resolution are first-order parameters in the JCAS results, and the power split and beam model in the JCAP results. Third, satellite specifics decide which regime applies: the two-way path loss and the self-interference of monostatic sensing make JCAS power-limited, the ranging bandwidth and the geometry make JCAP accuracy-limited, and the narrowband, low-power nature of navigation signals makes JSAP resolution-limited. Fourth, all results are simulation- or prototype-based and were obtained under heterogeneous assumptions; a common benchmark (orbit, band, bandwidth, array, user and target sets, channel model) and, ultimately, in-orbit demonstrations are needed before the configurations can be ranked quantitatively, which motivates the benchmark- and testbed-oriented research directions of Sections~\ref{sec:jcas_directions}--\ref{sec:jcsap_directions}.}

\subsection{Techno-Economic Assessment}
\label{sec:techno_economic}
\revi{rev2.1.1}{The cost reduction promised by MFSS is often presented at an intuitive level, as a side effect of concentrating several functions on fewer platforms. MFSS economics, however, cannot be inferred from the payload count alone: the added complexity is not needed by every application, and it is borne by all the entities of the infrastructure, not only by the spacecraft. A meaningful comparison must therefore evaluate the incremental life-cycle cost of integration against the cost of delivering the same service portfolio with separate missions.} For an architecture $a$, a mission-level decision can be expressed as
\begin{equation}
\begin{aligned}
C_{\mathrm{LC}}(a)={}&C_{\mathrm{NRE}}(a)+N_a C_{\mathrm{rec}}(a)
+C_{\mathrm{launch}}(a)\\
&+C_{\mathrm{ground}}(a)+C_{\mathrm{ops}}(a)+C_{\mathrm{risk}}(a),
\end{aligned}
\label{eq:lifecycle}
\end{equation}
where non-recurring engineering and qualification, recurring spacecraft production for $N_a$ platforms, launch, ground-segment, operations, and risk-adjusted expected-loss costs are included over the mission lifetime. The economic benefit of MFSS is $\Delta C_{\mathrm{LC}}=C_{\mathrm{LC}}(\text{separate missions})-C_{\mathrm{LC}}(a)$, evaluated subject to equivalent coverage, capacity, sensing quality, PNT integrity, availability, and lifetime, which avoids crediting an integrated payload with savings obtained by silently relaxing a service requirement. Hardware reuse reduces recurring payload cost, mass, volume, harnessing, power conversion, thermal-control demand, launch mass, and the number of gateways or operations teams, and software-defined reconfiguration increases revenue utilization by reallocating time, bandwidth, beams, and compute among services; conversely, deeper coupling raises non-recurring engineering and qualification expenditure, requires wider-band and higher-linearity RF hardware, tighter calibration and synchronization, more capable processors, cross-domain verification, cybersecurity assurance, and specialized user terminals, and creates common-mode failures whose mitigation by redundancy can offset the apparent saving.

\begin{table*}[!t]
\begin{revblock}{rev2.1.2}
\centering
\caption{Techno-economic implications of MFSS integration: (a) by integration depth and (b) by entity of a JCSAP infrastructure, compared with separate single-function missions.}
\label{tab:techno_economic}
\scriptsize
\setlength{\tabcolsep}{3pt}
\begin{tabularx}{\textwidth}{L{2.1cm}XXX}
\toprule
\multicolumn{4}{l}{\textbf{(a) By integration depth}} \\
\textbf{Architecture} & \textbf{Potential savings} & \textbf{Incremental cost and complexity} & \textbf{Dominant economic risk / most credible use condition} \\
\midrule
Separate missions & Independent qualification, procurement, and failure domains; simple service certification. & Duplicated payload/bus resources, launches, gateways, operations, and spectrum coordination. & Underutilized assets and costly cross-mission data fusion / requirements, lifetimes, or owners are incompatible and failure independence is essential. \\
Cooperative MFSS & Shared bus, aperture, selected RF/compute resources, launch, and operations. & Interface engineering, scheduling, isolation, and multi-mission qualification. & One function monopolizes shared power, thermal, beam, or compute resources / services are co-located and can be scheduled with limited waveform changes. \\
Integrated MFSS & Higher asset utilization through common resource grids, calibration, and processing. & Dynamic orchestration, tighter synchronization, interference control, and upgraded processors. & Verification scope and operational coupling erode recurring savings / demand is variable and cross-service scheduling produces measurable utilization gains. \\
Joint MFSS/JCSAP & Maximum reuse of waveform, spectrum, aperture, RF chain, timing, and processing. & Highest waveform/RF co-design, algorithm, terminal, certification, security, and qualification cost. & Common-mode failure and redesign can outweigh launch/hardware savings / high-volume or high-value missions where quantified synergy exceeds added NRE and risk. \\
\midrule
\multicolumn{4}{l}{\textbf{(b) By entity of a JCSAP infrastructure}} \\
\textbf{Entity} & \textbf{Where the savings arise} & \textbf{Where the additional complexity and cost arise} & \textbf{Affected terms of (\ref{eq:lifecycle})} \\
\midrule
Space segment (payload and bus) & One aperture, RF chain, clock, processor, and power/thermal subsystem instead of three; fewer platforms for a fixed service portfolio; shared TT\&C. & Wider-band, higher-linearity, higher-power RF hardware; stable clocks and calibration for PNT; self-interference isolation for monostatic sensing; larger processors; multi-function flight software; redundancy against common-mode failure. & $C_{\mathrm{NRE}}$, $C_{\mathrm{rec}}$, $C_{\mathrm{risk}}$ \\
Launch and constellation & Fewer launches and orbital slots; lower debris and collision-avoidance burden. & Heavier and more power-hungry platforms; orbit and constellation design must satisfy sensing revisit, PNT geometry, and communication coverage simultaneously. & $C_{\mathrm{launch}}$, $C_{\mathrm{rec}}$ \\
Ground segment & Shared gateways, mission control, data processing, and mission planning. & Precise orbit and clock determination and integrity monitoring for PNT; sensing product processing and archiving; cross-service scheduling and conflict resolution; multi-service network management and cybersecurity. & $C_{\mathrm{ground}}$, $C_{\mathrm{ops}}$ \\
User segment & One terminal receives data, positioning, and sensing products; positioning available without a GNSS chipset in denied environments. & Multi-service receivers (joint synchronization, DSSS/PRS processing, echo or reflectometry processing); higher terminal cost and power for users who need only one service. & $C_{\mathrm{rec}}$ (terminals), adoption risk \\
Spectrum and regulation & One filing for a shared band; better spectrum utilization. & Regulatory recognition of dual-use emissions (active sensing in communication bands, EPFD limits); coordination with GNSS and EESS allocations; certification of PNT integrity delivered by a communication operator. & $C_{\mathrm{NRE}}$, schedule risk \\
Operations and business & Revenue from three services on one asset; in-orbit re-tasking as demand changes. & Service-level agreements for services with different lifetimes, availability, and liability regimes; operator competence in three domains; verification of cross-service performance before each reconfiguration. & $C_{\mathrm{ops}}$, $C_{\mathrm{risk}}$ \\
\bottomrule
\end{tabularx}
\end{revblock}
\end{table*}

\revi{rev2.1.4}{Table~\ref{tab:techno_economic} organizes these effects by integration depth (part (a)) and attributes them to the entities of a JCSAP infrastructure (part (b)). The entity view shows that the savings are concentrated in the space segment and in launch, whereas the additional complexity is distributed across the ground segment (orbit and clock determination, integrity monitoring, sensing product processing, cross-service scheduling), the user segment (multi-service receivers that cost more than single-service ones), and the regulatory and operational domains (dual-use spectrum, service-level agreements with different liability regimes, multi-domain operator competence). For a user community that needs only one service---a broadband subscriber, a weather agency, or a timing customer---a JCSAP infrastructure is therefore not automatically cheaper, because the shared cost of the other functions is carried by every terminal and every ground-segment process; this is why the complexity of JCSAP is not needed for all applications.}

\revi{rev2.1.5}{The preferred architecture is consequently mission dependent. A cooperative design is often the economically robust entry point, because it captures bus and selected payload reuse while preserving functional independence. Integrated or joint designs become attractive when (i) the shared resources have high duty-cycle complementarity (e.g., sensing during communication off-peak periods or over oceans), (ii) the same signal or observation measurably improves more than one service, as in the results of Table~\ref{tab:tradeoff_summary}, (iii) production volume amortizes the additional non-recurring engineering, as in mega-constellations, and (iv) common-mode risk remains within service-availability and PNT-integrity constraints. A dedicated payload remains preferable when only one function is needed, when regulatory or safety certification requires isolation, when service lifetimes differ markedly, or when the required redundancy would duplicate most of the shared hardware. In terms of (\ref{eq:lifecycle}), JCSAP pays off when the avoided duplication in $C_{\mathrm{rec}}$, $C_{\mathrm{launch}}$, and $C_{\mathrm{ground}}$, together with the revenue enabled by in-orbit re-tasking, exceeds the growth of $C_{\mathrm{NRE}}$, of the terminal cost, and of $C_{\mathrm{risk}}$ across all entities. Thus, JCSAP is the upper end of a design space, not the default economic optimum.}

\section{Conclusions} \label{sezione_9}
\revi{revAE.9}{This survey has examined the convergence of communications, sensing, and PNT into multi-functional payloads for 6G satellite systems. After consolidating the fundamentals of the three single-functional system classes, it introduced a two-dimensional taxonomy that separates the functional configuration (JCAS, JCAP, JSAP, and JCSAP) from the integration depth (cooperative, integrated, and joint payloads), described the three integration levels, and devoted a dedicated section to each configuration, covering its architecture, classification, state of the art, advantages, achievable performance, key challenges, and research directions. Representative results from the literature, compared side by side in a single table, showed what each configuration can achieve today and at what cost for the other functions: RSMA-based joint beamforming keeps the communication rate high while sensing a target, a DSSS overlay on the 5G NTN downlink delivers centimeter-level ranging bounds for a fraction of the beam power, co-designed navigation/remote-sensing frames trade the navigation error bound against the sensing SAINR, and three-function integration improves all service metrics in a vehicular case study, while none of these results has yet been validated in flight. The comparison with terrestrial multi-functional networks and the life-cycle techno-economic framework showed that deeper integration improves utilization and functional synergy but is not automatically the technical or economic optimum: the added engineering, qualification, terminal, and common-mode-risk costs are borne by all entities of the infrastructure, and a cooperative payload is often the robust entry point. The identified challenges and research directions---waveform co-design, synchronization, interference control, resource management, onboard processing, security, and common benchmarks---are intended to guide the development of resilient, scalable, and sustainable MFSS for the 6G space era.}
{\footnotesize
\def\baselinestretch{0.85}
}

\end{document}